\documentclass{aa}  
\usepackage{graphicx}
\usepackage{url}
\usepackage[rightcaption,wide]{sidecap}
\usepackage[varg]{txfonts}
\usepackage{natbib,twoopt}
\usepackage{flushend}\usepackage[colorlinks,linkcolor=blue,urlcolor=blue,citecolor=black]{hyperref} 

\bibpunct{(}{)}{;}{a}{}{,}             

\makeatletter
  \newcommandtwoopt{\citeads}[3][][]{\href{http://ui.adsabs.harvard.edu/abs/#3}%
    {\def\hyper@linkstart##1##2{}%
     \let\hyper@linkend\@empty\citealp[#1][#2]{#3}}}
  \newcommandtwoopt{\citepads}[3][][]{\href{http://ui.adsabs.harvard.edu/abs/#3}%
    {\def\hyper@linkstart##1##2{}%
     \let\hyper@linkend\@empty\citep[#1][#2]{#3}}}
  \newcommandtwoopt{\citetads}[3][][]{\href{http://ui.adsabs.harvard.edu/abs/#3}%
    {\def\hyper@linkstart##1##2{}%
     \let\hyper@linkend\@empty\citet[#1][#2]{#3}}}
   \newcommandtwoopt{\citeauthorads}[3][][]%
    {\href{http://ui.adsabs.harvard.edu/abs/#3}
    {\def\hyper@linkstart##1##2{}%
     \let\hyper@linkend\@empty\citeauthor[#1][#2]{#3}}}
  \newcommandtwoopt{\citeyearads}[3][][]%
    {\href{http://ui.adsabs.harvard.edu/abs/#3}
    {\def\hyper@linkstart##1##2{}%
     \let\hyper@linkend\@empty\citeyear[#1][#2]{#3}}}
  \renewcommand*\aa@pageof{, page \thepage{} of \pageref*{LastPage}} 
\makeatother
\usepackage[varg]{txfonts}

\begin{document}

   \title{Archives of Photographic PLates for Astronomical USE (APPLAUSE)}

   \subtitle{Digitisation of astronomical plates and their integration into the  International Virtual Observatory 
}

   \author{Harry Enke 
          \inst{1}
          \and
          Taavi Tuvikene
          \inst{1,4}
          \and
          Detlef Groote
          \inst{2}
          \and
          Heinz Edelmann
          \inst{3}
          \and 
          Ulrich Heber\thanks{corresponding author}
          \inst{3} 
          }
\institute{
Leibniz Institute for Astrophysics Potsdam (AIP),
An der Sternwarte 16, 14482 Potsdam, Germany
\and
Hamburger Sternwarte, Universit\"at Hamburg, Gojenbergsweg 112, 21029 Hamburg, Germany
\and
Dr. Karl Remeis Observatory \& ECAP, Friedrich-Alexander University Erlangen-N\"{u}rnberg,
Sternwartstr. 7, 96049 Bamberg, Germany\\
\email{Ulrich.heber@fau.de}
\and 
Tartu Observatory, University of Tartu, Observatooriumi 1, 61602 T\~{o}ravere, Estonia
}

\date{}

 
  \abstract
   {}
   {The Archives of Photographic PLates for Astronomical USE (APPLAUSE)
   project is aimed at digitising astronomical photographic plates from three major German plate collections, making them accessible through integration into the International Virtual Observatory (IVO).}
   {Photographic plates and related materials (logbooks, envelopes, etc.) were scanned with commercial flatbed scanners. Astrometric and photometric calibrations were carried out with the developed PyPlate software, using \textit{Gaia} EDR3 data as a reference. The APPLAUSE data publication complies with IVO standards.}
  {The latest data release contains images and metadata from 27 plate collections from the partner institutes in Hamburg, Bamberg, and Potsdam, along with digitised archives provided by Tautenburg, Tartu, and Vatican
observatories. Altogether, over two billion calibrated measurements extracted from about 70,000 direct photographic plates can readily be used to create long-term light curves. 
For instance, we constructed the historic light curve 
of the enigmatic dipping star KIC\,8462852. We found no evidence of previously assumed variations on timescales of decades in our light curve. Potential uses of APPLAUSE images for transient sources can be appreciated by following the development of the nova shell of GK\,Per (1901) over time and the change in brightness of two extragalactic supernovae. The database holds about 10,000 spectral plates. We made use of objective prism plates to follow the temporal changes of Nova DN Gem through 1912 and 1913, highlighting an outburst in early 1913.
}
   {}

\titlerunning{Archives of Photographic PLates for Astronomical USE (APPLAUSE)}
\authorrunning{Enke et al. }
   \maketitle
%

\section{Introduction}

Over more than a century photographic plates (mostly glass plates with photographic emulsions) were used to record information on celestial objects. They were utilised both as detectors of light collected by astronomical cameras and telescopes as well as data storage devices. Applications range from direct imaging, via photometry using transmission filters, to recording the spectra of celestial objects.
Wide-field cameras were used to survey sections of the sky for moving objects such as asteroids and comets as well as for light variable and transient sources. Objective prisms were used to record low-resolution spectra of many stars in wide fields mostly with Schmidt cameras and telescopes. On small plate scales, photographic imaging is important to investigate the structure of extended celestial objects and measure accurate positions. High-resolution spectra of single objects have also been recorded on photographic plates.

 The variability information is stored in many plate archives around the world, but many are inaccessible to the scientific community.
The largest plate collection in the world at Harvard College Observatory hosts half a million astronomical plates taken over a time span of more than 100 years.
In Germany, well-kept archives still exist at many astronomical institutions including the Leibniz Institut für Astrophysik Potsdam, the Hamburg Observatory, and the Dr. Remeis Observatory in Bamberg. Other major ones (Sonneberg, Heidelberg, Tautenburg, and Jena) have already been digitised at least partly, but are not publicly available or  merely provide the images of the plates with some metadata.

Compared to modern CCD detectors, photographic plates are very large: typical sizes range from 16$\times$16\,cm$^2$ to 30$\times$30\,cm$^2$ for Schmidt plates and they contain tens or even hundreds of thousands of stellar images. This enormous amount of information can only be harvested by digitisation. Scanning such plates at the resolution of 2400\,PPI yields digital images with up to 800 megapixels. To make use of this information, metadata from plate envelopes and logbooks needs to be available and incorporated into the database. Recent status reports about international digitisation initiatives have been given by \citet{2018AN....339..408H,lswst2019_Hudec,2019AN....340..690H}.

The information contained on photographic plates is an important tool to study the long-term behaviour and variability of celestial objects, from asteroids to quasars. 
Hence, the unique scientific value lies in long-term time-domain astronomy, in particular, for irregular and very long-period (decades) phenomena and optical transients \citepads[see e.g.][]{2002ESASP.506..841F,2013AcPol..53c..27H,2015ApJ...812..133S,2016MNRAS.460.1233S,2016ApJ...822L..34S,2023MNRAS.524.3146S,2016ApJ...825...73H,2017AN....338..103W,
2019AstBu..74..490A,2022MNRAS.516.2095Y}.

In addition, many plate archives host historical spectral plates from 
objective prism surveys 
\citepads[e.g.][]{1988ASPC....2..143E,1983Afz....19..639M,1983ApJS...51..171P,1971AJ.....76..338S}, as well as  high-resolution slit spectra. In addition, there are archives with historic observations recorded on photographic plates with images of the solar disk, for instance, from the Einstein tower in Potsdam \citepads{2020AN....341..575P}.

\section{Access to plate archives by digitisation} 

The construction of long-term light curves through historical times is a key science driver for the digitisation of photographic plates. Photometric measurements from photographic plates from archives such as at Harvard College Observatory \citepads{2010AJ....140.1062L}, the Byurakan astronomical Observatory \citepads{2019ASPC..520..117M}, Sonneberg Observatory \citepads{2009chao.conf..311K}, and the APPLAUSE partners' archives in Bamberg, Hamburg, and Potsdam can provide time coverage for decades -- up to more than hundred years for studies of stellar variability.

Because of the sheer size of the data sets, data handling, and processing still is a challenge. The raw data need to be processed, cleaned from artifacts and calibrated both astrometrically as well as photometrically. The multitude of astronomical applications, however, implies that there is no silver bullet for the calibration problem of the data sets.

Internationally, strong efforts have been undertaken; for example, in Great Britain, the Royal Observatory Edinburgh (ROE) has operated a specialised machine to digitize photographic plates since 1967. It was upgraded to become the COSMOS (COordinates, Sizes, Magnitudes, Orientations and Shapes) machine and replaced by a new SuperCOSMOS machine in 1993. These machines were used to digitize the major large Schmidt surveys and a data archive was developed which proved to be a very important research resource, easily accessible through the SuperCOSMOS Science Archive maintained at ROE \citepads{2001MNRAS.326.1315H,2001MNRAS.326.1295H,2001MNRAS.326.1279H}\footnote{For high precision astrometry, dedicated measurement machines have been developed in addition to the COSMOS and DASCH machines, e.g. the StarScan Plate Measuring Machine at U.S. Naval Observatory  \citepads{2008PASP..120..644Z}. More recently, additional high precision measuring machines \citepads[e.g. DAMIAN and NAROO, ][]{2011MNRAS.415..701R,2021A&A...652A...3R} have been installed in Europe for special astrometric applications such as solar system objects \citepads{2012ASPC..461..315D,2023A&A...680A..41P}. The high astrometric precision is not required for the APPLAUSE project, because we do not aim at proper motion measurements, but make use of \textit{Gaia} proper motions in the process of cross matching.}. The pioneering project DASCH (\textrm{Digital Access to a Sky Century at Harvard}) has been running at the Harvard College Observatory for two decades \citepads{2009ASPC..410..101G,2010AJ....140.1062L,2012IAUS..285...29G,2013PASP..125..857T}, making the calibrated data from its archive available to the public.

The final goal is to digitize plate archives, extract the huge amount of information and make the data publicly available. In Germany, the APPLAUSE consortium has been formed by astronomers  
at the Leibniz Institute for Astrophysics Potsdam (AIP), Hamburg Observatory, and Dr. Remeis-Observatory in Bamberg
to digitise their own archives, include more from other observatories where possible, and integrate them into a database that complies with the standards of the international Virtual Observatory. This database holds digital images from direct and spectral photographic plates as well as from observations of the solar disk (Einstein tower).

 The Harvard College Observatory built an expensive scanner \citepads{2006SPIE.6312E..17S}  to cope with the huge amount of material, while the German team uses commercial flatbed scanners, which have been proven to deliver accurate enough resolution for scans to process the digitised images further at much lower costs (see Sect. \ref{subsect:sources}). Such scanners are used successfully by many similar projects \citepads[e.g. ][]{2004BaltA..13..665B,2016MNRAS.457.2900Y,2017RAA....17...28Y,2017arXiv171204672S,2020SoSyR..54..344K,2023PASJ...75..811J,UBAI23}.  Subsequent post-processing of the digitised plates with modern software will deliver crucial catalogued information. 

\paragraph{Cultural Heritage aspects:}

\begin{SCfigure*}
 \includegraphics{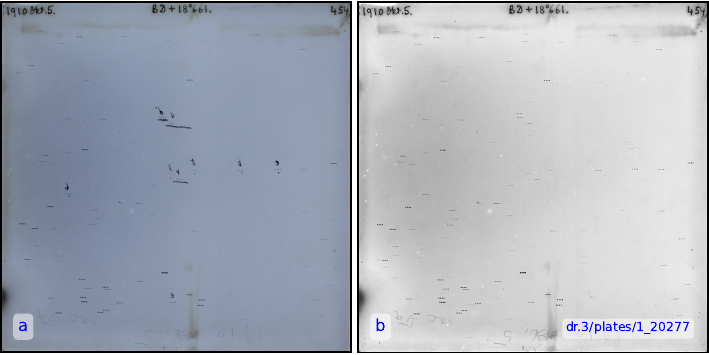}
 \hspace*{0.01\textwidth} \caption{Digitisation of the Potsdam Zeiss-Triplet plate \#454 from October 5, 1910: a)~preview image with original annotations, b)~full scan of the cleaned plate.
 The plate has four 3-minute exposures of the field around the star BD+18 661. The observer was Ejnar~Hertzsprung. The APPLAUSE Identifier (AID) is shown in the lower right of panel~\textit{b}, allowing for construction of the DOI of the digitised plate in the following way: \texttt{doi:10.17876/plate/\{AID\}}.}
 \label{fig:digitization_plates}
\end{SCfigure*}

Besides their scientific value, historical astronomical archives of photographic plates are considered a part of cultural heritage to be preserved. For example, the Markarian Survey at the Byurakan Astronomical Observatory and its digitised version (DFBS) were included in UNESCO Documentary Heritage ''Memory of the World'' International Register in 2011 as one of the rare heritage items from science \citepads{2021CoBAO..68..390M}. Thus, we have chosen to collaborate with the Europeana,  a virtual library created and maintained by the European Union, in order to make the cultural heritage of Europe more accessible to the public.
The Europeana collects digitised objects from archives, libraries, museums and other sources from all over Europe  
and makes them available to the public. All these objects are collected by the Europeana under the notion of being (digital) Cultural Heritage Objects (CHO). Since all our published plates document a period of scientific work in astronomy, we are committed to making all plate images, logbooks, and envelopes available to Europeana. The astronomers' annotations on the original plates are also preserved on low-resolution scans or photos that were done before cleaning the plates for the final high-resolution scan (see Fig. \ref{fig:digitization_plates}). This required implementing the data model of Europeana (EDM) and supporting with the APPLAUSE website the OAI-PMH protocol for harvesting our metadata on the published objects.

\paragraph{Science-ready data:}
To analyse digital images of direct plates, a data reduction pipeline \citepads[PyPlate software,][]{2014aspl.conf..127T,lswst2019_tuvikene} has been devised. A Virtual Observatory-compliant data publication (APPLAUSE\footnote{\url{https://www.plate-archive.org/applause/}}) has been set up to integrate and publish the calibrated data obtained.

\section{Plate archives of the APPLAUSE collaboration}

In response to the recommendation of the advisory board (Wissenschaftsrat) of the German federal government to secure  scientifically important collections via digitisation in 2009, the German Science Foundation (DFG) issued a call for proposals (Erschlie\ss ung und Digitalisierung von objektbezogenen wissenschaftlichen Sammlungen). In this context, a joint application by the AIP (Potsdam), Hamburg, and Bamberg observatories was successful and funded for six years
\citep[see also][]{lswst2019_heber,lswst2019_enke}. 
The telescopes, instruments, and plates used for the observation differ enormously in size, field coverage, image quality and limiting magnitudes (see Sect. \ref{subsect:sources} and \ref{subsect:photometry}). Small cameras are used for wide-field sky patrols (e.g. the Bamberg surveys were mostly carried out with camera objective lenses of 10-cm diameter on 16$\times$16-cm plates) providing shallow observations, while larger telescopes such as the Schmidt telescopes at Hamburg Observatory (80-cm aperture) or the Tautenburg Schmidt (138-cm aperture) provide deep images of high quality. Larger telescopes were also equipped with objective prisms as well as spectrographs to secure optical spectra of star fields (see Sect. \ref{sect:spectra}) or target specific objects (see Sect. \ref{sect:nova}). Some of the observations were taken in the context of international collaborations such as the Carte du Ciel (see Sect. \ref{sect:cdc}) and Kapteyn's selected areas projects (see Sect. \ref{sect:kapteyn}). Their images of originally annotated (uncleaned) plates contain valuable information for historians and are available through APPLAUSE.
No less than 27 archives have been published  to date by the APPLAUSE project. Given their inhomogeneity, it is useful to describe the plate collections and their historic  origin in some detail (see Sections \ref{sect:aip_archives} to \ref{sect:vo}). Table \ref{dr4archives} lists all digitised archives in the APPLAUSE database as of data release 4 (DR4).

\begin{table*}
\caption{Digitised plate archives in APPLAUSE DR4}
\centering
\begin{tabular}{ r  l  l  c r  r r  c }
\hline\hline
Archive   &Archive name  & Institute & Period & Plates & Scans & Processed & Pixel scale\\
ID        & (instrument) &           &        &        &       &      scans & (arcsec)\\
\hline
1       & Zeiss Triplet (P) &    AIP    & 1909--1938    &4921   &4915 & 2097    & $1.46 \ldots 12.5$\\                                    
2       & Carte du Ciel  (P) &   AIP    & 1893--1924 &  979 &   953     & 0       & ---\\                                 
3       & Great Schmidt Camera  (P)  &   AIP &   1952--1970 &   508 &   508     & 413 &    0.38, 1.27\\                                           
4       & Small Schmidt Camera  (P) & AIP  &     1949--1967 & 113       & 101     & 14    & 2.93\\                                                
5       & Ross Camera  (P) &     AIP  &  1956--1956 &   64 &    62 &    59      & 2.37\\                                          
6       & Einstein Turm Solar Plates$^a$ (P) &   AIP  &  1943--1991     & 3613    & 3613  & 3613  & ---\\
101 &  Lippert-Astrograph (HH)  &  HS  &  1912--1972  & 8750 & 13764 & 11289 &  $0.63 \ldots 1.44$\\
102 &  Grosser Schmidt-Spiegel (HH)  &  HS  &  1954--1975  & 5323 & 7463 & 5146  &  0.91\\
103 &  1m-Spiegelteleskop (HH)  &  HS  &  1911--1972  & 7643 & 7538 & 5409  &  0.74\\
104 &  HSS (Calar Alto)  &  HS  &  1980--1998  & 3255 & 4388 & 2644 &  0.91\\
105 &  AG-Teleskop (Bonn)  & HS  &  1926--1958  & 754 & 1454 & 1309 &   1.06\\
106 &  AG-Teleskop (HH)  & HS  &  1927--1973  & 3529 & 7019 & 6848 &   1.06\\
107 &  Doppel-Reflektor (HH)  &  HS  &  1934--1957  & 1642 & 2055 & 1897  &  0.73\\
108 &  ESO telescopes (La Silla)  &  HS  &  1974--1998  & 2099 & 2713 & 918 &  $0.18 \ldots 0.71$\\
109 &  Grosser Refraktor (HH)  &  HS  &  1914--1985  & 2602 & 2757 & 938 &   $0.24 \ldots 15.6$\\
110 &  Kleiner Schmidt-Spiegel II (HH)  &  HS  &  1952--1979  & 1566 & 1152 & 852 &   3.47\\
111 &  Schmidtsches Spiegelteleskop (HH)  &  HS  &  1931--1968  & 1746 & 1985 & 933 &  3.51\\
202 &  Metcalf Telescope (South Africa)  &  DROB  &  1963--1972  & 2718 & 2705 & 2701 &  1.77\\
203 &  BSSP (South Africa)  &  DROB  &  1963--1970  & 12565 & 12513 & 12503 &  3.57\\
204 &  BSSP (New Zealand)  &  DROB  &  1967--1976  & 5266 & 5243 & 5242 &  3.58\\
205 &  BSSP (Argentina)  &  DROB  &  1969--1972  & 870 & 824 & 823 &  3.52\\
206 &  Zeiss Objective (South Africa)  &  DROB  &  1964--1970  & 694 & 690 & 676 &  8.66\\
207 &  Ross B Camera (South Africa)  &  DROB  &  1964--1966  & 698 & 696 & 694 & 4.17\\
208 &  BNSP Astrograph  &  DROB  &  1912--1968  & 18865 & 17036 & 15993 &   $1.62 \ldots 17.1$\\
301 &  Tartu Old Observatory  &  TO  &  1947--1988  & 2275 & 2275 & 2227 &   $3.65 \ldots 10.8$\\
401 &  Tautenburg Schmidt Telescope  &  TLS  &  1960--1998  & 4101 & 20505 & 19192 &   0.51\\
501 &  Zeiss Double Astrograph  &  VO  &  1951--1999  & 544 & 544 & 532 &  1.62\\
\hline
\end{tabular}

\tablefoot{
The abbreviations denote:
AIP~= Leibniz Institute for Astrophysics, Potsdam; 
HS~= Hamburg Observatory;
DROB~= Dr. Remeis Observatory, Bamberg;
TO~= Tartu Observatory, University of Tartu, Estonia;
TLS~= Thuringian State Observatory;
VO~= Vatican Observatory;
BSSP~= Bamberg Southern Sky Patrol;
BNSP~= Bamberg Northern Sky Patrol;
HSS~= Hamburger Schmidt-Spiegel.\\
\tablefoottext{a}{Separate data release DR3s.}
}
\label{dr4archives}
\end{table*}

\begin{table}
\caption{Major digitised spectral archives from the Hamburg Observatory available in the APPLAUSE DR4. The
number of these plates are given for both objective-prism plates and single spectra as well as the period of time
in which they were observed.}
\label{hamburg_stock} 
\centering
\begin{tabular}{c l  r  r  c   
}
\hline\hline
Archive ID & Telescope  
& Prism & Single & Period \\
                     &       ID       
                     & spectra & spectra &  \\
\hline
101 & LA 
& 2067 &   & 1912--56 \\
111 & SS 
& 343  &   & 1932--54 \\
103 & 1m 
&  & 1990   & 1947--72 \\
110 & KS 
& 225  &   & 1954--68 \\
102 &  GS   
& 2139 &    & 1955--75 \\
104 & HS         
& 1720 &   & 1980--98 \\
108 & ES         
& 738 & 398  & 1974--98 \\
&Total  
& 7232 & 2388 & 1912--98 \\
\hline
\end{tabular}

\tablefoot{
Abbreviations used:
LA = Lippert-Astrograph; 1m~= 1\,m – Spectroscope; SS~= Original Schmidt'sches Spiegelteleskop; KS~= Kleiner Schmidtspiegel II; GS~= Gro\ss er Schmidtspiegel (80\,cm) at Hamburg Observatory, HS~= same Schmidt telescope at Calar Alto; ES~= ESO Schmidt telescope at ESO (La Silla).
}
\end{table}

\subsection{Leibniz Institute for Astrophysics Potsdam  (AIP)}\label{sect:aip_archives}

The collection at the Leibniz Institute for Astrophysics (AIP) holds  photographic plates from the beginning of their use as the medium for optical astronomy up to early 1970s, spanning one century of observational heritage and data. The collection includes plates exposed by astronomers such as Karl Schwarzschild or Ejnar Hertzsprung. A substantial part of the Carte du Ciel is also in the collection.
For instance, the AIP plate collection includes the Potsdam survey of the Northern stars of the Bonner Durchmusterung. New observational techniques, such as the application of objective prisms, observations of dark nebulae, and eclipsing binary stars has also taken place at Potsdam.

The earliest optical observations with different instruments at the Astrophysical Observatory Potsdam (AOP) at Telegrafenberg date back to the late 1870s.
Shortly after the foundation of the AOP in 1874 Oswald Lohse built the camera for the early 0.3-m Gro\ss e Refraktor and used it to experiment with several dry emulsion plates
for imaging celestial objects such as the Orion nebula, open, and globular clusters.
His collection taken between 1879 and 1889 consisted of 217 plates, of which 67 survived\footnote{They are not yet included in APPLAUSE}  the world wars in good condition \citepads{1999AN....320...63T}. 
Lohse's plates form the oldest collection of all plates of the team's archives \citepads{2004JHA....35..447S}.

The idea to construct a huge astronomical telescope (in those days) had already been discussed shortly after the founding of the observatory. In 1889 a double refractor, with apertures of 32.5\,cm and 23.5\,cm and a focal length of 3.4\,m, had just been mounted in the dome to the west of the main observatory building. It was used mainly for cartography of northern-hemisphere stars as part of the observatory’s contribution to the Carte du Ciel (CdC). 

\subsubsection{The Potsdam Carte du Ciel (CdC) plate archive}\label{sect:cdc}

The Carte du Ciel was the first international survey in astronomy using photographic plates. It had been planned by the Paris congress of astronomers in 1887 \citepads{2000A&G....41e..16J}. 
Other parts of the CdC plates have already been digitised, for instance, the
Bordeaux zone \citepads{2006A&A...449..435R}, Toulouse zone \citepads{2003A&A...402..395L}, Helsinki zone \citepads{2018A&A...616A.185L,2023A&A...671A..16L}, and San Fernando zone \citepads{2007A&A...471.1077V,2010A&A...509A..62V,2014RMxAC..43...59A}. Hence, the Potsdam CdC data, available through APPLAUSE, add towards a full digitisaton of the CdC. 
The Potsdam CdC zone (+32$^\circ$ to +39$^\circ$) was divided into 1232 areas and about 2200 plates were obtained within the framework of the CdC project. However, only 977 plates (45\% of all) have survived at AIP, the others got lost during the Second World War. The plates for the first epoch measurements had been obtained during the period May 1893 ~to February 1900. The plates for the second epoch (August 1913 to February 1924) were taken in two time intervals from 1913 August till 1914 July, and from 1916 February to 1924 February \citepads{2009AN....330..878T}.

\subsubsection{Other Potsdam plate archives}
In addition to the CdC archive there are nine further archives from AOP 
\citepads{2005PASRB...5..309T,2006IAUSS...6E...9B}.
and one from the Berlin-Babelsberg observatory (the Toepfer 40-cm telescope). From AIP we have 9175 plates from 6 archives in APPLAUSE (see also Sect. \ref{sect:kapteyn}), and there is a considerable amount of mainly spectral plates ($\sim$5000) which could not be included. Many more plates of the AOP (from 1969 Zentralinstitut für Astrophysik (ZIAP) of the Academy of Sciences of the GDR) are in Sonneberg. Also, due to former custom of scientific exchange, many plates are at other institutes, for instance, at Leiden Observatory. 
\subsection{Hamburg Observatory}\label{sect:hamburg_archives}

At Hamburg Observatory photographic plates have been in use throughout the twentieth
century. In total there are more than 45,000 records available obtained with a variety
of telescopes, not all of which were taken in Hamburg\footnote{Additionally, Hamburg Observatory provides its own web pages accessible
under \url{(https://plate-archive.hs.uni-hamburg.de/index.php/en/} presenting all this material as well as other historic material such as photos, documents, and constructional drawings, including annual reports by the Hamburg Observatory  (in German) from 1877 to 1949 which are not available elsewhere. Also a complete Bernhard Schmidt Archive is provided. This material, altogether, is complete up to 90\% and provides a unique insight into the technical and observational development in astronomy, especially concerning the upcoming astrophysics in the 20$^{th}$ century.
An overview of the history of the Hamburg observatory is provided by \citetads{2004JBAA..114...78A}. The Hamburg part of the APPLAUSE project was already described by \citeads{2014aspl.conf...53G}, \citet{groote2014} and \citet{groote2018}.
A lot of additional information is provided on the web site
\url{https://www.plate-archive.org/cms/info/hs-history/}.}.
The Hamburg plate collection also includes several spectroscopic archives, as detailed in Table~\ref{hamburg_stock}. 

Most of the telescopes listed in Table~\ref{hamburg_stock} with the exception
of the ESO and Vatican  telescopes are or were located on
the premises of Hamburg Observatory; 
the Lippert-Astrograph, the 1-m telescope, and the ''Gro\ss er Refraktor'' are still usable.
The 80-cm Schmidt telescope was operated in Hamburg from 1955--1975, whereupon
it was moved to Calar Alto, Spain, starting operation in 1980 \citepads{1984ASSL..110..203B}, and was used for the project 
''Hamburg Quasar Survey'' between 1984--1998 \citepads[HS, ][]{1988ASPC....2..143E,1995A&AS..111..195H}. Objective prism plates for the Hamburg ESO survey \citepads[HES, ][]{1996A&AS..115..227W} were taken with the 1 m ESO Schmidt telescope. The plates used for the
Bonn contribution towards the AGK catalogue were taken at the old observatory
in Bonn.

\paragraph{Kapteyn's selected areas}\label{sect:kapteyn}

To investigate the structure of the Milky Way, Kapteyn in 1906 started an international collaboration to observe 206 carefully selected areas \citep{1963ASPL....9...89L}, which are still used as  reference areas in modern astronomy \citepads{2010ApJ...725.2290C}. Many observatories across the world joined this effort, including the Hamburg and Potsdam observatories. The APPLAUSE database holds
about 1000 direct plates and 700 objective-prism plates from the Hamburg archives and 225 direct plates from Potsdam archives. 

\subsection{Archives of the Dr. Remeis-Observatory in Bamberg}\label{sect:bamberg_archives}

Photographic observations at the Dr. Remeis Observatory started in 1913 \citepads[see][]{1939VeBam...4....1Z}. Routine surveillance of the Northern Sky was initiated in the mid-1920s.
The Northern Sky patrol was carried out at the Bamberg site from 1926 to the early 1960s with a break during the Second World War.
It started with the so-called ”Felderplan”  (1926--1939) in collaboration with the observatories at Sonneberg \citep[see][]{lswst2019_kroll} and Berlin-Babelsberg. All stations were equipped with the same cameras of 13.5\,cm aperture, providing quite homogeneous data sets.
From the 1950s
the observing efficiency was greatly enhanced by constructing a mount to carry 
multiple cameras (up to 10) adjusted to observe adjacent fields on the sky. 
Details are given in \citetads{2006vopc.conf..109T} and \citetads{2018A&AT...30..467T}.

Another survey targeted the Southern Sky and was carried out at the Boyden (South Africa), Mount St.~John (New Zealand), and San Miguel (Argentina) observatories, from 1962 to 1976. 
This collection of plates is of great scientific importance because the Southern Sky was not routinely monitored during the 1960s from elsewhere \citepads[see][for details]{2018A&AT...30..467T}; in addition, the Harvard College Observatory had halted their monitoring programme for about that decade (i.e. the so-called Menzel gap\footnote{\url{https://hco.cfa.harvard.edu/about/}}). 

All Southern stations were equipped with identical Aero Ektar cameras (aperture 10\,cm). 
The major Southern station operated by Bamberg staff was set up at Boyden Observatory,  equipped with ten cameras on a single mount, 
while four cameras were operated from 1967 to 1976 at Mount John station 
and six at San Miguel from 1969--1972. 
Harvard instrumentations at Boyden Observatory such as a ROSS\,B and Metcalf 
were also used. Details can be found in \citetads{2005PASRB...5..303T}. Because identical cameras were used at the three stations, the corresponding archives (IDs 203, 204, and 205) provide rather homogeneous data sets.

A total of 41,676 plates (22,811 of the Southern Sky and 18,865 of the Northern Sky) were found in the Bamberg plate collection.  Only 0.8\% of the Southern plates found were of insufficient quality for digital processing, while 15\% of the Northern plates turned out not useful for processing. 
The sky coverage is indicated by the Mollweide diagrams shown in the appendix (see Fig. \ref{fig:mollweide_appendix}).

\subsection{Other archives}
Over the course of our digitisation project, the Tartu, Tautenburg, and Vatican observatories asked us to include their already scanned plate data into our database. 

\subsubsection{Digitised plates from the Tartu Observatory}

The Tartu archive  holds about 4,000 photographic direct-image plates taken between 1950 and 1987 with the Petzval
astrograph at the Tartu Old Observatory. 
The plate sizes are about 120 $\times$ 90
mm$^2$. 
Tartu's project to digitise those plates started in 2005 using an Epson
Expression 10\,000\,XL Pro flatbed scanner \citepads{2007ASPC..376..201A}. 

\subsubsection{Digitised plates from the Karl Schwarzschild Observatory at Thuringian State Observatory}

The Karl-Schwarzschild-Observatorium  was founded in 1960 as an affiliated institute of the former Deutsche Akademie der Wissenschaften zu Berlin and named in honour of the famous astronomer and physicist Karl Schwarzschild (1873-1916).
The Tautenburg archive \citepads{1994IAUS..161..367Z,1999AcHA....6..223M} hosts 9,213 plates mostly taken between 1960 and 1989 covering various fields in the Northern Sky
with the 2-m Schmidt telescope (see Mollweide diagram in Fig. \ref{fig:mollweide_appendix}) . The large plates (24$\times$24 cm$^2$) have been scanned with a special custom-made scanner \citepads{2008A&A...477...67H}.

\subsubsection{Digitised plates from the Vatican Observatory}\label{sect:vo}

The Vatican Observatory (Specula Vatican) took part in the Carte du Ciel project, securing 540 plates starting from 1894.
Since 1930, the Vatican Observatory has been located at the Papal Palace of Castel Gandolfo. Two Zeiss telescopes started operation in 1935: The Double Astrograph hosts a four-lens objective with an aperture of 40~cm, and a 60-cm reflector. The former was used to image the sky on photographic plates up to a size of 30$\times$30 cm$^2$ at an image scale 1.42\arcmin/mm, while the latter was restricted to smaller plates. The Double Astrograph was operated until 1974. In addition, a 65-cm Schmidt telescope was installed and operated from 1957 to 1986. The archives of the Vatican Observatory hold almost 10,000 plates \citepads{2004BaltA..13..665B}.

Another archive of plates from the Vatican Observatory is located at the Hamburg Observatory. The plates have been digitised in Hamburg, 
but are not yet included in APPLAUSE.
  
\begin{figure*}
\includegraphics[width=\textwidth]{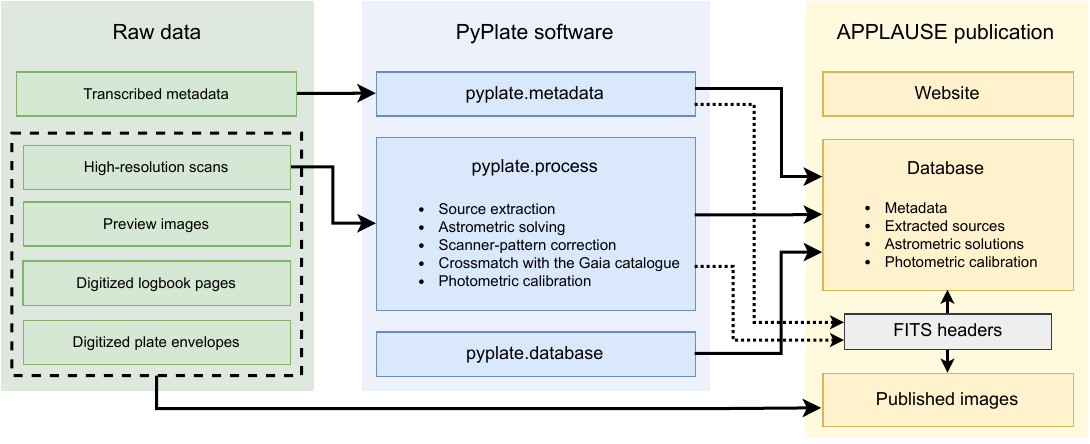}
\caption{Schematic overview of the data processing and publishing workflow.}\label{fig:workflow}
\end{figure*}

\section{Digitisation and data processing}\label{digitization}

In this section, we describe the digitisation of photographic plates and related materials, and processing of metadata, images, and extracted data. The data processing workflow is shown in Fig.~\ref{fig:workflow}.

\subsection{Digitisation of photographic plates}\label{sec:workflow}

The first step in the digitisation process was to create low-resolution preview images of the plates, preserving annotations on the plate for historic context. Previews were obtained either by scanning or taking photographs of the plates before removing the annotations and dust.

High-resolution scans were made after cleaning, using Epson 10\,000 Pro XL flatbed scanners at the resolution of 2400 pixel per inch (PPI). At Hamburg, a fraction of direct-image plates were scanned twice by rotating the plates 90 degrees between the scans.  
The high-resolution images were converted to FITS format \citepads{1981A&AS...44..363W}. 

\begin{SCfigure*}
 \includegraphics{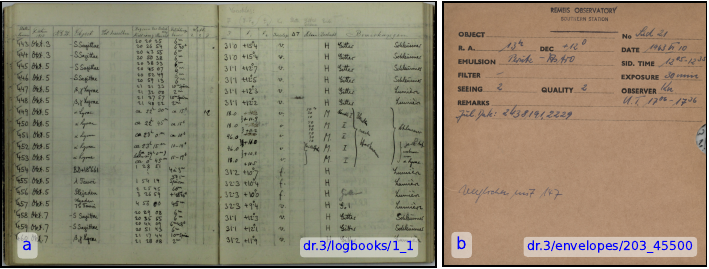}
\hspace*{0.011\textwidth} 
\caption{Digitisation of logbooks and plate envelopes: a)~image of a logbook page from the Potsdam Zeiss-Triplet archive; b)~image of a plate envelope from the Bamberg South archive. The APPLAUSE Identifier (AID) is shown in the lower right in each panel (for DOI construction see Fig.~\ref{fig:digitization_plates}).}
 \label{fig:digitization_logs}
\end{SCfigure*}

\subsection{Metadata transcription}\label{subsect:transcription}

Plate metadata consists of information on observations, plates as physical objects and scans, telescopes and methods used for observations, start and end times of exposures, weather conditions at the time of observation, plate emulsions and development, information on physical states of the plates at the time of scanning, and  scanning data.
Manually transcriptions were undertaken for the metadata from plate envelopes, logbooks, and observer notebooks throughout the entire period of our operation (a time-consuming process prone to errors).

The transcription of plate metadata in Bamberg and Potsdam started in the 1990s, as the initiative of K.~Tsvetkova and M.~Tsvetkov,  subsequently leading to the implementation of the Wide-Field Plate Database (WFPDB) based in Sofia \citepads{2012IAUS..285..417T}. 
As a result of the inventory of the AIP wide-field photographic observations, 11 archives of the Potsdam Observatory from 1879--1970 were included into the Catalogue of Wide-Field Plate Archives \citepads{2005PASRB...5..309T}.

\subsection{Processing of metadata and raw images}\label{subsect:metadata}

In addition to the plates (when available), the plate envelopes, logbook pages, and observer notes were also digitised (see Fig. \ref{fig:digitization_logs}).
Metadata and raw images were divided into 'archives'. We define an archive as a set of plates and related materials that were obtained with a specific telescope at a specific location; for example: plates and logbooks of the observations with the Gro\ss er Schmidt-Spiegel at the Hamburg Observatory constitute one archive (ID 102), but plates obtained with the same telescope at Calar Alto\footnote{named Hamburger Schmidt-Spiegel there} form another (ID 104).

Metadata processing started with ingesting metadata from input files and linking data to individual plates in the archive.
Raw observation times were processed in order to calculate exposure start-, end-, mid-exposure times and to provide them as Universal Time, Julian date, and decimal year. By combining observation times, locations, and telescope pointings, the heliocentric Julian date and air mass for the mid-exposure could be calculated.
All metadata, organised by individual plates, were then written to database tables and into the FITS headers of the high-resolution scans.
For the extended metadata of the FITS files of the digitised plates
we developed an enhanced FITS header structure with suggested keywords\footnote{\url{https://www.ivoa.net/documents/Notes/plateheaders/index.html}}, which has been approved as a FITS convention\footnote{\url{https://fits.gsfc.nasa.gov/registry/photoplates.html}}.

\subsection{Source extraction and astrometric calibration}\label{subsect:sources}

Sources were extracted from images with the Source-Extractor
program \citepads{1996A&AS..117..393B}. Extracted sources were classified as true sources or artifacts, based on a machine-learning algorithm \citep{lswst2019_matijevic}. Initial astrometric solutions were obtained with the Astrometry.net package \citepads{2008ASPC..394...27H} without any knowledge about telescope pointings or field of view. Reference stars were taken from the \textit{Gaia} EDR3 catalogue \citepads{2016A&A...595A...1G, 2021A&A...649A...1G} and their sky positions were calculated for the observation epoch, using the proper motions and  observation times from plate metadata published in the literature.

The initial solutions were refined with the SCAMP program that minimises the quadratic sum of differences in position between extracted sources and reference stars \citepads{2006ASPC..351..112B}. We allowed SCAMP to use the third-order polynomials to characterise image distortions. As an output, SCAMP provides astrometric solutions in the form of the FITS World Coordinate System (WCS)\footnote{\url{https://fits.gsfc.nasa.gov/fits_wcs.html}}, and error estimates in sky coordinates.

Considering that a significant fraction of direct plates had multiple exposures, we carried out astrometric solving in a loop. When a solution was found, the sources that matched reference stars were removed from consideration and another solution was attempted with the remainder of sources. The process was repeated until no further solutions were found. This way we found multiple solutions for 7,558 plates, with the largest number of astrometric solutions of 41 from a single scan.

\begin{figure}
\includegraphics[width=\columnwidth]{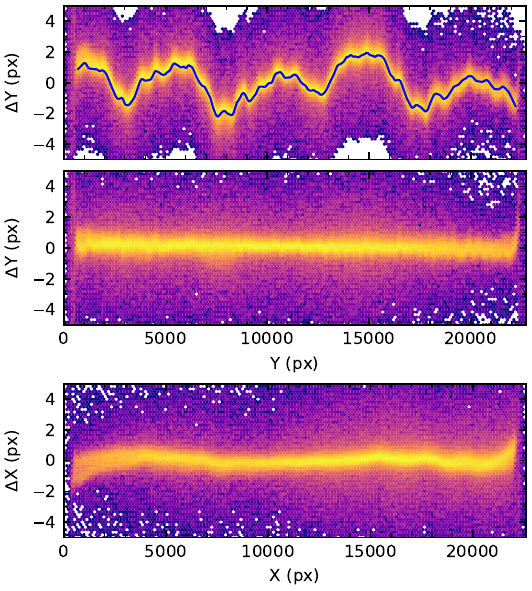}
\caption{Scanner pattern along the scan direction (top) and residuals after subtracting the pattern (middle panel) for a single plate scan. The bottom panel shows the image distortions along the scanner arm.}\label{fig:scanner_astrometry}
\end{figure}

Flatbed scanners introduce a hacksaw pattern in image coordinates, due to non-uniform movement of the scanner arm.  After the initial astrometric solving, we cross-matched the extracted sources with reference stars and determined the scanner pattern along both image axes. The mean pattern was then subtracted from source coordinates along the scan direction. Figure~\ref{fig:scanner_astrometry} shows the scanner pattern (upper panel) and residuals after subtraction of the pattern (middle panel). As can be seen from the figure, the resulting scatter is of similar magnitude as in the direction along the scanner arm (bottom panel). The SCAMP-refined astrometric solution was determined for the whole image, using the pattern-corrected source coordinates.

\begin{figure}
\includegraphics[width=\columnwidth]{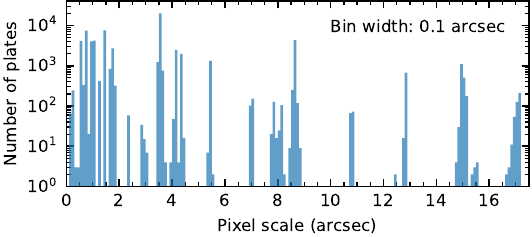}
\caption{Histogram of pixel scales of all astrometrically solved plates in APPLAUSE DR4.}\label{fig:plate_scales}
\end{figure}

\begin{figure}
\includegraphics[width=\columnwidth]{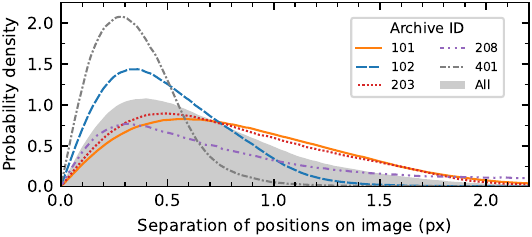}
\caption{Distributions of separations between extracted sources and reference stars for a sample of individual archives and all archives combined.}\label{fig:astrom_accuracy}
\end{figure}

Refined astrometric solutions were written  into FITS headers and to the APPLAUSE database. The solutions were also used to transform image coordinates of extracted sources to sky coordinates. Their accuracy depends on the plate scale of the individual archives (see Table~\ref{dr4archives} and Fig.~\ref{fig:plate_scales}) and the number of reference stars available. In Fig.~\ref{fig:astrom_accuracy} we show the distribution of separations between extracted sources and reference stars for all archives combined, as well as for selected individual archives. The best accuracy is achieved for plates obtained with Schmidt telescopes (archive ID 401, 102), while plates with a very wide field of view (e.g. ID 208) exhibit long tails in the distribution.
In conclusion, for most archives, an average accuracy of better than 0.5 pixels was achieved, allowing for a reliable cross-matching of extracted sources with modern catalogues.

\subsection{Photometric calibration}\label{subsect:photometry}

\begin{figure}
\includegraphics[width=\columnwidth]{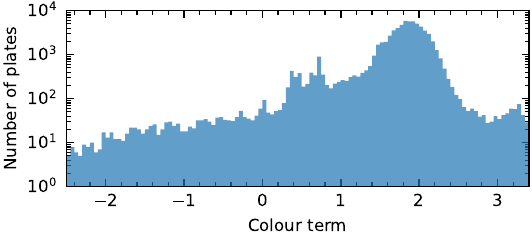}
\caption{Histogram of colour terms based on all successfully processed plates.}\label{fig:colour_term}
\end{figure}

\begin{figure}
\includegraphics[width=\columnwidth]{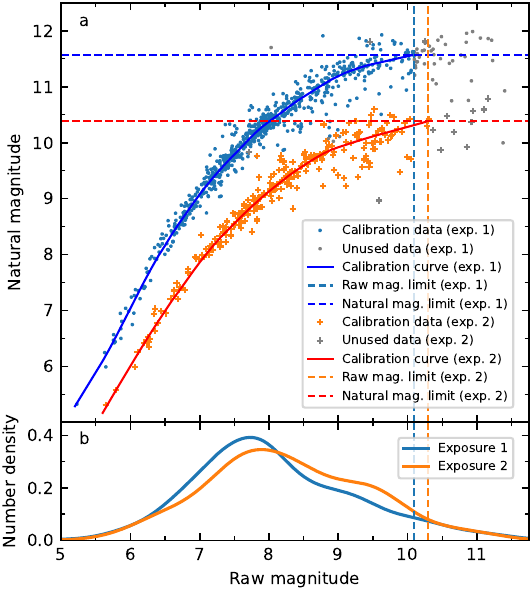}
\caption{Photometric calibration of magnitudes extracted from a two-exposure plate (archive ID 1, plate ID 20149). The observation was made with the Zeiss-Triplet telescope on May 23, 1910; exposure~1 lasted 610\,s and exposure~2 was 190\,s. Panel~\textit{a} shows the relation between raw and natural magnitudes for two exposures. Fitted calibration curves are drawn with solid lines. Dashed lines denote limiting magnitudes in the raw and natural magnitude scales. Panel~\textit{b} shows the distribution of plate magnitudes for both exposures.}\label{fig:calib}
\end{figure}

Photometric calibration was carried out for all plate scans that had astrometric calibration successfully completed. We used \textit{Gaia} EDR3 as our reference photometry.

Each photographic plate has a unique colour response that can be characterised with a colour term \citepads{2010AJ....140.1062L}. Magnitudes in the plate natural photometric system can be calculated with the following relation:
\begin{equation}
    m_{\mathrm{nat}} = G_{\mathrm{RP}} + C_{\mathrm{Gaia}} (G_{\mathrm{BP}}-G_{\mathrm{RP}}),
\end{equation}
where $C_{\mathrm{Gaia}}$ is the colour term. The value of zero corresponds to the \textit{Gaia} $G_\mathrm{RP}$ passband and the value of one to the $G_\mathrm{BP}$ passband.

The distribution of colour terms for all successfully processed plates is shown in Fig.~\ref{fig:colour_term}. It can be seen that most of the plates have bluer colour response than the \textit{Gaia} $G_\mathrm{BP}$ passband.

After determining the colour term, calibration curves were derived and raw magnitudes were transformed to the plate natural system. For sources with known colour indices, we also calculated the $G_{\mathrm{BP}}$ and $G_{\mathrm{RP}}$ magnitudes. Figure~\ref{fig:calib} (panel~\textit{a}) depicts calibration curves for a two-exposure plate.
 
\begin{figure}
\includegraphics[width=\columnwidth]{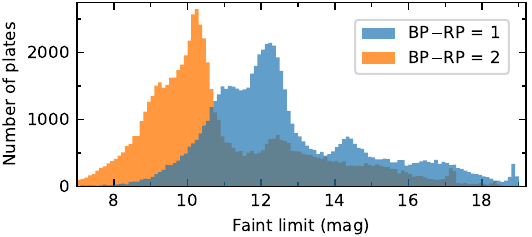}
\caption{Histogram of limiting magnitudes in the APPLAUSE database for blue objects ($G_{\mathrm{BP}}-G_{\mathrm{RP}}=1$) and red ones ($G_{\mathrm{BP}}-G_{\mathrm{RP}}=2$).}\label{fig:plate_limits}
\end{figure}

Limiting magnitudes were determined by analysing the distribution of raw magnitudes of extracted sources belonging to each astrometric solution (exposure). The magnitude at which the numerical density of magnitudes dropped below 0.2 of the peak density was adopted as the faint limit of the exposure (see Fig.~\ref{fig:calib}b). Using the colour term, faint limits can be calculated for sources with different colour indices. The distributions of faint limits for objects with \textit{Gaia} colours $G_{\mathrm{BP}}-G_{\mathrm{RP}}=1$\,mag and $G_{\mathrm{BP}}-G_{\mathrm{RP}}=2$\,mag are shown in Fig. \ref{fig:plate_limits}.
Because most photographic emulsions were blue sensitive, the limiting magnitudes for bluer objects are fainter (blue histogram) than for red objects (yellow histogram) by as much as 2\,mag.

\subsection{PyPlate software}\label{subsect:pyplate}

To handle very inhomogeneous sets of images and metadata and to build the APPLAUSE database, we developed a software package PyPlate and applied it via a dedicated pipeline (see Fig.~\ref{fig:workflow}).
For source extraction from images and for astrometric solving, PyPlate calls for external software, as described in Sect.~\ref{subsect:sources}.
The PyPlate package is written in Python, is open source and is available from a GitHub repository\footnote{\url{https://github.com/astrotuvi/pyplate}}. A comprehensive description of the PyPlate software will be published in a separate paper.
PyPlate has already been used by other digitisation projects, for instance, with Asiago plates, as described by \citetads{2017MmSAI..88..444N,2018RMxAA..54..341N} and \citet{lswst2019_nesci}, as well as for Sonneberg plates  \citepads{2016arXiv161000265S}.

\begin{figure}
\includegraphics[width=\columnwidth]{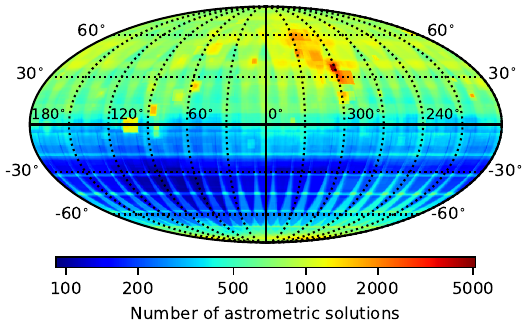}
\caption{Sky coverage of all astrometric solutions in APPLAUSE DR4 in equatorial coordinates.}\label{fig:equatorial}
\end{figure}

\begin{figure*}
\includegraphics[width=\textwidth]{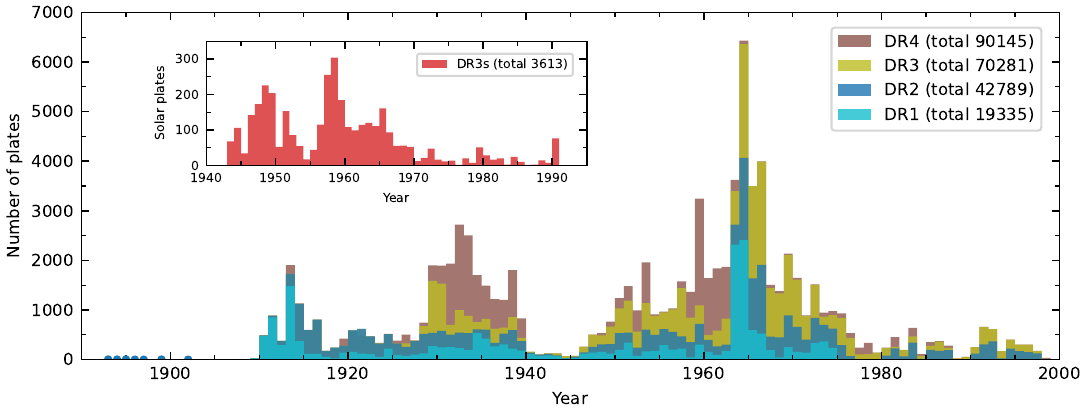}
\caption{Time coverage of scanned plates in APPLAUSE data releases from the first (DR1) in 2015 to the last (DR4) in 2022. The inset shows the histogram of solar plates in APPLAUSE (DR3s).}\label{fig:timeline}
\end{figure*}

\section{ APPLAUSE data publication}

The APPLAUSE data publication consists of four data releases of digitised photographic images of celestial objects with accompanying metadata and 
extracted tabular data and one release of digitised solar full-disk images \citepads[see][for details]{2020AN....341..575P}. Each release denotes a fixed suite of processes to create the published information.

\subsection{Data Release of digitised astronomical photographic plate}\label{subsect:DR4}

The latest APPLAUSE data release\footnote{\url{https://www.plate-archive.org/metadata/applause_dr4/}} of digitised photographic plates was published in June 2022, along with metadata about the plates (gathered by transcription of available material), digitised observation logbooks and notes,  and table data with extracted values from the PyPlate processing.
Images of spectral plates have been integrated in the database, but not yet calibrated.

The release contains 
images and metadata from 27 plate archives and collections from the Hamburg, Bamberg, Potsdam, Tautenburg, Tartu, and Vatican observatories,
including:

\begin{itemize} 
    \item 113,960 images of logbook pages and plate covers
    \item 219 logbooks and observer notebooks
    \item 105,854 preview images 
    \item 121,855 scans of 90,161 plates (and 3,613 plates/scans of solar observations); 
    \item metadata for 94,090 plates with a total of 139,539 exposures
    \item astrometric solutions and photometric calibration for processed plates
    \item 4.486 billion extracted sources (individual detections), 
    of which 2.132 billion with matched  \textit{Gaia} EDR3 designations, 1.256 billion unique matches. 
\end{itemize} 

Fig. \ref{fig:equatorial} shows the distribution of astrometrically solved plate exposures across the sky. It can be seen that the whole sky is covered.
The time coverage is shown in Fig. \ref{fig:timeline}. The oldest scanned plate is from 1893, belonging to the Potsdam Carte du Ciel archive. The youngest plate is from 1998, obtained with the ESO 1-metre Schmidt telescope at La Silla, Chile.

Access to images is provided through a query interface\footnote{\url{https://www.plate-archive.org/cms/home/}}.
If the result table contains links to images, the files can be downloaded either individually or  grouped as selected columns or rows. APPLAUSE provides JPEG or PNG images of digitised logbook pages, plate covers and envelopes, and other related material. As with plate images, the log-page images can be accessed via the query interface.
The source catalogue can also be queried using the Table Access Protocol (TAP), implemented in, for instance, Astropy \citepads{2022ApJ...935..167A} and the Tool for OPerations on Catalogues And Tables (TOPCAT) \citepads{2005ASPC..347...29T}.

\subsection{Digitised spectral plates}\label{sect:spectra}

Scanned images of spectral plates (both objective-prism spectra and single-object spectra) are also included in APPLAUSE and accessible through the query interface. They are accompanied by previews of  logbooks, envelopes, and notes  (as available) linked to their scans. 

The calibration of spectral plates was beyond the scope of the project funding. Therefore, 
 7,372 objective-prism plates and 2,574 single-object spectra (slit or prism) from the Hamburg Observatory have been digitised, but not yet calibrated. To this end, a spectral calibration 
procedure needs to be adapted to our data, making use of procedures developed in Hamburg for the 
Hamburg Quasar Survey (HQS) \citepads{1995A&AS..111..195H} and 
Hamburg ESO Survey (HES)
\citepads{1996A&AS..115..227W}.

\subsection{Archives to be processed}

While the major parts of the digitised archives from Potsdam, Hamburg, Bamberg, Tartu, and Tautenburg have been processed and integrated in the APPLAUSE database, there are still scanned images from various archives awaiting their integration into APPLAUSE. These include 
smaller archives of direct and spectral plates from the Hamburg and Potsdam Observatory,
additional digitised plates from the Vatican Observatory, some 
of which from the CdC, and
digitised plates from Jena \citepads{jena_2014}.
Smaller collections of older plates are kept in Potsdam, but require very intense search for the metadata, since logs or envelopes are only partially available. 

\subsection{Data publication framework}

Using the Daiquiri\footnote{Daiquiri: An Open Source framework for the publication of scientific databases, \url{https://github.com/django-daiquiri}} data publication framework developed by AIP, the VO standard data access methods are available for all APPLAUSE data, in addition to the browser interface.

The minting of the Digital Object Identifiers (DOI)
has been done for the plate archive taking guidance from the Cultural Heritage Object (CHO) as suggested by Europeana\footnote{\url{https://www.europeana.eu/}}
and the \mbox{DataCite}\footnote{(\url{https://schema.datacite.org/}} kernel. 
The CHO notion resulted in minting separate DOI for each digitised material object: photoplate, envelope, logbook (and similar written notebooks), but not single pages from such a book. From this we constructed the Applause Identifier (AID) consisting of (object type, archive\_id, object\_id) in the corresponding tables, e.g AID = "envelopes/101\_8092" resolves to the image of the envelope 8092 from the Hamburg Lippert-Astrograph archive with archive\_id 101. These AID also provide part of the path to access the object via the REST API of the publication framework\footnote{In this example case using \url{https://www.plate-archive.org/objects/dr.3/envelopes/101\_8092/}, the image of the envelope is retrieved and shown in your browser.} 
The DOI of the object uses the AID likewise, with the DOI prefix of AIP, the data release signifier, and the AID\footnote{\url{https://doi.org/10.17876/plate/dr.3/envelopes/101\_8092}.}. A DOI requires a set of metadata information and a persistent landing page to which the DOI link resolves. Using the standard metadata harvesting API (OAI-PMH) one can also retrieve the metadata for this object\footnote{\url{https://www.plate-archive.org/cms/documentation/oaipmh/}.}.

\begin{figure*}
 \includegraphics[width=\textwidth]{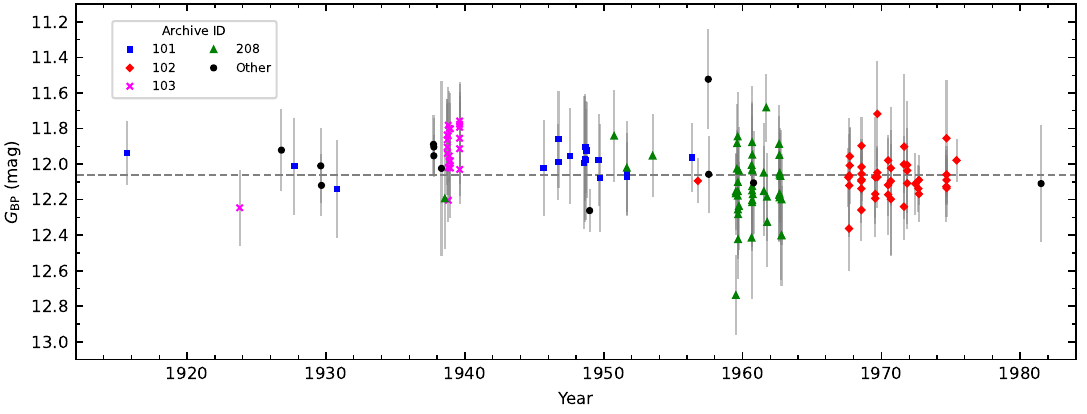} 
 \hspace*{0.01\textwidth}\caption{Light curve of KIC\,8462852 with 1-$\sigma$ errors in the \textit{Gaia} $G_\mathrm{BP}$ photometric passband, based on data in APPLAUSE DR4. Dashed line shows the $G_\mathrm{BP}$ magnitude in the \textit{Gaia} DR3 catalogue.}\label{fig:boyajian}
\end{figure*}

\section{Light curves from historical photographic plates}

The long-term light light curves of variable stars can be constructed from APPLAUSE data. As an example, we constructed the APPLAUSE light curve  
of the enigmatic dipping star KIC\,8462852.
We note that the long-term fading of stars has also been reported from analyses of photographic plates \citepads[e.g.][]{2002ESASP.506..841F,2015ApJ...812..133S,2016ApJ...822L..34S,2019RNAAS...3...77S}.

 A particularly interesting object is the F-type star \object{KIC\,8462852} (also known as Boyajian’s  star), which is unique among known variable stars because of its sudden, erratic light variations that are unlike those seen for any other known star. \citetads{2016MNRAS.457.3988B} suggested that the star is orbited by dust clumps.
The dips in the light curves
 led to coining of the term "dipper star" to describe this novel type of variability \citepads{2019ApJ...880L...7S}.
 
 Light curves of KIC\,8462852 from historical photographic plates have also been investigated extensively to search for long-term trends. 
\citetads{2016ApJ...825...73H}, \citetads{lund2016stability}, and \citetads{2016ApJ...822L..34S} used digitised magnitudes from photographic plates from the Harvard College Observatory. 
 While \citetads{2016ApJ...822L..34S} concluded that the star has dimmed by 0.164 mag between 1890 and 1990, \citetads{2016ApJ...825...73H} did not find evidence in support of long-term dimming. \citetads{2017ApJ...837...85H} studied plates from Sonneberg Observatory taken between 1934--1995 and Sternberg Observatory Moscow taken between 1895 and 1995, but could not confirm the dimming. They concluded that the brightness of the star did not decrease in a hundred years to a limit of 0.03 mag per century. The results of \citetads{2016ApJ...822L..34S} and \citepads{2017ApJ...837...85H} are incompatible at the 5$\sigma$ level. \citetads{2018JAVSO..46...33C} used archival photographic plates from the Maria Mitchell Observatory taken between 1922 and 1991 to study the long-term light curve of KIC\,8462852.
The light curve revealed a decrease of brightness by 0.12 $\pm$ 0.02 magnitude per century,  somewhat less than derived from the Harvard light curve \citepads{2016ApJ...822L..34S} but inconsistent with the result of \citetads{2017ApJ...837...85H} and \citetads{lund2016stability}. The latter investigation pointed at a systematic difference between DASCH data taken before 1953 and past 1969 (i.e. the 'Menzel' gap) and concluded that no long-term trend is apparent when this offset is taken into account.
 \citetads{2018ApJ...854L..11H} investigated high-precision  (few mmag) digital survey data from  ASAS (SN, V, I), Kepler,  \textit{Gaia}, SuperWASP, and citizen-scientist observations (AAVSO, HAO, and Burke-Gaffney) covering the years 2006--2017.

 The 11-years long ASAS light curve starting 2006 appeared not to be monotonic but showed two brightening events that may hint at a magnetic activity cycle. However, spectroscopic follow-ups showed no evidence for a sufficiently strong magnetic field nor for large spot coverage of the stellar surface \citepads{2019MNRAS.486..236M}. It is important to emphasise that such small variations \citepads[see Fig.\,1 in ][]{2018ApJ...854L..11H} are  far too low in amplitude to be detectable on digitised historic plates, such as those available in APPLAUSE or DASCH. 

Models of destroyed or infalling objects can explain the observed dips, but fail to explain long-term fading. 
These and other controversies reported in the literature motivated us to inspect the historic long-term light curve of KIC\,8462852
from the APPLAUSE database.

\begin{figure}
\includegraphics[width=\columnwidth]{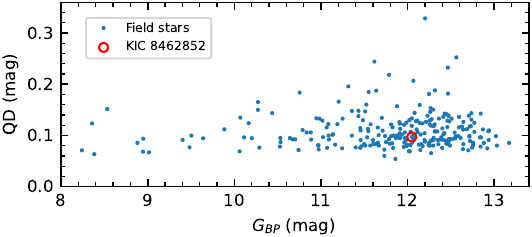}
\caption{ Quartile deviations (QD = 0.5 IQR) 
of magnitudes as a function of the \textit{Gaia} $G_{\mathrm{BP}}$ magnitude for stars in a one-degree radius of KIC\,8462852.}\label{fig:phot_accuracy}
\end{figure}

Figure 12 
shows the light curve spanning the time from 1915 to 1981. 
Relatively dense time coverage was achieved at Hamburg Observatory in the late 1930s (1m-Spiegelteleskop) and the 1970s (Gro\ss er Schmidtspiegel), as well as  by the Bamberg Northern Patrol in the 1960s\footnote{
During that period of time the patrol was carried out with six cameras differing in the size of the lenses, the plate size and the field of view \citepads[see ][for details]{2018A&AT...30..467T}; hence, the data appear quite inhomogeneous. 
}. 
We compared the scatter in the KIC\,8462852 light curve with light curves of stars in a one-degree radius of KIC\,8462852. The quartile deviations (half of the interquartile ranges) of magnitudes are shown in Fig.~\ref{fig:phot_accuracy}. The lower part of the plot is occupied by constant stars or stars with low-amplitude variability. KIC\,8462852 has a QD value close to constant stars of the same brightness, suggesting that potential erratic dipping variability must have been too low in amplitude to be detectable in our data.
The light curve suggests that the star was systematically brighter than the \textit{Gaia} DR3 magnitude from late 1930s to mid-1950s, but we cannot rule out systematic errors in photometric calibrations of various archives, leading to small offsets in magnitudes. Therefore, we refrained from a statistical analysis of the data because an in-depth investigation of systematic uncertainties has to be carried out before that is possible. This is beyond the scope of this paper.

\section{Exploiting direct images}

The vast majority of published direct images in APPLAUSE have been astrometrically solved and the mapping between pixel and sky coordinates (World Coordinate System, WCS) has been embedded in FITS headers. This enables the search for objects with known sky coordinates in images and then transforming the image coordinates to sky coordinates.

Most of the photographic plates have been taken in the course of sky patrols. However, some observations targeted special objects. The APPLAUSE database holds images dedicated to the study of individual celestial objects such as comets, asteroids, moons, planets, nebulae, novae, supernovae, and galaxies. APPLAUSE can be queried for such object types, with details given on the APPLAUSE web pages\footnote{\url{https://www.plate-archive.org/applause/}} . 

APPLAUSE offers the possibility to study the spatial evolution of nova shells in time on the digitised images. 
The APPLAUSE database holds 1034 images that targeted novae, including images of the Nova \object{GK\,Per} (1901) (Sect.~\ref{sect:nova}). In addition, 43 plates that have targeted supernovae are available including 
SN\,1939A and SN\,1959D
(see Sect. \ref{sect:sn}). 

\subsection{Novae}\label{sect:nova}

Nova outbursts are transient phenomena caused by nuclear reactions 
on the surface of a white dwarf. This occurs in cataclysmic systems, where a white dwarf 
accretes gas from a secondary companion via Roche lobe overflow until the accumulated envelope becomes sufficiently  
hot and dense to ignite hydrogen burning in an electron degenerate gas. The energy released leads to an ejection of the envelope, which expands
into a nova shell \citepads[for a recent comprehensive review see ][]{2021ARA&A..59..391C}.

Nova GK Per (1901)
is one of the brightest (as well as the second-closest) nova known to date \citepads{1901ApJ....13..170P,1901ApJ....13..173H,2018MNRAS.481.3033S}. As a recurrent dwarf nova, it undergoes outbursts on short recurrence time scales of 1 to 2 years, which seem to increase over time \citepads{2002A&A...382..910S}. This makes it an ideal target for X-ray monitoring, for instance, with Swift \citepads{2024MNRAS.tmp..690P}. 
The early light curve has been published by \citetads{1919MNRAS..79..362W} and recently \citetads{2020BAVSR..69..181G} added magnitudes measured on plates taken by K\"ustner between 9/1902 and 2/1904 at Bonn. 
No less than eight optical outbursts were recorded between 1970 to 2000 \citepads{2002A&A...382..910S}.

\begin{figure*}
 \includegraphics[width=\textwidth]{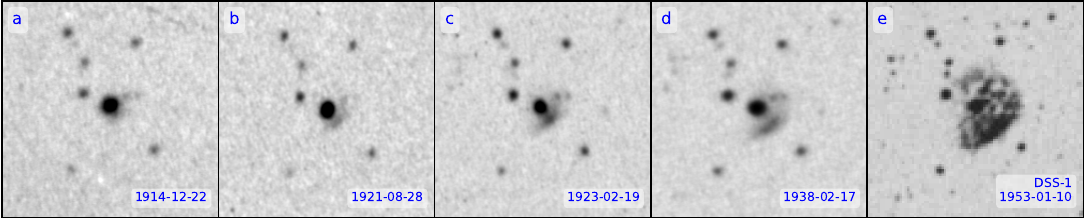}
 \caption{Expanding ejecta of GK Per (Nova Persei 1901). Panels \textit{a--d} show cutouts of plates obtained with the Hamburg 1-m telescope. For comparison, the FDSS image is shown in panel \textit{e}.} 
 \label{fig:GK_Per}
\end{figure*}

 Figure~\ref{fig:GK_Per} shows APPLAUSE images at four epochs depicting the development of the nova shell from 1914 to 1938 and compares them to the DSS image taken in 1953. The expansion of the shell is clearly visible.
 The expansion of the nova shell knots has recently been modelled from modern imaging and spectroscopy (\citeads{2012ApJ...761...34L}, \citeads{2016A&A...595A..64H}).
However, the geometry of the shell is still not uniquely explained. Therefore, the APPLAUSE images might help to constrain the model as they trace the early evolution of the shell.

\begin{figure*}
 \includegraphics[width=\textwidth]{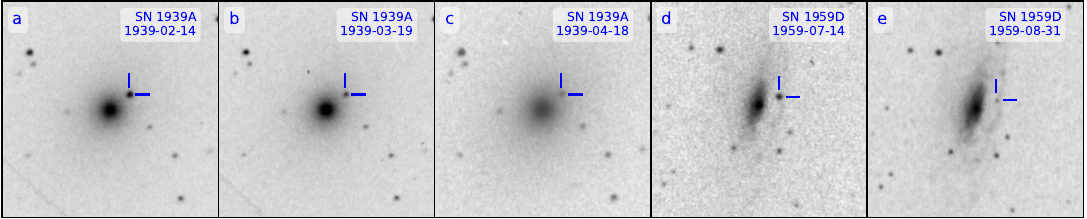}
 \caption{Sample of known supernovae in APPLAUSE images: SN 1939A is shown in panels \textit{a--c} and SN 1959D in panels \textit{d} and \textit{e}.} 
 \label{fig:SN_known}
\end{figure*}

\subsection{Extragalactic supernovae: SN\,1939A and SN\,1959D
}\label{sect:sn}

Supernova explosions arise either from the core collapse of massive stars at the end of their nuclear evolution (Typ II, Ib and Ic) or from the 
thermonuclear explosion of a white dwarf \citepads[see ][for a comprehensive overview of supernova explosion]{2017suex.book.....B}.

The APPLAUSE database contains
images of the Type Ia supernova \object{SN\,1939A} \citepads{1939AN....268..341B} in \object{NGC\,4636} and the Type IIL \object{SN\,1959D} in \object{NGC\,7331} \citepads{1961ApJ...133..883A}. SN\,1939A was  discovered by Wachmann at Hamburg Observatory  on January 18, 1939 well before it reached its maximum light \citepads[see ][for the light curve through its light maximum]{1939BHarO.911...41H}. SN\,1959D was discovered by Humason on June 28, 1959, past its brightness maximum. 
Their light curves have been traced for a long time, in particular the evolution of X-ray emission of the latter has been tracked through decades \citepads{2008ApJ...683..767S,2020ApJ...901..119R}. 

Three APPLAUSE images of SN\,1939A  taken at intervals of about 1 month: the first one at three weeks after discovery; then, two for SN\,1959D taken 16 days past its discovery,  with a dimming seen about six weeks later for both SNe (see Fig. \ref{fig:SN_known}). Hence, digital APPLAUSE image can be used to search for yet unknown transient objects on different time scales, for instance, by \citetads{2008A&A...477...67H} who discovered 22 novae unknown before from Tautenburg Schmidt plates. In addition to supernovae and novae, flare stars and afterglows from gamma ray bursts \citepads{2013AcPol..53c..27H} may be discovered in retrospect.

\begin{figure*}
 \includegraphics[width=\textwidth]{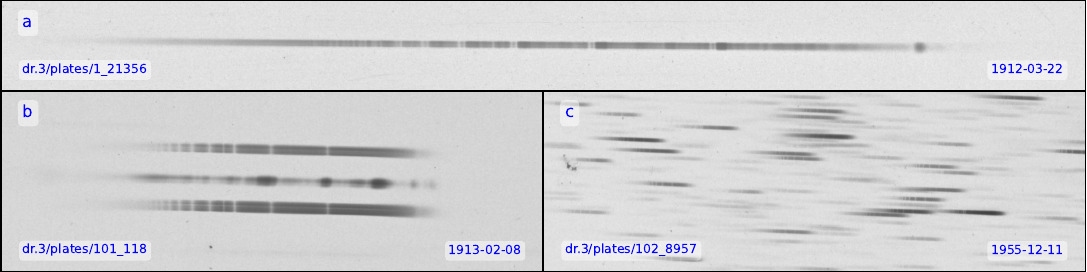}
 \caption{Cut-outs from scans of objective-prism spectral plates.
 Panel \textit{a} shows a spectrum of Nova Gem 1912 taken with the Potsdam Zeiss-Triplet telescope (archive ID 1), panel \textit{b} a cut-out of a triply exposed spectral plate centered on Nova Gem 1912 (for details see text), and panel \textit{c} spectral image of a 36.4$\times$12.1-arcmin region in Cassiopeia taken with the Hamburg Gro\ss er Schmidt Spiegel (ID 101). Date of observation and APPLAUSE Identifier (AID) are given in each panel (for DOI construction see Fig.~\ref{fig:digitization_plates}).}
 \label{fig:spectra_images}
\end{figure*}

\section{APPLAUSE spectra}

The APPLAUSE database holds about 10,000 digitised spectral plates from Hamburg (see Table \ref{hamburg_stock}) and Potsdam archives. 
In the plate archives of the Hamburg Observatory, there are substantial
numbers of photographic plates containing spectra in form of objective-prism 
plates or as plates with one to several single spectra and are listed in Table \ref{hamburg_stock} (examples are shown in Fig. \ref{fig:spectra_images}).  Panels \textit{a} and \textit{b} in Fig.~\ref{fig:spectra_images} show spectra of Nova Gem 1912 taken with the Zeiss-Triplet (archive ID 1, see Table \ref{dr4archives}) at Potsdam (panel \textit{a}) and
the Hamburg Lippert telescope (ID 101, panel \textit{b}) taken in 1912 and 1913, respectively. Panel \textit{c} is a cut-out of a field in  Cassiopeia taken with the Gro\ss er Schmidt Spiegel (archive ID 102) at the Hamburg Observatory. 
Panel \textit{b} shows a triple exposure. A technical construction was created at the Lippert telescope to place short-exposure spectra of a bright comparison star ($\delta$ Gem or $\beta$ Gem) on both sides of the nova spectrum for wavelength calibration \citepads[see][]{schorr1912}. Panel \textit{a} displays a spectrum in an early phase of the nova evolution taken 12 days after maximum light.

The post-outburst spectra result from a discrete shell and a continuous wind, in which the narrower
metal line spectrum is formed in the wind, while the broader line spectrum is formed in the gas ejected in the nova outburst.   
The early post-outburst spectra are dominated by emission lines of hydrogen, helium, and metals, most notably Fe {\sc ii}, occasionally P Cygni profiles occur. Depending on the expansion velocity of the nova shell, their spectral appearance and time evolution differ considerably with respect to degree of ionisation and time scale \citepads[for details see ][]{ 1992AJ....104..725W}. Once the density in the expanding shell declined to sufficiently low values, forbidden lines occur in the spectrum of the nova \citepads[see ][]{2006agna.book.....O}. Digitised spectral images of novae  allow us to trace their evolution over time. 

\begin{figure*}
\begin{center}
 \includegraphics[width=\textwidth]{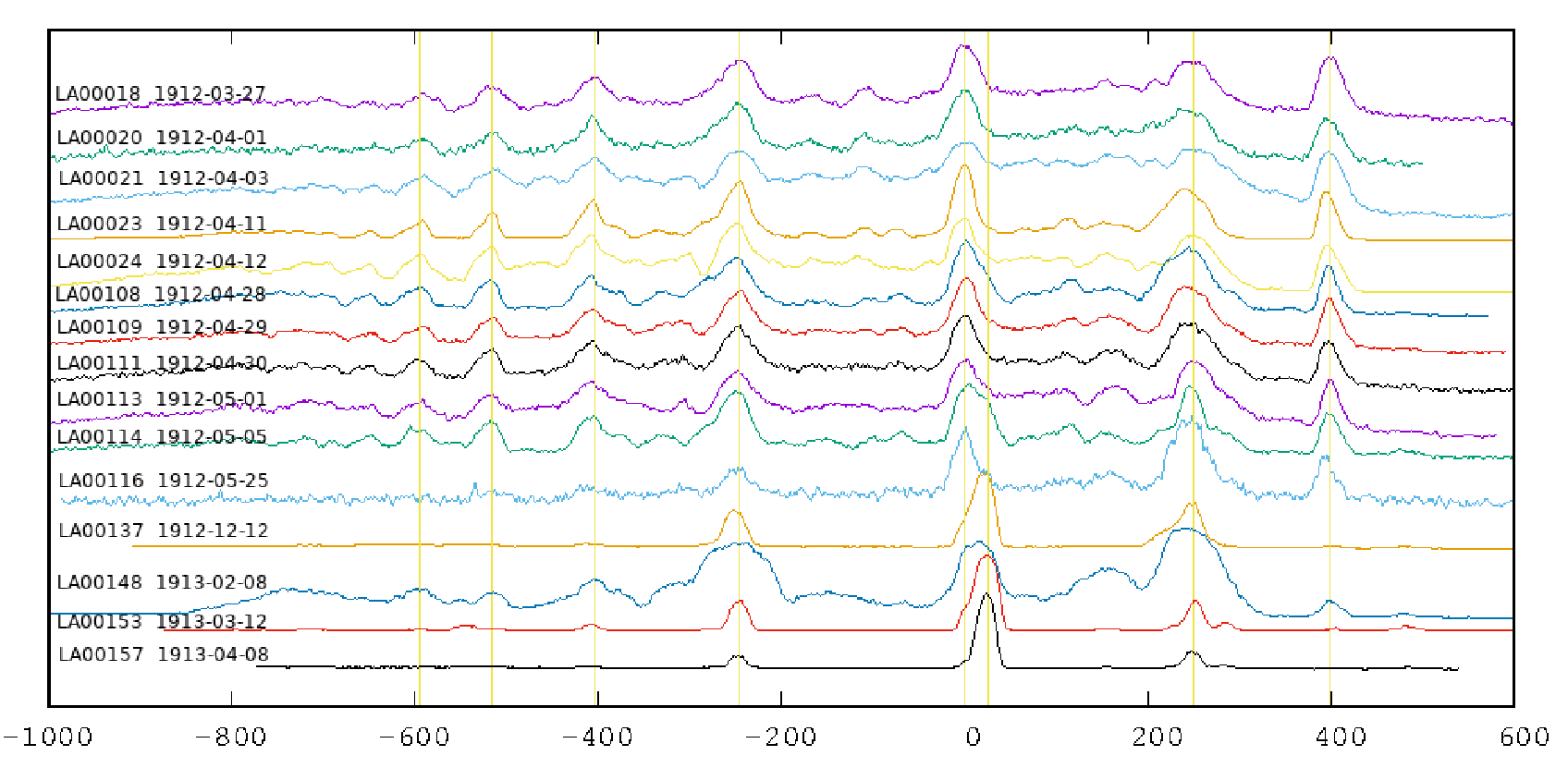}
\end{center}
\caption{Fifteen shell spectra of Nova Gem 1912, background subtracted, scaled and shifted to H-gamma at pixel position "0" and arbitrarily in y-direction ordered by date from top to bottom. The strongest emission line features are marked by vertical lines. Note: the wavelength scale is non-linear.
Adapted from \citetads{groote2014,groote2018}.}
\label{fig:spectra_novagem1912}
\end{figure*}

Nova Gem 1912 was discovered on March 12, 1912 before reaching maximum light \citepads{1912AN....191...65.} and attracted a lot of attention. Its spectral evolution was recorded at many observatories worldwide \citepads[e.g.][]{pickering1912,1912AN....191..167W,1913MNRAS..73..646C} including the Hamburg (\citeads{schorr1912} and \citeads{schwassmann1912}) and Potsdam observatories (\citeads{1912AN....192..117F} and 
\citeads{1912AN....192..123L}). APPLAUSE data can be used to track variations of spectral features over time.

\begin{figure*}
\includegraphics[width=\textwidth]{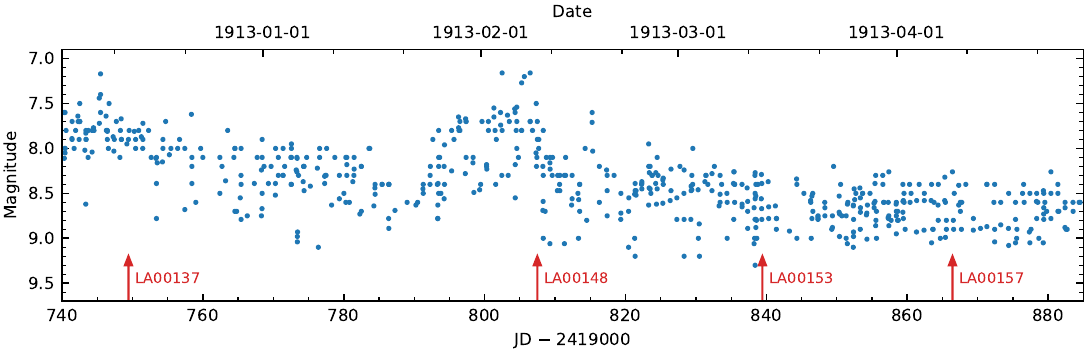}
\caption{Light curve of Nova Gem 1912 showing data from December 1912 to April 1913 as extracted from the AAVSO archive. The vertical arrows mark the date when the Lippert astrograph spectra were taken. Please note that LA00148 is the peculiar spectrum of Fig. \ref{fig:spectra_novagem1912}.}
\label{fig:lc_novagem1912}
\end{figure*}

A selection of extracted spectra of Nova Gem 1912 are shown in Fig. \ref{fig:spectra_novagem1912}.
In the early phase from March to May 1912, the shell is dense and the emission spectrum is dominated by the strongly broadened Balmer series, showing up as P-Cygni profiles. The emission lines are formed in
the already expanded shell around the star, while the optically thick shell also absorbs photons from the stellar surface visible as blue-shifted absorption. 
From December 1912, the shell has expanded, the Balmer lines are weakened, and nebular lines such as [O III], 4364 \AA\ become strong. The spectrum of February 1913 is peculiar because very broad Balmer emission lines reappear pointing to a shell expanding at much higher velocity and, therefore, to another outburst;
meanwhile, a month later, the Balmer emission has disappeared again. To our knowledge, this phenomenon has gone unnoticed in the literature. It coincides with a brightening by 0.8 to 1.0 mag in the AAVSO\footnote{\url{https://www.aavso.org}} light curve 
from end of January 1912 to mid-February 1913 (see Fig.~\ref{fig:lc_novagem1912}).
At the time of the observation, such phenomena were not well understood -- whereas they can be modelled nowadays (after calibration), even in earlier phases when the stellar surface dominates the spectrum.

\section{Summary }

Here, we describe the
Archives of Photographic PLates for Astronomical USE
(APPLAUSE) project, aimed
at building up a database of digitised astronomical photographic plates integrated into the international Virtual Observatory. Twenty four archives from the plate collections at the Leibniz Institut for Astrophysics (AIP) in Potsdam, Hamburg Observatory, and Dr. Remeis Observatory in Bamberg were complemented with digitised plate archives from the Tartu Observatory, Thuringian State Observatory, and Vatican Observatory. The plates were taken with diverse telescopes, cameras, and photographic emulsions. We note that the plate scales  
vary considerably. Wide-field surveys were carried out with small cameras at shallow limiting magnitudes, while larger Schmidt telescopes (e.g. the Hamburg Schmidt Telescope) at excellent observing sites provided the deepest images. The plate archives hold celestial images, mostly from sky patrol missions taken between 1893 and 1999, solar full-disk images as well as photographic objective prism and slit spectra. 
Flatbed scanners were used to produce digital images of celestial plates, which were then calibrated astrometrically and photometrically using  \textit{Gaia} EDR3 positions and magnitudes. Each photographic plate has a unique wavelength dependent sensitivity, which is accounted for by deriving colour terms using the  \textit{Gaia} $G_{\mathrm{BP}}-G_{\mathrm{RP}}$ colours. 
Sources were extracted and classified making use of a machine learning technique to reduce the impact of artifacts. To handle these data reduction procedures, the PyPlate software package was developed. Metadata such as plate envelopes, log books, and evaluation documents were digitised
and associated with each corresponding digital image and incorporated into the APPLAUSE database. To preserve the cultural heritage information, a preview image
of each plate was taken
to secure annotations and markings before cleaning the plates. 

The latest data release of APPLAUSE contains fully calibrated data from 66,328 direct photographic plates and more than 3,500 full-disk solar images.   Additionally, it features uncalibrated material from more than 10,000 spectral plates. The main scientific impact 
of the database lies with studies of stellar variability on different, mostly long-term scales as well as transient objects. The 2.132 billion measurements of \textit{Gaia} matched sources are available in the latest data release and can readily be used to create 
light curves. We constructed the APPLAUSE light curve 
of the enigmatic KIC\,8462852,  
a key object for research into extrasolar planetary systems. Claimed variations on timescales of decades are not obvious in our light curve.
Besides the source catalogue's digitised images hold a lot of information on the evolution of objects, such as the occurrence of nova and supernova outbursts and the expansion of nova shells; we demonstrated this for the cases of two historic supernovae and the evolution of the remnant of Nova GK Per 1901. Although the set is still uncalibrated, the digitised spectral plates offer a treasure store of information. We studied the time evolution of the spectrum of Nova Gem 1912 extracted from Hamburg plates and identified a yet unnoticed outburst in the beginning of 1913. 

The data publication hosted at the AIP complies with the standards of the International Virtual Observatory Alliance and offers various option to query it. This will allow us to combine the historic data in APPLAUSE with digital data from a multitude of modern astronomical catalogues.  

Since the  APPLAUSE database and software tools are established and tested, they can now be extended to incorporate additional digitised archives. A re-calibration of the photometry using the latest \textit{Gaia} spectrophotometry and the calibration of spectra would be promising tasks useful for improving quantitative exploration of APPLAUSE in the future.

\begin{acknowledgements}

This work would not have been possible without the decade-long preparatory work of Katya Tsvetkova and Milcho Tsvetkov. Their painstaking acquisition and transcription of metadata turned the WFPDB into the seed database for metadata of the APPLAUSE project.
The work was carried out within the project “Digitalisierung astronomischer Fotoplatten und ihre Integration in das internationale Virtual Observatory”.  Funding for APPLAUSE has been provided by DFG (German Research Foundation, grant numbers EN926/3-1/2, HE1356/63-1/2, GR969/4-1 and SCHM1032/552), the Leibniz Institute for Astrophysics Potsdam (AIP), the Erlangen Centre for Astroparticle Physics (ECAP), the University of Erlangen-N\"urnberg and its Universit\"atsbund, the Hamburger Observatory (University of Hamburg) and Tartu Observatory. We thank Prof. Dr. Matthias Steinmetz for providing generous resources at AIP to process and publish APPLAUSE. We thank Prof. Dr. J\"urgen Schmitt of Hamburg Observatory for his initiation and continuous support of the Hamburg project, especially for his involvement in the DFG application, as well as Profs. Drs. Joern Wilms and Horst Drechsel of Dr. Remeis Observatory (Bamberg) for their continuous support of the project and Dr. Ingo Kreykenbohm for maintaining the local computer resources for the project. Norbert Jansen thoroughly and unremittingly carried out the scanning procedure of the Bamberg plates from the very first day until the last plate was scanned. The project was supported by many volunteers and colleagues, see table at \url{https://www.plate-archive.org/cms/project/description/}. 
We are indebted to Gal Matijevic for developing a machine-learning tool for classification of artifacts from plate scans. We appreciate various discussions with Rene Hudec and the organisation of two important conferences held in Prague (ASTROPLATE I, 2014, and II, 2016).
Digitised plate material has also been made available from Thuringian State Observatory (TLS), and from the archives of the Vatican Observatory. We are grateful to Helmut Meusinger (TLS) and Richard D'Souza (Vatican Observatory) for providing us with their digitised data sets.
TT acknowledges support by the ETAg CoE grant “Foundations of the Universe" (TK202) and by the ETAg project PRG1006.
This work has made use of data from the European Space Agency (ESA) mission
{\textit{Gaia}} (\url{https://www.cosmos.esa.int/gaia}), processed by the {\it  \textit{Gaia}}
Data Processing and Analysis Consortium (DPAC,
\url{https://www.cosmos.esa.int/web/gaia/dpac/consortium}). Funding for the DPAC
has been provided by national institutions, in particular the institutions
participating in the {\textit{Gaia}} Multilateral Agreement.

Part of this work is based on photographic data of the National Geographic Society -- Palomar Observatory Sky Survey (NGS-POSS) obtained using the Oschin Telescope on Palomar Mountain. The NGS-POSS was funded by a grant from the National Geographic Society to the California Institute of Technology. The plates were processed into the present compressed digital form with their permission. The Digitized Sky Survey was produced at the Space Telescope Science Institute under US Government grant NAG W-2166.
We acknowledge with thanks the variable star observations from the AAVSO International Database contributed by observers worldwide and used in this research.

\end{acknowledgements}

\bibliographystyle{aa}
\bibliography{applause.bib}

\begin{thebibliography}{125}
\expandafter\ifx\csname natexlab\endcsname\relax\def\natexlab#1{#1}\fi

\bibitem[{{Abad}(2014)}]{2014RMxAC..43...59A}
{Abad}, C. 2014, in Revista Mexicana de Astronomia y Astrofisica Conference
  Series, Vol.~43, 59--62

\bibitem[{Abdumannapovich {et~al.}(2023)Abdumannapovich, Khabibullaevich,
  Jurakobilovich, Zhao, Yang, Tang, \& Yu}]{UBAI23}
Abdumannapovich, E.~S., Khabibullaevich, Y.~Q., Jurakobilovich, T.~S., {et~al.}
  2023, Journal for the History of Astronomy, 54, 456

\bibitem[{{Anderson} \& {Engels}(2004)}]{2004JBAA..114...78A}
{Anderson}, S.~R. \& {Engels}, D. 2004, Journal of the British Astronomical
  Association, 114, 78

\bibitem[{{Annuk}(2007)}]{2007ASPC..376..201A}
{Annuk}, K. 2007, in Astronomical Data Analysis Software and Systems XVI, ed.
  R.~A. {Shaw}, F.~{Hill}, \& D.~J. {Bell}, Vol. 376, 201

\bibitem[{{Arp}(1961)}]{1961ApJ...133..883A}
{Arp}, H. 1961, \apj, 133, 883

\bibitem[{{Astropy Collaboration} {et~al.}(2022){Astropy Collaboration},
  {Price-Whelan}, {Lim}, {Earl}, {Starkman}, {Bradley}, {Shupe}, {Patil},
  {Corrales}, {Brasseur}, {N{\"o}the}, {Donath}, {Tollerud}, {Morris},
  {Ginsburg}, {Vaher}, {Weaver}, {Tocknell}, {Jamieson}, {van Kerkwijk},
  {Robitaille}, {Merry}, {Bachetti}, {G{\"u}nther}, {Aldcroft},
  {Alvarado-Montes}, {Archibald}, {B{\'o}di}, {Bapat}, {Barentsen},
  {Baz{\'a}n}, {Biswas}, {Boquien}, {Burke}, {Cara}, {Cara}, {Conroy},
  {Conseil}, {Craig}, {Cross}, {Cruz}, {D'Eugenio}, {Dencheva}, {Devillepoix},
  {Dietrich}, {Eigenbrot}, {Erben}, {Ferreira}, {Foreman-Mackey}, {Fox},
  {Freij}, {Garg}, {Geda}, {Glattly}, {Gondhalekar}, {Gordon}, {Grant},
  {Greenfield}, {Groener}, {Guest}, {Gurovich}, {Handberg}, {Hart},
  {Hatfield-Dodds}, {Homeier}, {Hosseinzadeh}, {Jenness}, {Jones}, {Joseph},
  {Kalmbach}, {Karamehmetoglu}, {Ka{\l}uszy{\'n}ski}, {Kelley}, {Kern},
  {Kerzendorf}, {Koch}, {Kulumani}, {Lee}, {Ly}, {Ma}, {MacBride}, {Maljaars},
  {Muna}, {Murphy}, {Norman}, {O'Steen}, {Oman}, {Pacifici}, {Pascual},
  {Pascual-Granado}, {Patil}, {Perren}, {Pickering}, {Rastogi}, {Roulston},
  {Ryan}, {Rykoff}, {Sabater}, {Sakurikar}, {Salgado}, {Sanghi}, {Saunders},
  {Savchenko}, {Schwardt}, {Seifert-Eckert}, {Shih}, {Jain}, {Shukla}, {Sick},
  {Simpson}, {Singanamalla}, {Singer}, {Singhal}, {Sinha}, {Sip{\H{o}}cz},
  {Spitler}, {Stansby}, {Streicher}, {{\v{S}}umak}, {Swinbank}, {Taranu},
  {Tewary}, {Tremblay}, {Val-Borro}, {Van Kooten}, {Vasovi{\'c}}, {Verma}, {de
  Miranda Cardoso}, {Williams}, {Wilson}, {Winkel}, {Wood-Vasey}, {Xue},
  {Yoachim}, {Zhang}, {Zonca}, \& {Astropy Project
  Contributors}}]{2022ApJ...935..167A}
{Astropy Collaboration}, {Price-Whelan}, A.~M., {Lim}, P.~L., {et~al.} 2022,
  \apj, 935, 167

\bibitem[{{Barbieri} {et~al.}(2004){Barbieri}, {Blanco}, {Bucciarelli},
  {Coluzzi}, {Di Paola}, {Lanteri}, {Li Causi}, {Marilli}, {Massimino},
  {Mottola}, {Nesci}, {Omizzolo}, {Pedichini}, {Rampazzi}, {Rossi}, {Sclavi},
  {Stagni}, {Tsvetkov}, \& {Viotti}}]{2004BaltA..13..665B}
{Barbieri}, C., {Blanco}, C., {Bucciarelli}, B., {et~al.} 2004, Baltic
  Astronomy, 13, 665

\bibitem[{{Bertin}(2006)}]{2006ASPC..351..112B}
{Bertin}, E. 2006, in Astronomical Society of the Pacific Conference Series,
  Vol. 351, Astronomical Data Analysis Software and Systems XV, ed.
  C.~{Gabriel}, C.~{Arviset}, D.~{Ponz}, \& S.~{Enrique}, 112

\bibitem[{{Bertin} \& {Arnouts}(1996)}]{1996A&AS..117..393B}
{Bertin}, E. \& {Arnouts}, S. 1996, \aaps, 117, 393

\bibitem[{{Beyer}(1939)}]{1939AN....268..341B}
{Beyer}, M. 1939, Astronomische Nachrichten, 268, 341

\bibitem[{{Birkle}(1984)}]{1984ASSL..110..203B}
{Birkle}, K. 1984, in Astrophysics and Space Science Library, Vol. 110, IAU
  Colloq. 78: Astronomy with Schmidt-Type Telescopes, ed. M.~{Capaccioli}, 203

\bibitem[{{Boehm} {et~al.}(2006){Boehm}, {Steinmetz}, {Tsvetkov}, \&
  {Tsvetkova}}]{2006IAUSS...6E...9B}
{Boehm}, P., {Steinmetz}, M., {Tsvetkov}, M., \& {Tsvetkova}, K. 2006, IAU
  Special Session, 6, 9

\bibitem[{{Boyajian} {et~al.}(2016){Boyajian}, {LaCourse}, {Rappaport},
  {Fabrycky}, {Fischer}, {Gandolfi}, {Kennedy}, {Korhonen}, {Liu}, {Moor},
  {Olah}, {Vida}, {Wyatt}, {Best}, {Brewer}, {Ciesla}, {Cs{\'a}k}, {Deeg},
  {Dupuy}, {Handler}, {Heng}, {Howell}, {Ishikawa}, {Kov{\'a}cs}, {Kozakis},
  {Kriskovics}, {Lehtinen}, {Lintott}, {Lynn}, {Nespral}, {Nikbakhsh},
  {Schawinski}, {Schmitt}, {Smith}, {Szabo}, {Szabo}, {Viuho}, {Wang},
  {Weiksnar}, {Bosch}, {Connors}, {Goodman}, {Green}, {Hoekstra}, {Jebson},
  {Jek}, {Omohundro}, {Schwengeler}, \& {Szewczyk}}]{2016MNRAS.457.3988B}
{Boyajian}, T.~S., {LaCourse}, D.~M., {Rappaport}, S.~A., {et~al.} 2016,
  \mnras, 457, 3988

\bibitem[{{Branch} \& {Wheeler}(2017)}]{2017suex.book.....B}
{Branch}, D. \& {Wheeler}, J.~C. 2017, {Supernova Explosions}

\bibitem[{{Carlin} {et~al.}(2010){Carlin}, {Casetti-Dinescu}, {Grillmair},
  {Majewski}, \& {Girard}}]{2010ApJ...725.2290C}
{Carlin}, J.~L., {Casetti-Dinescu}, D.~I., {Grillmair}, C.~J., {Majewski},
  S.~R., \& {Girard}, T.~M. 2010, \apj, 725, 2290

\bibitem[{{Castelaz} \& {Barker}(2018)}]{2018JAVSO..46...33C}
{Castelaz}, M. \& {Barker}, T. 2018, Jaavso, 46, 33

\bibitem[{{Chomiuk} {et~al.}(2021){Chomiuk}, {Metzger}, \&
  {Shen}}]{2021ARA&A..59..391C}
{Chomiuk}, L., {Metzger}, B.~D., \& {Shen}, K.~J. 2021, \araa, 59, 391

\bibitem[{{Cortie}(1913)}]{1913MNRAS..73..646C}
{Cortie}, A.~L. 1913, \mnras, 73, 646

\bibitem[{{De Cuyper} {et~al.}(2012){De Cuyper}, {de Decker}, {Laux}, {Winter},
  \& {Zacharias}}]{2012ASPC..461..315D}
{De Cuyper}, J., {de Decker}, G., {Laux}, U., {Winter}, L., \& {Zacharias}, N.
  2012, in Astronomical Society of the Pacific Conference Series, Vol. 461,
  Astronomical Data Analysis Software and Systems XXI, ed. P.~{Ballester},
  D.~{Egret}, \& N.~P.~F. {Lorente}, 315

\bibitem[{{Enebo}(1912)}]{1912AN....191...65.}
{Enebo}, S. 1912, Astronomische Nachrichten, 191, 65

\bibitem[{{Engels} {et~al.}(1988){Engels}, {Groote}, {Hagen}, \&
  {Reimrs}}]{1988ASPC....2..143E}
{Engels}, D., {Groote}, D., {Hagen}, H.~J., \& {Reimrs}, D. 1988, in
  Astronomical Society of the Pacific Conference Series, Vol.~2, Optical
  Surveys for Quasars, ed. P.~{Osmer}, M.~M. {Phillips}, R.~{Green}, \&
  C.~{Foltz}, 143

\bibitem[{{Enke}(2019)}]{lswst2019_enke}
{Enke}, H. 2019, {The APPLAUSE archive: Concept, building blocks, features of
  the APPLAUSE Archive}, {Large Surveys with Small Telescope: Past, Present and
  Future (Astroplate III), 2019 March 11 – 13, Bamberg, Germany},
  {DOI:10789/plate/lswst/012}

\bibitem[{{Fr{\"o}hlich} \& {R{\"u}diger}(2002)}]{2002ESASP.506..841F}
{Fr{\"o}hlich}, H.-E. \& {R{\"u}diger}, G. 2002, in ESA Special Publication,
  Vol.~2, Solar Variability: From Core to Outer Frontiers, ed. A.~{Wilson},
  841--842

\bibitem[{{Furuhjelm}(1912)}]{1912AN....192..117F}
{Furuhjelm}, R. 1912, Astronomische Nachrichten, 192, 117

\bibitem[{{Gaia Collaboration} {et~al.}(2021){Gaia Collaboration}, {Brown},
  {Vallenari}, {Prusti}, {de Bruijne}, {Babusiaux}, {Biermann}, {Creevey},
  {Evans}, {Eyer}, {Hutton}, {Jansen}, {Jordi}, {Klioner}, {Lammers},
  {Lindegren}, {Luri}, {Mignard}, {Panem}, {Pourbaix}, {Randich}, {Sartoretti},
  {Soubiran}, {Walton}, {Arenou}, {Bailer-Jones}, {Bastian}, {Cropper},
  {Drimmel}, {Katz}, {Lattanzi}, {van Leeuwen}, {Bakker}, {Cacciari},
  {Casta{\~n}eda}, {De Angeli}, {Ducourant}, {Fabricius}, {Fouesneau},
  {Fr{\'e}mat}, {Guerra}, {Guerrier}, {Guiraud}, {Jean-Antoine Piccolo},
  {Masana}, {Messineo}, {Mowlavi}, {Nicolas}, {Nienartowicz}, {Pailler},
  {Panuzzo}, {Riclet}, {Roux}, {Seabroke}, {Sordo}, {Tanga}, {Th{\'e}venin},
  {Gracia-Abril}, {Portell}, {Teyssier}, {Altmann}, {Andrae}, {Bellas-Velidis},
  {Benson}, {Berthier}, {Blomme}, {Brugaletta}, {Burgess}, {Busso}, {Carry},
  {Cellino}, {Cheek}, {Clementini}, {Damerdji}, {Davidson}, {Delchambre},
  {Dell'Oro}, {Fern{\'a}ndez-Hern{\'a}ndez}, {Galluccio}, {Garc{\'\i}a-Lario},
  {Garcia-Reinaldos}, {Gonz{\'a}lez-N{\'u}{\~n}ez}, {Gosset}, {Haigron},
  {Halbwachs}, {Hambly}, {Harrison}, {Hatzidimitriou}, {Heiter},
  {Hern{\'a}ndez}, {Hestroffer}, {Hodgkin}, {Holl}, {Jan{\ss}en}, {Jevardat de
  Fombelle}, {Jordan}, {Krone-Martins}, {Lanzafame}, {L{\"o}ffler}, {Lorca},
  {Manteiga}, {Marchal}, {Marrese}, {Moitinho}, {Mora}, {Muinonen}, {Osborne},
  {Pancino}, {Pauwels}, {Petit}, {Recio-Blanco}, {Richards}, {Riello},
  {Rimoldini}, {Robin}, {Roegiers}, {Rybizki}, {Sarro}, {Siopis}, {Smith},
  {Sozzetti}, {Ulla}, {Utrilla}, {van Leeuwen}, {van Reeven}, {Abbas}, {Abreu
  Aramburu}, {Accart}, {Aerts}, {Aguado}, {Ajaj}, {Altavilla}, {{\'A}lvarez},
  {{\'A}lvarez Cid-Fuentes}, {Alves}, {Anderson}, {Anglada Varela}, {Antoja},
  {Audard}, {Baines}, {Baker}, {Balaguer-N{\'u}{\~n}ez}, {Balbinot}, {Balog},
  {Barache}, {Barbato}, {Barros}, {Barstow}, {Bartolom{\'e}}, {Bassilana},
  {Bauchet}, {Baudesson-Stella}, {Becciani}, {Bellazzini}, {Bernet}, {Bertone},
  {Bianchi}, {Blanco-Cuaresma}, {Boch}, {Bombrun}, {Bossini}, {Bouquillon},
  {Bragaglia}, {Bramante}, {Breedt}, {Bressan}, {Brouillet}, {Bucciarelli},
  {Burlacu}, {Busonero}, {Butkevich}, {Buzzi}, {Caffau}, {Cancelliere},
  {C{\'a}novas}, {Cantat-Gaudin}, {Carballo}, {Carlucci}, {Carnerero},
  {Carrasco}, {Casamiquela}, {Castellani}, {Castro-Ginard}, {Castro Sampol},
  {Chaoul}, {Charlot}, {Chemin}, {Chiavassa}, {Cioni}, {Comoretto}, {Cooper},
  {Cornez}, {Cowell}, {Crifo}, {Crosta}, {Crowley}, {Dafonte}, {Dapergolas},
  {David}, {David}, {de Laverny}, {De Luise}, {De March}, {De Ridder}, {de
  Souza}, {de Teodoro}, {de Torres}, {del Peloso}, {del Pozo}, {Delbo},
  {Delgado}, {Delgado}, {Delisle}, {Di Matteo}, {Diakite}, {Diener},
  {Distefano}, {Dolding}, {Eappachen}, {Edvardsson}, {Enke}, {Esquej}, {Fabre},
  {Fabrizio}, {Faigler}, {Fedorets}, {Fernique}, {Fienga}, {Figueras},
  {Fouron}, {Fragkoudi}, {Fraile}, {Franke}, {Gai}, {Garabato},
  {Garcia-Gutierrez}, {Garc{\'\i}a-Torres}, {Garofalo}, {Gavras}, {Gerlach},
  {Geyer}, {Giacobbe}, {Gilmore}, {Girona}, {Giuffrida}, {Gomel}, {Gomez},
  {Gonzalez-Santamaria}, {Gonz{\'a}lez-Vidal}, {Granvik},
  {Guti{\'e}rrez-S{\'a}nchez}, {Guy}, {Hauser}, {Haywood}, {Helmi}, {Hidalgo},
  {Hilger}, {H{\l}adczuk}, {Hobbs}, {Holland}, {Huckle}, {Jasniewicz},
  {Jonker}, {Juaristi Campillo}, {Julbe}, {Karbevska}, {Kervella}, {Khanna},
  {Kochoska}, {Kontizas}, {Kordopatis}, {Korn}, {Kostrzewa-Rutkowska},
  {Kruszy{\'n}ska}, {Lambert}, {Lanza}, {Lasne}, {Le Campion}, {Le Fustec},
  {Lebreton}, {Lebzelter}, {Leccia}, {Leclerc}, {Lecoeur-Taibi}, {Liao},
  {Licata}, {Lindstr{\o}m}, {Lister}, {Livanou}, {Lobel}, {Madrero Pardo},
  {Managau}, {Mann}, {Marchant}, {Marconi}, {Marcos Santos}, {Marinoni},
  {Marocco}, {Marshall}, {Martin Polo}, {Mart{\'\i}n-Fleitas}, {Masip},
  {Massari}, {Mastrobuono-Battisti}, {Mazeh}, {McMillan}, {Messina},
  {Michalik}, {Millar}, {Mints}, {Molina}, {Molinaro}, {Moln{\'a}r},
  {Montegriffo}, {Mor}, {Morbidelli}, {Morel}, {Morris}, {Mulone}, {Munoz},
  {Muraveva}, {Murphy}, {Musella}, {Noval}, {Ord{\'e}novic}, {Orr{\`u}},
  {Osinde}, {Pagani}, {Pagano}, {Palaversa}, {Palicio}, {Panahi}, {Pawlak},
  {Pe{\~n}alosa Esteller}, {Penttil{\"a}}, {Piersimoni}, {Pineau}, {Plachy},
  {Plum}, {Poggio}, {Poretti}, {Poujoulet}, {Pr{\v{s}}a}, {Pulone}, {Racero},
  {Ragaini}, {Rainer}, {Raiteri}, {Rambaux}, {Ramos}, {Ramos-Lerate}, {Re
  Fiorentin}, {Regibo}, {Reyl{\'e}}, {Ripepi}, {Riva}, {Rixon}, {Robichon},
  {Robin}, {Roelens}, {Rohrbasser}, {Romero-G{\'o}mez}, {Rowell}, {Royer},
  {Rybicki}, {Sadowski}, {Sagrist{\`a} Sell{\'e}s}, {Sahlmann}, {Salgado},
  {Salguero}, {Samaras}, {Sanchez Gimenez}, {Sanna}, {Santove{\~n}a},
  {Sarasso}, {Schultheis}, {Sciacca}, {Segol}, {Segovia}, {S{\'e}gransan},
  {Semeux}, {Shahaf}, {Siddiqui}, {Siebert}, {Siltala}, {Slezak}, {Smart},
  {Solano}, {Solitro}, {Souami}, {Souchay}, {Spagna}, {Spoto}, {Steele},
  {Steidelm{\"u}ller}, {Stephenson}, {S{\"u}veges}, {Szabados}, {Szegedi-Elek},
  {Taris}, {Tauran}, {Taylor}, {Teixeira}, {Thuillot}, {Tonello}, {Torra},
  {Torra}, {Turon}, {Unger}, {Vaillant}, {van Dillen}, {Vanel}, {Vecchiato},
  {Viala}, {Vicente}, {Voutsinas}, {Weiler}, {Wevers}, {Wyrzykowski}, {Yoldas},
  {Yvard}, {Zhao}, {Zorec}, {Zucker}, {Zurbach}, \&
  {Zwitter}}]{2021A&A...649A...1G}
{Gaia Collaboration}, {Brown}, A.~G.~A., {Vallenari}, A., {et~al.} 2021, \aap,
  649, A1

\bibitem[{{Gaia Collaboration} {et~al.}(2016){Gaia Collaboration}, {Prusti},
  {de Bruijne}, {Brown}, {Vallenari}, {Babusiaux}, {Bailer-Jones}, {Bastian},
  {Biermann}, {Evans}, {Eyer}, {Jansen}, {Jordi}, {Klioner}, {Lammers},
  {Lindegren}, {Luri}, {Mignard}, {Milligan}, {Panem}, {Poinsignon},
  {Pourbaix}, {Randich}, {Sarri}, {Sartoretti}, {Siddiqui}, {Soubiran},
  {Valette}, {van Leeuwen}, {Walton}, {Aerts}, {Arenou}, {Cropper}, {Drimmel},
  {H{\o}g}, {Katz}, {Lattanzi}, {O'Mullane}, {Grebel}, {Holland}, {Huc},
  {Passot}, {Bramante}, {Cacciari}, {Casta{\~n}eda}, {Chaoul}, {Cheek}, {De
  Angeli}, {Fabricius}, {Guerra}, {Hern{\'a}ndez}, {Jean-Antoine-Piccolo},
  {Masana}, {Messineo}, {Mowlavi}, {Nienartowicz}, {Ord{\'o}{\~n}ez-Blanco},
  {Panuzzo}, {Portell}, {Richards}, {Riello}, {Seabroke}, {Tanga},
  {Th{\'e}venin}, {Torra}, {Els}, {Gracia-Abril}, {Comoretto},
  {Garcia-Reinaldos}, {Lock}, {Mercier}, {Altmann}, {Andrae}, {Astraatmadja},
  {Bellas-Velidis}, {Benson}, {Berthier}, {Blomme}, {Busso}, {Carry},
  {Cellino}, {Clementini}, {Cowell}, {Creevey}, {Cuypers}, {Davidson}, {De
  Ridder}, {de Torres}, {Delchambre}, {Dell'Oro}, {Ducourant}, {Fr{\'e}mat},
  {Garc{\'\i}a-Torres}, {Gosset}, {Halbwachs}, {Hambly}, {Harrison}, {Hauser},
  {Hestroffer}, {Hodgkin}, {Huckle}, {Hutton}, {Jasniewicz}, {Jordan},
  {Kontizas}, {Korn}, {Lanzafame}, {Manteiga}, {Moitinho}, {Muinonen},
  {Osinde}, {Pancino}, {Pauwels}, {Petit}, {Recio-Blanco}, {Robin}, {Sarro},
  {Siopis}, {Smith}, {Smith}, {Sozzetti}, {Thuillot}, {van Reeven}, {Viala},
  {Abbas}, {Abreu Aramburu}, {Accart}, {Aguado}, {Allan}, {Allasia},
  {Altavilla}, {{\'A}lvarez}, {Alves}, {Anderson}, {Andrei}, {Anglada Varela},
  {Antiche}, {Antoja}, {Ant{\'o}n}, {Arcay}, {Atzei}, {Ayache}, {Bach},
  {Baker}, {Balaguer-N{\'u}{\~n}ez}, {Barache}, {Barata}, {Barbier}, {Barblan},
  {Baroni}, {Barrado y Navascu{\'e}s}, {Barros}, {Barstow}, {Becciani},
  {Bellazzini}, {Bellei}, {Bello Garc{\'\i}a}, {Belokurov}, {Bendjoya},
  {Berihuete}, {Bianchi}, {Bienaym{\'e}}, {Billebaud}, {Blagorodnova},
  {Blanco-Cuaresma}, {Boch}, {Bombrun}, {Borrachero}, {Bouquillon}, {Bourda},
  {Bouy}, {Bragaglia}, {Breddels}, {Brouillet}, {Br{\"u}semeister},
  {Bucciarelli}, {Budnik}, {Burgess}, {Burgon}, {Burlacu}, {Busonero}, {Buzzi},
  {Caffau}, {Cambras}, {Campbell}, {Cancelliere}, {Cantat-Gaudin}, {Carlucci},
  {Carrasco}, {Castellani}, {Charlot}, {Charnas}, {Charvet}, {Chassat},
  {Chiavassa}, {Clotet}, {Cocozza}, {Collins}, {Collins}, {Costigan}, {Crifo},
  {Cross}, {Crosta}, {Crowley}, {Dafonte}, {Damerdji}, {Dapergolas}, {David},
  {David}, {De Cat}, {de Felice}, {de Laverny}, {De Luise}, {De March}, {de
  Martino}, {de Souza}, {Debosscher}, {del Pozo}, {Delbo}, {Delgado},
  {Delgado}, {di Marco}, {Di Matteo}, {Diakite}, {Distefano}, {Dolding}, {Dos
  Anjos}, {Drazinos}, {Dur{\'a}n}, {Dzigan}, {Ecale}, {Edvardsson}, {Enke},
  {Erdmann}, {Escolar}, {Espina}, {Evans}, {Eynard Bontemps}, {Fabre},
  {Fabrizio}, {Faigler}, {Falc{\~a}o}, {Farr{\`a}s Casas}, {Faye}, {Federici},
  {Fedorets}, {Fern{\'a}ndez-Hern{\'a}ndez}, {Fernique}, {Fienga}, {Figueras},
  {Filippi}, {Findeisen}, {Fonti}, {Fouesneau}, {Fraile}, {Fraser}, {Fuchs},
  {Furnell}, {Gai}, {Galleti}, {Galluccio}, {Garabato}, {Garc{\'\i}a-Sedano},
  {Gar{\'e}}, {Garofalo}, {Garralda}, {Gavras}, {Gerssen}, {Geyer}, {Gilmore},
  {Girona}, {Giuffrida}, {Gomes}, {Gonz{\'a}lez-Marcos},
  {Gonz{\'a}lez-N{\'u}{\~n}ez}, {Gonz{\'a}lez-Vidal}, {Granvik}, {Guerrier},
  {Guillout}, {Guiraud}, {G{\'u}rpide}, {Guti{\'e}rrez-S{\'a}nchez}, {Guy},
  {Haigron}, {Hatzidimitriou}, {Haywood}, {Heiter}, {Helmi}, {Hobbs},
  {Hofmann}, {Holl}, {Holland }, {Hunt}, {Hypki}, {Icardi}, {Irwin}, {Jevardat
  de Fombelle}, {Jofr{\'e}}, {Jonker}, {Jorissen}, {Julbe}, {Karampelas},
  {Kochoska}, {Kohley}, {Kolenberg}, {Kontizas}, {Koposov}, {Kordopatis},
  {Koubsky}, {Kowalczyk}, {Krone-Martins}, {Kudryashova}, {Kull}, {Bachchan},
  {Lacoste-Seris}, {Lanza}, {Lavigne}, {Le Poncin-Lafitte}, {Lebreton},
  {Lebzelter}, {Leccia}, {Leclerc}, {Lecoeur-Taibi}, {Lemaitre}, {Lenhardt},
  {Leroux}, {Liao}, {Licata}, {Lindstr{\o}m}, {Lister}, {Livanou}, {Lobel},
  {L{\"o}ffler}, {L{\'o}pez}, {Lopez-Lozano}, {Lorenz}, {Loureiro},
  {MacDonald}, {Magalh{\~a}es Fernandes}, {Managau}, {Mann}, {Mantelet},
  {Marchal}, {Marchant}, {Marconi}, {Marie}, {Marinoni}, {Marrese},
  {Marschalk{\'o}}, {Marshall}, {Mart{\'\i}n-Fleitas}, {Martino}, {Mary},
  {Matijevi{\v{c}}}, {Mazeh}, {McMillan}, {Messina}, {Mestre}, {Michalik},
  {Millar}, {Miranda}, {Molina}, {Molinaro}, {Molinaro}, {Moln{\'a}r},
  {Moniez}, {Montegriffo}, {Monteiro}, {Mor}, {Mora}, {Morbidelli}, {Morel},
  {Morgenthaler}, {Morley}, {Morris}, {Mulone}, {Muraveva}, {Musella},
  {Narbonne}, {Nelemans}, {Nicastro}, {Noval}, {Ord{\'e}novic},
  {Ordieres-Mer{\'e}}, {Osborne}, {Pagani}, {Pagano}, {Pailler}, {Palacin},
  {Palaversa}, {Parsons}, {Paulsen}, {Pecoraro}, {Pedrosa}, {Pentik{\"a}inen},
  {Pereira}, {Pichon}, {Piersimoni}, {Pineau}, {Plachy}, {Plum}, {Poujoulet},
  {Pr{\v{s}}a}, {Pulone}, {Ragaini}, {Rago}, {Rambaux}, {Ramos-Lerate},
  {Ranalli}, {Rauw}, {Read}, {Regibo}, {Renk}, {Reyl{\'e}}, {Ribeiro},
  {Rimoldini}, {Ripepi}, {Riva}, {Rixon}, {Roelens}, {Romero-G{\'o}mez},
  {Rowell}, {Royer}, {Rudolph}, {Ruiz-Dern}, {Sadowski}, {Sagrist{\`a}
  Sell{\'e}s}, {Sahlmann}, {Salgado}, {Salguero}, {Sarasso}, {Savietto},
  {Schnorhk}, {Schultheis}, {Sciacca}, {Segol}, {Segovia}, {Segransan},
  {Serpell}, {Shih}, {Smareglia}, {Smart}, {Smith}, {Solano}, {Solitro},
  {Sordo}, {Soria Nieto}, {Souchay}, {Spagna}, {Spoto}, {Stampa}, {Steele},
  {Steidelm{\"u}ller}, {Stephenson}, {Stoev}, {Suess}, {S{\"u}veges}, {Surdej},
  {Szabados}, {Szegedi-Elek}, {Tapiador}, {Taris}, {Tauran}, {Taylor},
  {Teixeira}, {Terrett}, {Tingley}, {Trager}, {Turon}, {Ulla}, {Utrilla},
  {Valentini}, {van Elteren}, {Van Hemelryck}, {van Leeuwen}, {Varadi},
  {Vecchiato}, {Veljanoski}, {Via}, {Vicente}, {Vogt}, {Voss}, {Votruba},
  {Voutsinas}, {Walmsley}, {Weiler}, {Weingrill}, {Werner}, {Wevers},
  {Whitehead}, {Wyrzykowski}, {Yoldas}, {{\v{Z}}erjal}, {Zucker}, {Zurbach},
  {Zwitter}, {Alecu}, {Allen}, {Allende Prieto}, {Amorim},
  {Anglada-Escud{\'e}}, {Arsenijevic}, {Azaz}, {Balm}, {Beck}, {Bernstein},
  {Bigot}, {Bijaoui}, {Blasco}, {Bonfigli}, {Bono}, {Boudreault}, {Bressan},
  {Brown}, {Brunet}, {Bunclark}, {Buonanno}, {Butkevich}, {Carret}, {Carrion},
  {Chemin}, {Ch{\'e}reau}, {Corcione}, {Darmigny}, {de Boer}, {de Teodoro}, {de
  Zeeuw}, {Delle Luche}, {Domingues}, {Dubath}, {Fodor}, {Fr{\'e}zouls},
  {Fries}, {Fustes}, {Fyfe}, {Gallardo}, {Gallegos}, {Gardiol}, {Gebran},
  {Gomboc}, {G{\'o}mez}, {Grux}, {Gueguen}, {Heyrovsky}, {Hoar}, {Iannicola},
  {Isasi Parache}, {Janotto}, {Joliet}, {Jonckheere}, {Keil}, {Kim},
  {Klagyivik}, {Klar}, {Knude}, {Kochukhov}, {Kolka}, {Kos}, {Kutka}, {Lainey},
  {LeBouquin}, {Liu}, {Loreggia}, {Makarov}, {Marseille}, {Martayan},
  {Martinez-Rubi}, {Massart}, {Meynadier}, {Mignot}, {Munari}, {Nguyen},
  {Nordlander}, {Ocvirk}, {O'Flaherty}, {Olias Sanz}, {Ortiz}, {Osorio},
  {Oszkiewicz}, {Ouzounis}, {Palmer}, {Park}, {Pasquato}, {Peltzer}, {Peralta},
  {P{\'e}turaud}, {Pieniluoma}, {Pigozzi}, {Poels}, {Prat}, {Prod'homme},
  {Raison}, {Rebordao}, {Risquez}, {Rocca-Volmerange}, {Rosen}, {Ruiz-Fuertes},
  {Russo}, {Sembay}, {Serraller Vizcaino}, {Short}, {Siebert}, {Silva},
  {Sinachopoulos}, {Slezak}, {Soffel}, {Sosnowska}, {Strai{\v{z}}ys}, {ter
  Linden}, {Terrell}, {Theil}, {Tiede}, {Troisi}, {Tsalmantza}, {Tur},
  {Vaccari}, {Vachier}, {Valles}, {Van Hamme}, {Veltz}, {Virtanen}, {Wallut},
  {Wichmann}, {Wilkinson}, {Ziaeepour}, \& {Zschocke}}]{2016A&A...595A...1G}
{Gaia Collaboration}, {Prusti}, T., {de Bruijne}, J.~H.~J., {et~al.} 2016,
  \aap, 595, A1

\bibitem[{{Geffert}(2020)}]{2020BAVSR..69..181G}
{Geffert}, M. 2020, BAV Rundbrief - Mitteilungsblatt der Berliner
  Arbeits-gemeinschaft f\"ur Ver\"anderliche Sterne, 69, 181

\bibitem[{{Grindlay} {et~al.}(2012){Grindlay}, {Tang}, {Los}, \&
  {Servillat}}]{2012IAUS..285...29G}
{Grindlay}, J., {Tang}, S., {Los}, E., \& {Servillat}, M. 2012, in IAU
  Symposium, Vol. 285, New Horizons in Time Domain Astronomy, ed. E.~{Griffin},
  R.~{Hanisch}, \& R.~{Seaman}, 29--34

\bibitem[{{Grindlay} {et~al.}(2009){Grindlay}, {Tang}, {Simcoe}, {Laycock},
  {Los}, {Mink}, {Doane}, \& {Champine}}]{2009ASPC..410..101G}
{Grindlay}, J., {Tang}, S., {Simcoe}, R., {et~al.} 2009, in Astronomical
  Society of the Pacific Conference Series, Vol. 410, Preserving Astronomy's
  Photographic Legacy: Current State and the Future of North American
  Astronomical Plates, ed. W.~{Osborn} \& L.~{Robbins}, 101

\bibitem[{Groote(2014)}]{groote2014}
Groote, D. 2014, Nuncius Hamburgensis, 24, 453, {ISBN 978-3-8495-7967-8, Hg.
  Gudrun Wolfschmidt}

\bibitem[{Groote(2018)}]{groote2018}
Groote, D. 2018, {in 'Selene’s Two Faces: From 17th Century Drawings to
  Spacecraft Imaging, Nuncius Series: Studies and Sources in the Material and
  Visual History of Science}, 3, 206, {(ISBN 978-90-04-29886-6 und ISBN
  978-90-04-29887-3), Hg. Carmen Pérez Gonzáles}

\bibitem[{{Groote} {et~al.}(2014){Groote}, {Tuvikene}, {Edelmann}, {Arlt},
  {Heber}, \& {Enke}}]{2014aspl.conf...53G}
{Groote}, D., {Tuvikene}, T., {Edelmann}, H., {et~al.} 2014, in Astroplate
  2014, 53

\bibitem[{{Hagen} {et~al.}(1995){Hagen}, {Groote}, {Engels}, \&
  {Reimers}}]{1995A&AS..111..195H}
{Hagen}, H.~J., {Groote}, D., {Engels}, D., \& {Reimers}, D. 1995, \aaps, 111,
  195

\bibitem[{{Hale}(1901)}]{1901ApJ....13..173H}
{Hale}, G.~E. 1901, \apj, 13, 173

\bibitem[{{Hambly} {et~al.}(2001{\natexlab{a}}){Hambly}, {Davenhall}, {Irwin},
  \& {MacGillivray}}]{2001MNRAS.326.1315H}
{Hambly}, N.~C., {Davenhall}, A.~C., {Irwin}, M.~J., \& {MacGillivray}, H.~T.
  2001{\natexlab{a}}, \mnras, 326, 1315

\bibitem[{{Hambly} {et~al.}(2001{\natexlab{b}}){Hambly}, {Irwin}, \&
  {MacGillivray}}]{2001MNRAS.326.1295H}
{Hambly}, N.~C., {Irwin}, M.~J., \& {MacGillivray}, H.~T. 2001{\natexlab{b}},
  \mnras, 326, 1295

\bibitem[{{Hambly} {et~al.}(2001{\natexlab{c}}){Hambly}, {MacGillivray},
  {Read}, {Tritton}, {Thomson}, {Kelly}, {Morgan}, {Smith}, {Driver},
  {Williamson}, {Parker}, {Hawkins}, {Williams}, \&
  {Lawrence}}]{2001MNRAS.326.1279H}
{Hambly}, N.~C., {MacGillivray}, H.~T., {Read}, M.~A., {et~al.}
  2001{\natexlab{c}}, \mnras, 326, 1279

\bibitem[{{Harvey} {et~al.}(2016){Harvey}, {Redman}, {Boumis}, \&
  {Akras}}]{2016A&A...595A..64H}
{Harvey}, E., {Redman}, M.~P., {Boumis}, P., \& {Akras}, S. 2016, \aap, 595,
  A64

\bibitem[{{Heber}(2019)}]{lswst2019_heber}
{Heber}, U. 2019, {The APPLAUSE-Projekt: The plate archives}, {Large Surveys
  with Small Telescope: Past, Present and Future (Astroplate III), 2019 March
  11 – 13, Bamberg, Germany}, {DOI:10789/plate/lswst/007}

\bibitem[{{Henze} {et~al.}(2008){Henze}, {Meusinger}, \&
  {Pietsch}}]{2008A&A...477...67H}
{Henze}, M., {Meusinger}, H., \& {Pietsch}, W. 2008, \aap, 477, 67

\bibitem[{{Hippke} \& {Angerhausen}(2018)}]{2018ApJ...854L..11H}
{Hippke}, M. \& {Angerhausen}, D. 2018, \apjl, 854, L11

\bibitem[{{Hippke} {et~al.}(2016){Hippke}, {Angerhausen}, {Lund}, {Pepper}, \&
  {Stassun}}]{2016ApJ...825...73H}
{Hippke}, M., {Angerhausen}, D., {Lund}, M.~B., {Pepper}, J., \& {Stassun},
  K.~G. 2016, \apj, 825, 73

\bibitem[{{Hippke} {et~al.}(2017){Hippke}, {Kroll}, {Matthai}, {Angerhausen},
  {Tuvikene}, {Stassun}, {Roshchina}, {Vasileva}, {Izmailov}, {Samus},
  {Pastukhova}, {Bryukhanov}, \& {Lund}}]{2017ApJ...837...85H}
{Hippke}, M., {Kroll}, P., {Matthai}, F., {et~al.} 2017, \apj, 837, 85

\bibitem[{{Hoffleit}(1939)}]{1939BHarO.911...41H}
{Hoffleit}, D. 1939, Harvard College Observatory Bulletin, 911, 41

\bibitem[{{Hogg} {et~al.}(2008){Hogg}, {Blanton}, {Lang}, {Mierle}, \&
  {Roweis}}]{2008ASPC..394...27H}
{Hogg}, D.~W., {Blanton}, M., {Lang}, D., {Mierle}, K., \& {Roweis}, S. 2008,
  in Astronomical Society of the Pacific Conference Series, Vol. 394,
  Astronomical Data Analysis Software and Systems XVII, ed. R.~W. {Argyle},
  P.~S. {Bunclark}, \& J.~R. {Lewis}, 27

\bibitem[{{Hudec}(2018)}]{2018AN....339..408H}
{Hudec}, R. 2018, Astronomische Nachrichten, 339, 408

\bibitem[{{Hudec}(2019{\natexlab{a}})}]{lswst2019_Hudec}
{Hudec}, R. 2019{\natexlab{a}}, { Astronomical Photographic Archives: Past,
  Present, Future}, {Large Surveys with Small Telescope: Past, Present and
  Future (Astroplate III), 2019 March 11 – 13, Bamberg, Germany},
  {DOI:10789/plate/lswst/002}

\bibitem[{{Hudec}(2019{\natexlab{b}})}]{2019AN....340..690H}
{Hudec}, R. 2019{\natexlab{b}}, Astronomische Nachrichten, 340, 690

\bibitem[{{Hudec} {et~al.}(2013){Hudec}, {Kopel}, {Macsics}, {Hadwige},
  {Heber}, \& {Cay{\'e}}}]{2013AcPol..53c..27H}
{Hudec}, R., {Kopel}, F., {Macsics}, R., {et~al.} 2013, Acta Polytechnica, 53,
  27

\bibitem[{{Jia} {et~al.}(2023){Jia}, {Yang}, {Shang}, {Yu}, \&
  {Zhao}}]{2023PASJ...75..811J}
{Jia}, P., {Yang}, Z., {Shang}, Z., {Yu}, Y., \& {Zhao}, J. 2023, \pasj, 75,
  811

\bibitem[{{Jones}(2000)}]{2000A&G....41e..16J}
{Jones}, D. 2000, Astronomy and Geophysics, 41, 16

\bibitem[{{Khovrichev} {et~al.}(2020){Khovrichev}, {Narizhnaya}, {Vasil'eva},
  {Izmailov}, {Kulikova}, \& {Bikulova}}]{2020SoSyR..54..344K}
{Khovrichev}, M.~Y., {Narizhnaya}, N.~V., {Vasil'eva}, T.~A., {et~al.} 2020,
  Solar System Research, 54, 344

\bibitem[{{Kroll}(2009)}]{2009chao.conf..311K}
{Kroll}, P. 2009, in Cultural Heritage of Astronomical Observatories: From
  Classical Astronomy to Modern Astrophysics, 311--315

\bibitem[{{Kroll}(2019)}]{lswst2019_kroll}
{Kroll}, P. 2019, {Photographic and Digital Surveys at Sonneberg Observatory},
  {Large Surveys with Small Telescope: Past, Present and Future (Astroplate
  III), 2019 March 11 – 13, Bamberg, Germany}, {DOI:10789/plate/lswst/005}

\bibitem[{{Lamareille} {et~al.}(2003){Lamareille}, {Thi{\'e}vin}, {Fournis},
  {Grimault}, {Broquet}, \& {Davoust}}]{2003A&A...402..395L}
{Lamareille}, F., {Thi{\'e}vin}, J., {Fournis}, B., {et~al.} 2003, \aap, 402,
  395

\bibitem[{{Laycock} {et~al.}(2010){Laycock}, {Tang}, {Grindlay}, {Los},
  {Simcoe}, \& {Mink}}]{2010AJ....140.1062L}
{Laycock}, S., {Tang}, S., {Grindlay}, J., {et~al.} 2010, \aj, 140, 1062

\bibitem[{{Lehtinen} {et~al.}(2018){Lehtinen}, {Prusti}, {de Bruijne},
  {Lammers}, {Manara}, {Ness}, {Siddiqui}, {Markkanen}, {Poutanen}, \&
  {Muinonen}}]{2018A&A...616A.185L}
{Lehtinen}, K., {Prusti}, T., {de Bruijne}, J., {et~al.} 2018, \aap, 616, A185

\bibitem[{{Lehtinen} {et~al.}(2023){Lehtinen}, {Prusti}, {de Bruijne},
  {Lammers}, {Manara}, {Ness}, {Siddiqui}, {Poutanen}, {Muinonen}, \&
  {Morrison}}]{2023A&A...671A..16L}
{Lehtinen}, K., {Prusti}, T., {de Bruijne}, J., {et~al.} 2023, \aap, 671, A16

\bibitem[{{Liimets} {et~al.}(2012){Liimets}, {Corradi},
  {Santander-Garc{\'\i}a}, {Villaver}, {Rodr{\'\i}guez-Gil}, {Verro}, \&
  {Kolka}}]{2012ApJ...761...34L}
{Liimets}, T., {Corradi}, R.~L.~M., {Santander-Garc{\'\i}a}, M., {et~al.} 2012,
  \apj, 761, 34

\bibitem[{{Ludendorff}(1912)}]{1912AN....192..123L}
{Ludendorff}, H. 1912, Astronomische Nachrichten, 192, 123

\bibitem[{Lund {et~al.}(2016)Lund, Pepper, Stassun, Hippke, \&
  Angerhausen}]{lund2016stability}
Lund, M.~B., Pepper, J., Stassun, K.~G., Hippke, M., \& Angerhausen, D. 2016,
  The Stability of F-star Brightness on Century Timescales

\bibitem[{{Lynds}(1963)}]{1963ASPL....9...89L}
{Lynds}, B.~T. 1963, Leaflet of the Astronomical Society of the Pacific, 9, 89

\bibitem[{{Markarian} \& {Stepanian}(1983)}]{1983Afz....19..639M}
{Markarian}, B.~E. \& {Stepanian}, D.~A. 1983, Astrofizika, 19, 639

\bibitem[{{Mart{\'\i}nez Gonz{\'a}lez} {et~al.}(2019){Mart{\'\i}nez
  Gonz{\'a}lez}, {Gonz{\'a}lez-Fern{\'a}ndez}, {Asensio Ramos},
  {Socas-Navarro}, {Westendorp Plaza}, {Boyajian}, {Wright}, {Collier Cameron},
  {Gonz{\'a}lez Hern{\'a}ndez}, {Holgado}, {Kennedy}, {Masseron}, {Molinari},
  {Saario}, {Sim{\'o}n-D{\'\i}az}, \&
  {Toledo-Padr{\'o}n}}]{2019MNRAS.486..236M}
{Mart{\'\i}nez Gonz{\'a}lez}, M.~J., {Gonz{\'a}lez-Fern{\'a}ndez}, C., {Asensio
  Ramos}, A., {et~al.} 2019, \mnras, 486, 236

\bibitem[{{Matijevic}(2019)}]{lswst2019_matijevic}
{Matijevic}, G. 2019, {False-positives detection with convolutional neural
  networks}, {Large Surveys with Small Telescope: Past, Present and Future
  (Astroplate III), 2019 March 11 – 13, Bamberg, Germany},
  {DOI:10789/plate/lswst/012}

\bibitem[{{Meusinger} {et~al.}(1999){Meusinger}, {Brunzendorf}, {Froebrich}, \&
  {Krieg}}]{1999AcHA....6..223M}
{Meusinger}, H., {Brunzendorf}, J., {Froebrich}, D., \& {Krieg}, R. 1999, Acta
  Historica Astronomiae, 6, 223

\bibitem[{{Mickaelian} {et~al.}(2019){Mickaelian}, {Erastova}, {Gyulzadyan},
  {Ohanian}, {Gigoyan}, {Mikayelyan}, {Sinamyan}, \&
  {Paronyan}}]{2019ASPC..520..117M}
{Mickaelian}, A.~M., {Erastova}, L.~K., {Gyulzadyan}, M.~V., {et~al.} 2019, in
  Astronomical Society of the Pacific Conference Series, Vol. 520, Astronomical
  Heritage of the Middle East, ed. S.~V. {Farmanyan}, A.~M. {Mickaelian},
  J.~{McKim Malville}, \& M.~{Bagheri}, 117

\bibitem[{{Mickaelian} {et~al.}(2021){Mickaelian}, {Sargsyan}, {Mikayelyan},
  {Gigoyan}, {Nesci}, \& {Rossi}}]{2021CoBAO..68..390M}
{Mickaelian}, A.~M., {Sargsyan}, L.~A., {Mikayelyan}, G.~A., {et~al.} 2021,
  Communications of the Byurakan Astrophysical Observatory, 68, 390

\bibitem[{{Nesci}(2019)}]{lswst2019_nesci}
{Nesci}, R. 2019, {Records from the past: variable stars from the Asiago plate
  archive}, {Large Surveys with Small Telescope: Past, Present and Future
  (Astroplate III), 2019 March 11 – 13, Bamberg, Germany},
  {DOI:10789/plate/lswst/004}

\bibitem[{{Nesci} {et~al.}(2017){Nesci}, {Enke}, {Rossi}, {Tuvikene}, \&
  {Bagaglia}}]{2017MmSAI..88..444N}
{Nesci}, R., {Enke}, H., {Rossi}, C., {Tuvikene}, T., \& {Bagaglia}, M. 2017,
  \memsai, 88, 444

\bibitem[{{Nesci} {et~al.}(2018){Nesci}, {Tuvikene}, {Rossi}, {Gaudenzi},
  {Galleti}, {Ochner}, \& {Enke}}]{2018RMxAA..54..341N}
{Nesci}, R., {Tuvikene}, T., {Rossi}, C., {et~al.} 2018, \rmxaa, 54, 341

\bibitem[{{Osterbrock} \& {Ferland}(2006)}]{2006agna.book.....O}
{Osterbrock}, D.~E. \& {Ferland}, G.~J. 2006, {Astrophysics of gaseous nebulae
  and active galactic nuclei}

\bibitem[{{Pal} {et~al.}(2020){Pal}, {Verma}, {Rendtel}, {Gonz{\'a}lez
  Manrique}, {Enke}, \& {Denker}}]{2020AN....341..575P}
{Pal}, P.~S., {Verma}, M., {Rendtel}, J., {et~al.} 2020, Astronomische
  Nachrichten, 341, 575

\bibitem[{{Pei} {et~al.}(2024){Pei}, {Orio}, \& {Zhang}}]{2024MNRAS.tmp..690P}
{Pei}, S., {Orio}, M., \& {Zhang}, X. 2024, \mnras [\eprint[arXiv]{2403.05328}]

\bibitem[{{Perlbarg} {et~al.}(2023){Perlbarg}, {Desmars}, {Robert}, \&
  {Hestroffer}}]{2023A&A...680A..41P}
{Perlbarg}, A.~C., {Desmars}, J., {Robert}, V., \& {Hestroffer}, D. 2023, \aap,
  680, A41

\bibitem[{{Pesch} \& {Sanduleak}(1983)}]{1983ApJS...51..171P}
{Pesch}, P. \& {Sanduleak}, N. 1983, \apjs, 51, 171

\bibitem[{{Pickering}(1912)}]{pickering1912}
{Pickering}. 1912, Astronomische Nachrichten, 191, 101

\bibitem[{{Pickering}(1901)}]{1901ApJ....13..170P}
{Pickering}, E.~C. 1901, \apj, 13, 170

\bibitem[{Poghosyan {et~al.}(2014)Poghosyan, Pfau, Tsvetkova, Mugrauer,
  Tsvetkov, Hambaryan, Neuhaeuser, \& Kalaglarsky}]{jena_2014}
Poghosyan, A., Pfau, W., Tsvetkova, K., {et~al.} 2014, Astronomische
  Nachrichten, 335

\bibitem[{{Ramakrishnan} \& {Dwarkadas}(2020)}]{2020ApJ...901..119R}
{Ramakrishnan}, V. \& {Dwarkadas}, V.~V. 2020, \apj, 901, 119

\bibitem[{{Rapaport} {et~al.}(2006){Rapaport}, {Ducourant}, {Le Campion},
  {Fresneau}, {Argyle}, {Soubiran}, {Teixeira}, {Camargo}, {Colin}, {Daigne},
  {P{\'e}ri{\'e}}, \& {R{\'e}qui{\`e}me}}]{2006A&A...449..435R}
{Rapaport}, M., {Ducourant}, C., {Le Campion}, J.~F., {et~al.} 2006, \aap, 449,
  435

\bibitem[{{Robert} {et~al.}(2011){Robert}, {de Cuyper}, {Arlot}, {de Decker},
  {Guibert}, {Lainey}, {Pascu}, {Winter}, \& {Zacharias}}]{2011MNRAS.415..701R}
{Robert}, V., {de Cuyper}, J.~P., {Arlot}, J.~E., {et~al.} 2011, \mnras, 415,
  701

\bibitem[{{Robert} {et~al.}(2021){Robert}, {Desmars}, {Lainey}, {Arlot},
  {Perlbarg}, {Horville}, {Aboudarham}, {Etienne}, {Gu{\'e}rard}, {Ilovaisky},
  {Khovritchev}, {Le Poncin-Lafitte}, {Le Van Suu}, {Neiner}, {Pascu},
  {Poirier}, {Schneider}, {Tanga}, \& {Valls-Gabaud}}]{2021A&A...652A...3R}
{Robert}, V., {Desmars}, J., {Lainey}, V., {et~al.} 2021, \aap, 652, A3

\bibitem[{{Schaefer}(2016{\natexlab{a}})}]{2016MNRAS.460.1233S}
{Schaefer}, B.~E. 2016{\natexlab{a}}, \mnras, 460, 1233

\bibitem[{{Schaefer}(2016{\natexlab{b}})}]{2016ApJ...822L..34S}
{Schaefer}, B.~E. 2016{\natexlab{b}}, \apjl, 822, L34

\bibitem[{{Schaefer}(2018)}]{2018MNRAS.481.3033S}
{Schaefer}, B.~E. 2018, \mnras, 481, 3033

\bibitem[{{Schaefer}(2019)}]{2019RNAAS...3...77S}
{Schaefer}, B.~E. 2019, Research Notes of the American Astronomical Society, 3,
  77

\bibitem[{{Schaefer}(2023)}]{2023MNRAS.524.3146S}
{Schaefer}, B.~E. 2023, \mnras, 524, 3146

\bibitem[{{Schaefer} \& {Edwards}(2015)}]{2015ApJ...812..133S}
{Schaefer}, B.~E. \& {Edwards}, Z.~I. 2015, \apj, 812, 133

\bibitem[{{Schmidt}(2019)}]{2019ApJ...880L...7S}
{Schmidt}, E.~G. 2019, \apjl, 880, L7

\bibitem[{Schorr(1912)}]{schorr1912}
Schorr, R. 1912, Jahrbuch der Hamburgischen wissenschaftlichen Anstalten, XXX

\bibitem[{{Schwa\ss mann}(1912)}]{schwassmann1912}
{Schwa\ss mann}. 1912, Astronomische Nachrichten, 191, 101

\bibitem[{{Simcoe} {et~al.}(2006){Simcoe}, {Grindlay}, {Los}, {Doane},
  {Laycock}, {Mink}, {Champine}, \& {Sliski}}]{2006SPIE.6312E..17S}
{Simcoe}, R.~J., {Grindlay}, J.~E., {Los}, E.~J., {et~al.} 2006, in Society of
  Photo-Optical Instrumentation Engineers (SPIE) Conference Series, Vol. 6312,
  Applications of Digital Image Processing XXIX, ed. A.~G. {Tescher}, 631217

\bibitem[{{Slettebak} \& {Brundage}(1971)}]{1971AJ.....76..338S}
{Slettebak}, A. \& {Brundage}, R.~K. 1971, \aj, 76, 338

\bibitem[{{Sokolovsky} {et~al.}(2017){Sokolovsky}, {Zubareva}, {Kolesnikova},
  {Samus}, {Antipin}, \& {Belinski}}]{2017arXiv171204672S}
{Sokolovsky}, K.~V., {Zubareva}, A.~M., {Kolesnikova}, D.~M., {et~al.} 2017,
  arXiv e-prints, arXiv:1712.04672

\bibitem[{{Soria} \& {Perna}(2008)}]{2008ApJ...683..767S}
{Soria}, R. \& {Perna}, R. 2008, \apj, 683, 767

\bibitem[{{Spasovic} {et~al.}(2016){Spasovic}, {Dersch}, {Lange}, {Jovanovic},
  \& {Schrimpf}}]{2016arXiv161000265S}
{Spasovic}, M., {Dersch}, C., {Lange}, C., {Jovanovic}, D., \& {Schrimpf}, A.
  2016, arXiv e-prints, arXiv:1610.00265

\bibitem[{{Staubermann}(2004)}]{2004JHA....35..447S}
{Staubermann}, K. 2004, Journal for the History of Astronomy, 35, 447

\bibitem[{{Tang} {et~al.}(2013){Tang}, {Grindlay}, {Los}, \&
  {Servillat}}]{2013PASP..125..857T}
{Tang}, S., {Grindlay}, J., {Los}, E., \& {Servillat}, M. 2013, \pasp, 125, 857

\bibitem[{{Taylor}(2005)}]{2005ASPC..347...29T}
{Taylor}, M.~B. 2005, in Astronomical Society of the Pacific Conference Series,
  Vol. 347, Astronomical Data Analysis Software and Systems XIV, ed.
  P.~{Shopbell}, M.~{Britton}, \& R.~{Ebert}, 29

\bibitem[{{Tsvetkov} \& {Tsvetkova}(2012)}]{2012IAUS..285..417T}
{Tsvetkov}, M. \& {Tsvetkova}, K. 2012, in New Horizons in Time Domain
  Astronomy, ed. E.~{Griffin}, R.~{Hanisch}, \& R.~{Seaman}, Vol. 285, 417--419

\bibitem[{{Tsvetkov} {et~al.}(2005{\natexlab{a}}){Tsvetkov}, {Tsvetkova},
  {Borisova}, {Kalaglarsky}, {Bogdanovski}, {Heber}, {Bues}, {Drechsel}, \&
  {Knigge}}]{2005PASRB...5..303T}
{Tsvetkov}, M., {Tsvetkova}, K., {Borisova}, A., {et~al.} 2005{\natexlab{a}},
  Publications of the Astronomical Society ``Rudjer Boskovic'', 5, 303

\bibitem[{{Tsvetkov} {et~al.}(1999){Tsvetkov}, {Tsvetkova}, {Richter},
  {Scholz}, \& {B{\"o}hm}}]{1999AN....320...63T}
{Tsvetkov}, M., {Tsvetkova}, K., {Richter}, G., {Scholz}, G., \& {B{\"o}hm}, P.
  1999, Astronomische Nachrichten, 320, 63

\bibitem[{{Tsvetkov} {et~al.}(2005{\natexlab{b}}){Tsvetkov}, {Tsvetkova},
  {Stavrev}, {Richter}, {B{\"o}hm}, \& {Staubermann}}]{2005PASRB...5..309T}
{Tsvetkov}, M., {Tsvetkova}, K., {Stavrev}, K.~Y., {et~al.} 2005{\natexlab{b}},
  Publications of the Astronomical Society ``Rudjer Boskovic'', 5, 309

\bibitem[{{Tsvetkova} {et~al.}(2009){Tsvetkova}, {Tsvetkov}, {B{\"o}hm},
  {Steinmetz}, \& {Dick}}]{2009AN....330..878T}
{Tsvetkova}, K., {Tsvetkov}, M., {B{\"o}hm}, P., {Steinmetz}, M., \& {Dick},
  W.~R. 2009, Astronomische Nachrichten, 330, 878

\bibitem[{{Tsvetkova} {et~al.}(2018){Tsvetkova}, {Tsvetkov}, {Kirov},
  {Kalaglarsky}, {Edelmann}, \& {Heber}}]{2018A&AT...30..467T}
{Tsvetkova}, K., {Tsvetkov}, M., {Kirov}, N., {et~al.} 2018, Astronomical and
  Astrophysical Transactions, 30, 467

\bibitem[{{Tsvetkova} {et~al.}(2006){Tsvetkova}, {Tsvetkov}, {Heber},
  {Bogdanovski}, \& {Kalaglarsky}}]{2006vopc.conf..109T}
{Tsvetkova}, K.~P., {Tsvetkov}, M.~K., {Heber}, U., {Bogdanovski}, R.~G., \&
  {Kalaglarsky}, D.~G. 2006, in Virtual Observatory: Plate Content
  Digitization, Archive Mining and Image Sequence Processing, ed.
  M.~{Tsvetkov}, V.~{Golev}, F.~{Murtagh}, \& R.~{Molina}, 109--114

\bibitem[{{Tuvikene}(2019)}]{lswst2019_tuvikene}
{Tuvikene}, T. 2019, {PyPlate: a software package for processing digitized
  astronomical photographic plates}, {Large Surveys with Small Telescope: Past,
  Present and Future (Astroplate III), 2019 March 11 – 13, Bamberg, Germany},
  {DOI:10789/plate/lswst/011}

\bibitem[{{Tuvikene} {et~al.}(2014){Tuvikene}, {Edelmann}, {Groote}, \&
  {Enke}}]{2014aspl.conf..127T}
{Tuvikene}, T., {Edelmann}, H., {Groote}, D., \& {Enke}, H. 2014, in Astroplate
  2014, 127

\bibitem[{{Vicente} {et~al.}(2007){Vicente}, {Abad}, \&
  {Garz{\'o}n}}]{2007A&A...471.1077V}
{Vicente}, B., {Abad}, C., \& {Garz{\'o}n}, F. 2007, \aap, 471, 1077

\bibitem[{{Vicente} {et~al.}(2010){Vicente}, {Abad}, {Garz{\'o}n}, \&
  {Girard}}]{2010A&A...509A..62V}
{Vicente}, B., {Abad}, C., {Garz{\'o}n}, F., \& {Girard}, T.~M. 2010, \aap,
  509, A62

\bibitem[{{{\v{S}}imon}(2002)}]{2002A&A...382..910S}
{{\v{S}}imon}, V. 2002, \aap, 382, 910

\bibitem[{{{\v{S}}imon} \& {Edelmann}(2019)}]{2019AstBu..74..490A}
{{\v{S}}imon}, V. \& {Edelmann}, H. 2019, Astrophysical Bulletin, 74, 490

\bibitem[{{Wells} {et~al.}(1981){Wells}, {Greisen}, \&
  {Harten}}]{1981A&AS...44..363W}
{Wells}, D.~C., {Greisen}, E.~W., \& {Harten}, R.~H. 1981, \aaps, 44, 363

\bibitem[{{Wertz} {et~al.}(2017){Wertz}, {Horns}, {Groote}, {Tuvikene},
  {Czesla}, \& {Schmitt}}]{2017AN....338..103W}
{Wertz}, M., {Horns}, D., {Groote}, D., {et~al.} 2017, Astronomische
  Nachrichten, 338, 103

\bibitem[{{Williams}(1919)}]{1919MNRAS..79..362W}
{Williams}, A.~S. 1919, \mnras, 79, 362

\bibitem[{{Williams}(1992)}]{1992AJ....104..725W}
{Williams}, R.~E. 1992, \aj, 104, 725

\bibitem[{{Wisotzki} {et~al.}(1996){Wisotzki}, {Koehler}, {Groote}, \&
  {Reimers}}]{1996A&AS..115..227W}
{Wisotzki}, L., {Koehler}, T., {Groote}, D., \& {Reimers}, D. 1996, \aaps, 115,
  227

\bibitem[{{Wolf}(1912)}]{1912AN....191..167W}
{Wolf}, M. 1912, Astronomische Nachrichten, 191, 167

\bibitem[{{Yacob} {et~al.}(2022){Yacob}, {Berdnikov}, {Pastukhova}, {Kniazev},
  \& {Whitelock}}]{2022MNRAS.516.2095Y}
{Yacob}, A.~M., {Berdnikov}, L.~N., {Pastukhova}, E.~N., {Kniazev}, A.~Y., \&
  {Whitelock}, P.~A. 2022, \mnras, 516, 2095

\bibitem[{{Yan} {et~al.}(2016){Yan}, {Qiao}, {Dourneau}, {Yu}, {Zhang},
  {Cheng}, \& {Xi}}]{2016MNRAS.457.2900Y}
{Yan}, D., {Qiao}, R.~C., {Dourneau}, G., {et~al.} 2016, \mnras, 457, 2900

\bibitem[{{Yu} {et~al.}(2017){Yu}, {Zhao}, {Tang}, \&
  {Shang}}]{2017RAA....17...28Y}
{Yu}, Y., {Zhao}, J.-H., {Tang}, Z.-H., \& {Shang}, Z.-J. 2017, Research in
  Astronomy and Astrophysics, 17, 28

\bibitem[{{Zacharias} {et~al.}(2008){Zacharias}, {Winter}, {Holdenried}, {De
  Cuyper}, {Rafferty}, \& {Wycoff}}]{2008PASP..120..644Z}
{Zacharias}, N., {Winter}, L., {Holdenried}, E.~R., {et~al.} 2008, \pasp, 120,
  644

\bibitem[{{Ziener}(1994)}]{1994IAUS..161..367Z}
{Ziener}, R. 1994, in Astronomy from Wide-Field Imaging, ed. H.~T.
  {MacGillivray}, Vol. 161, 367

\bibitem[{{Zinner}(1939)}]{1939VeBam...4....1Z}
{Zinner}, E. 1939, Veroeffentlichungen der Remeis-Sternwarte zu Bamberg, 4, 1

\end{thebibliography}

\begin{appendix}

\onecolumn
\section{Mollweide maps of observations per archive} 
\label{sect:mollweide_appendix}
\begin{figure}[ht]
\begin{tabular}{cc}
        \includegraphics[width=0.49\textwidth]{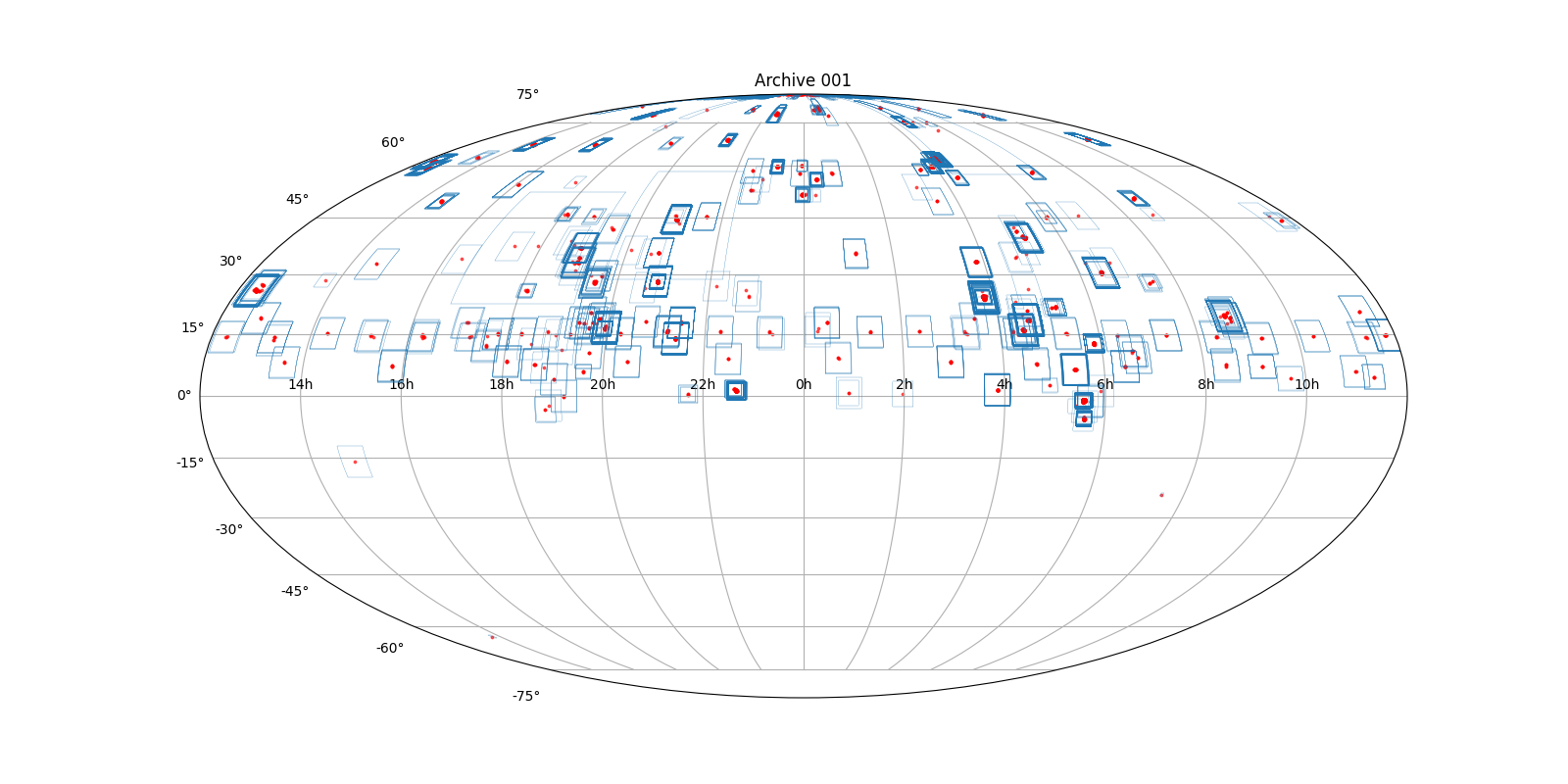} & 
        \includegraphics[width=0.49\textwidth]{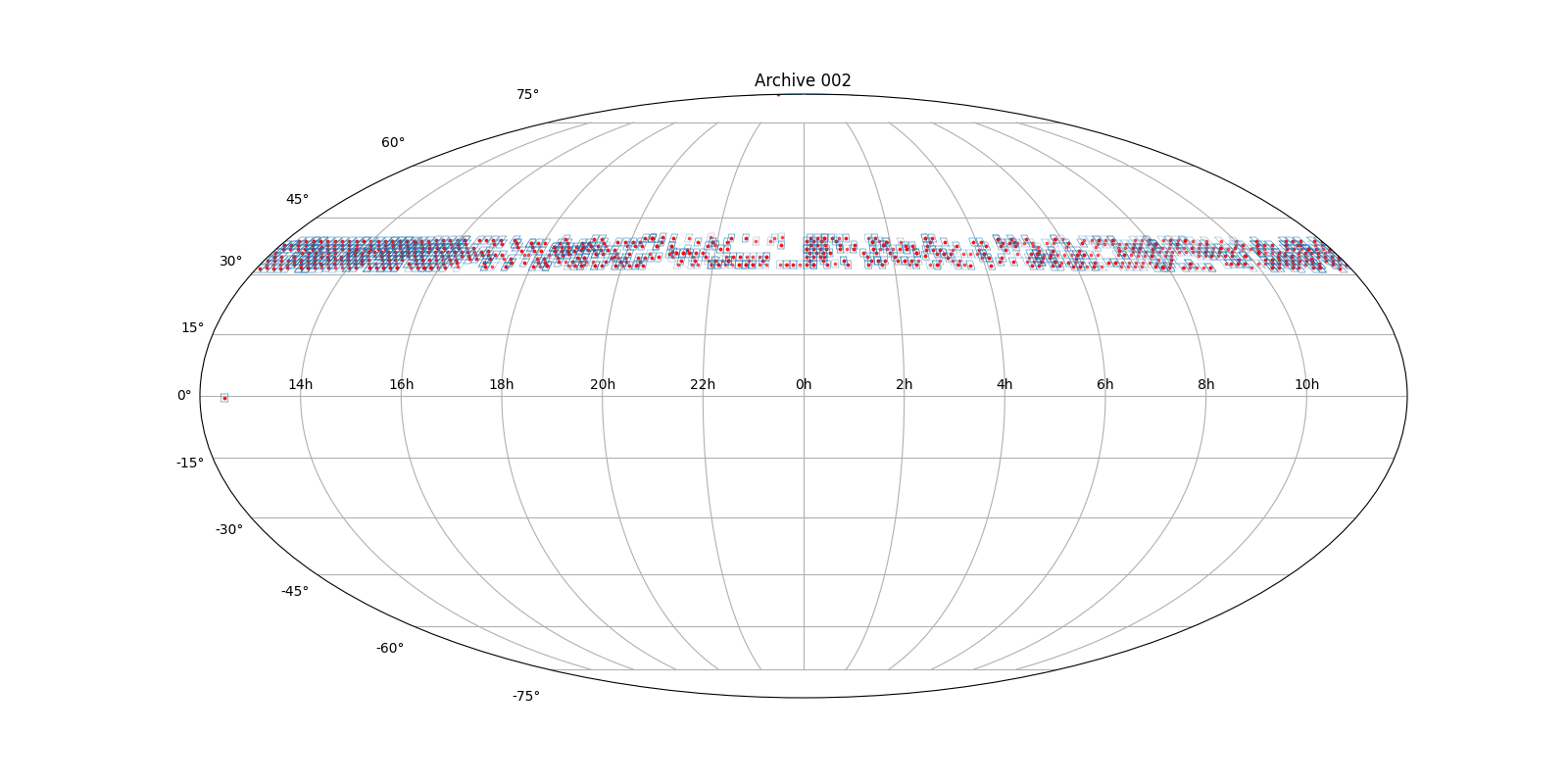}
 \\[-11pt]
        Zeiss Triplet, Potsdam, 4921 plates (ID 1) & Carte du Ciel, Potsdam,  979 plates (ID 2)
  \\[1pt]
        \includegraphics[width=0.49\textwidth]{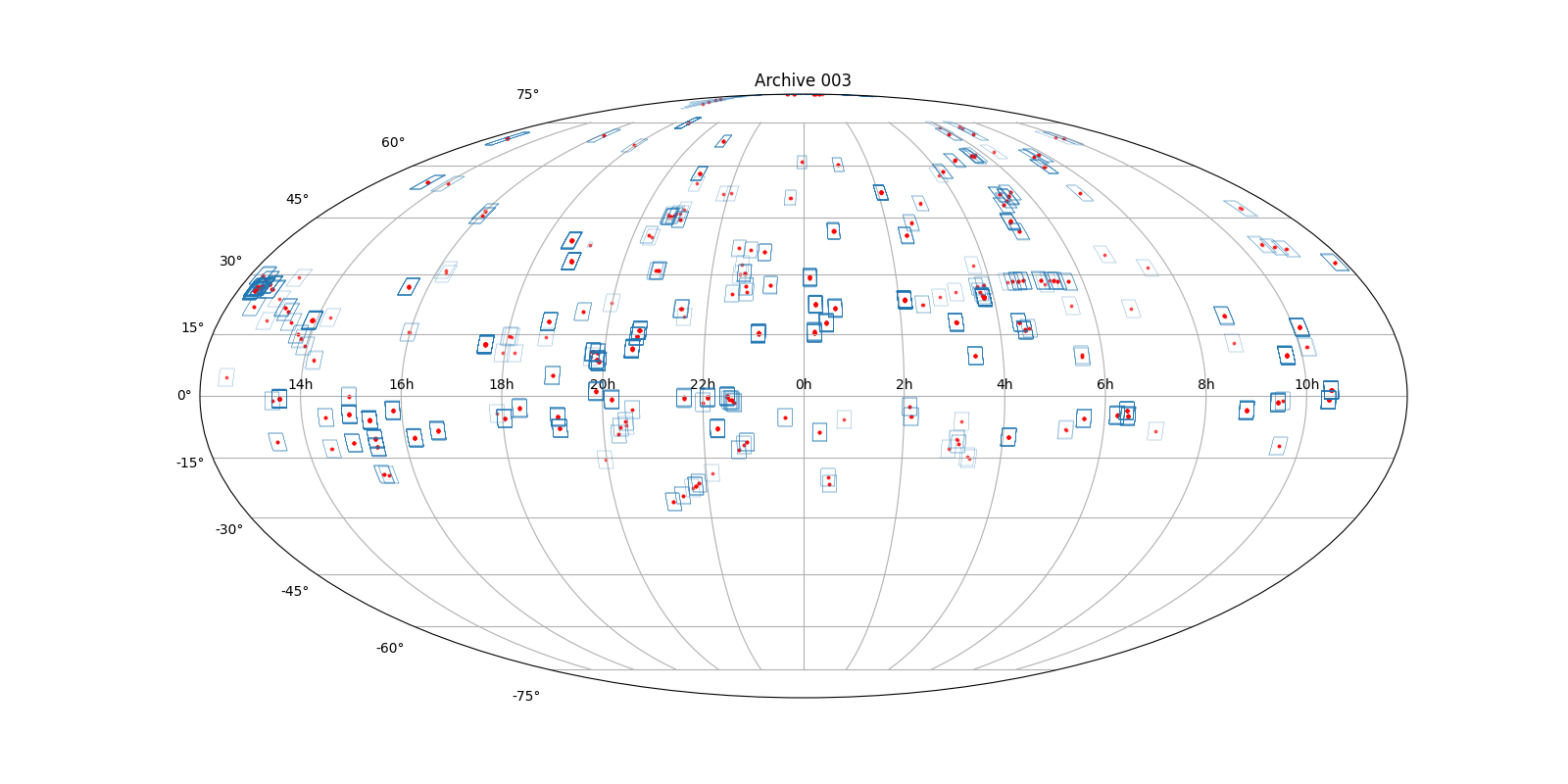} & 
        \includegraphics[width=0.49\textwidth]{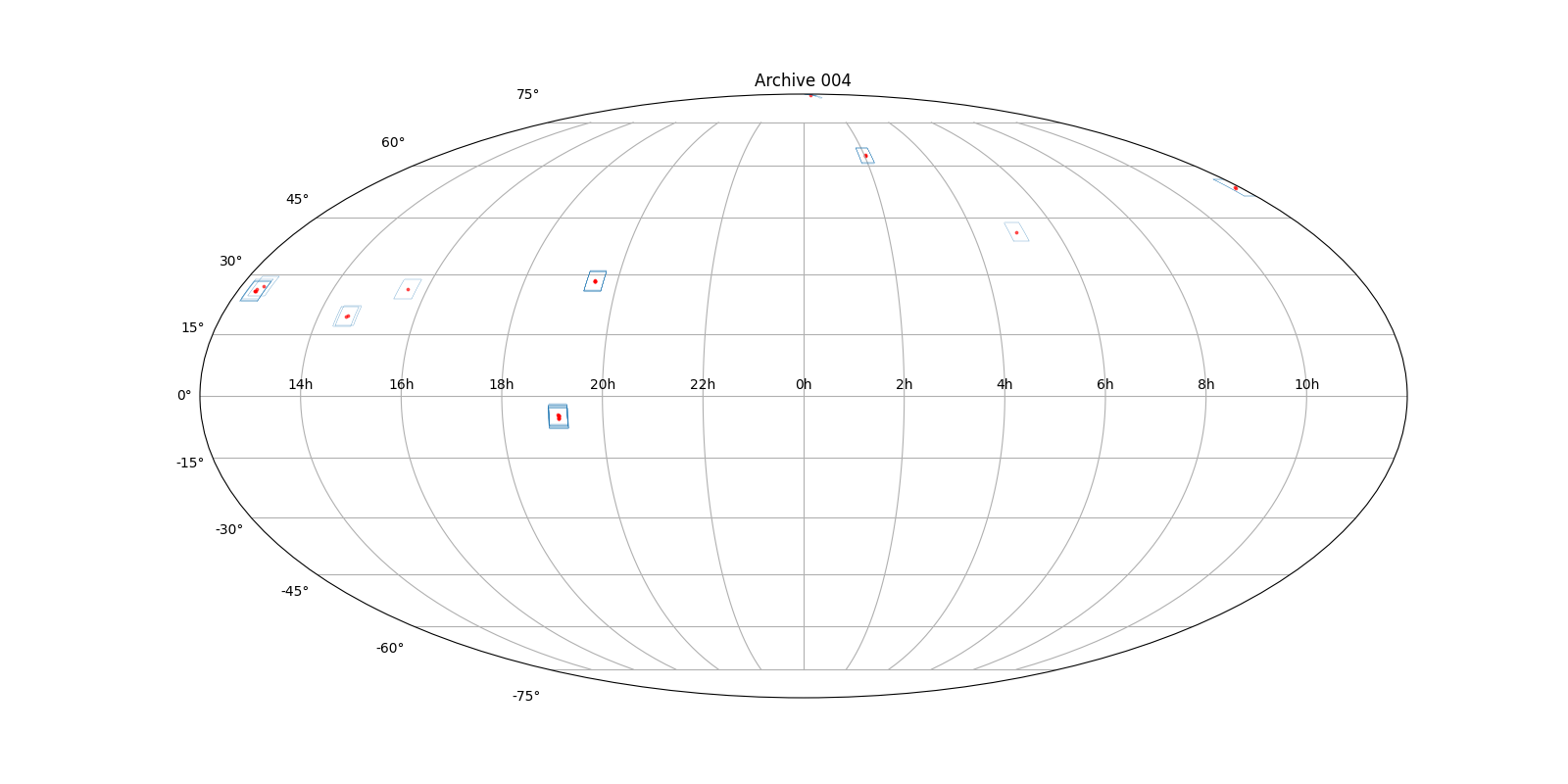}
 \\[-11pt]
        Great Schmidt Camera, Potsdam, 508 plates (ID 3)&  Small Schmidt Camera, Potsdam, 101 plates (ID 4)
  \\[1pt]
        \includegraphics[width=0.49\textwidth]{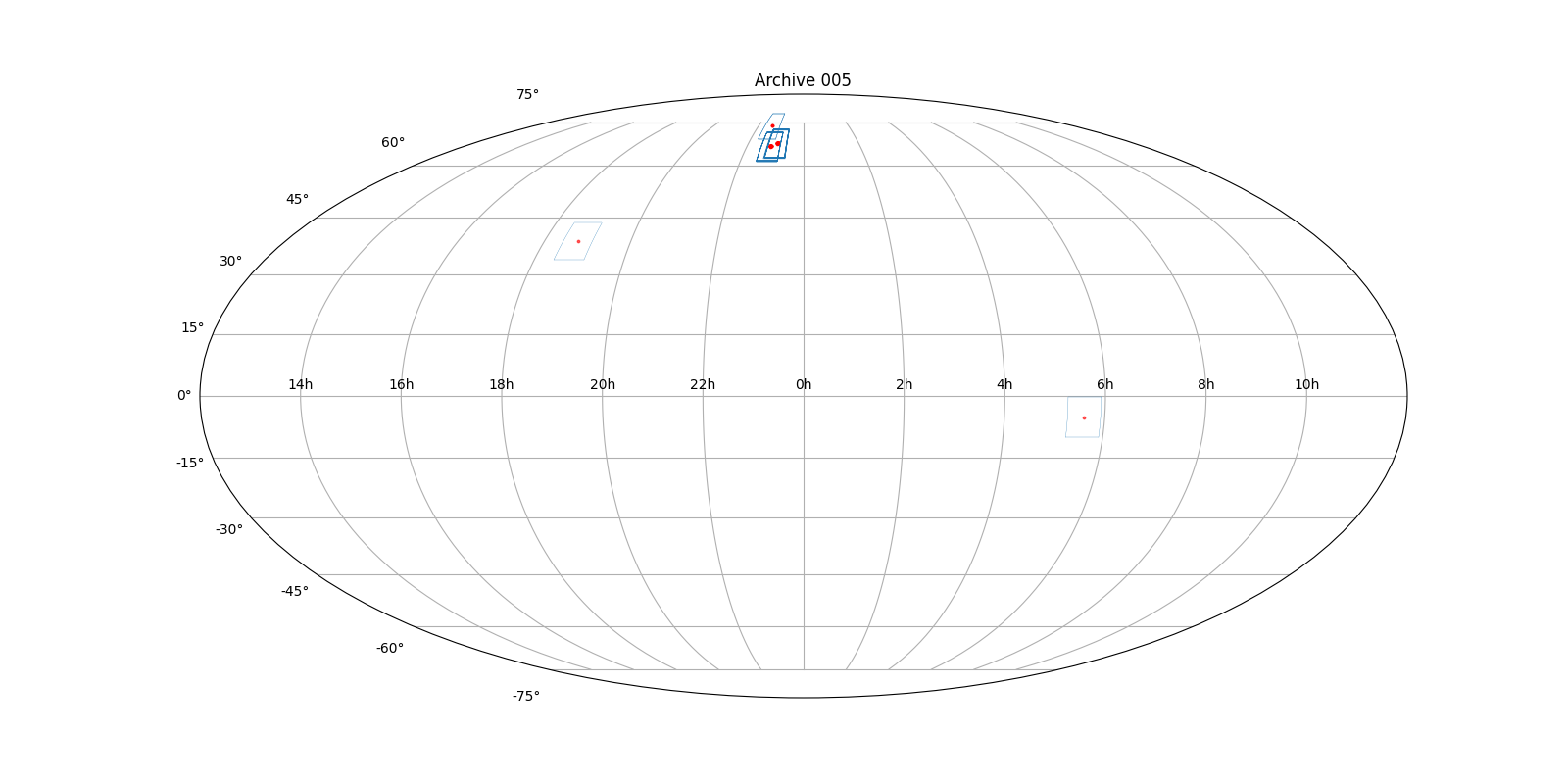} & 
    \includegraphics[width=0.5\textwidth]{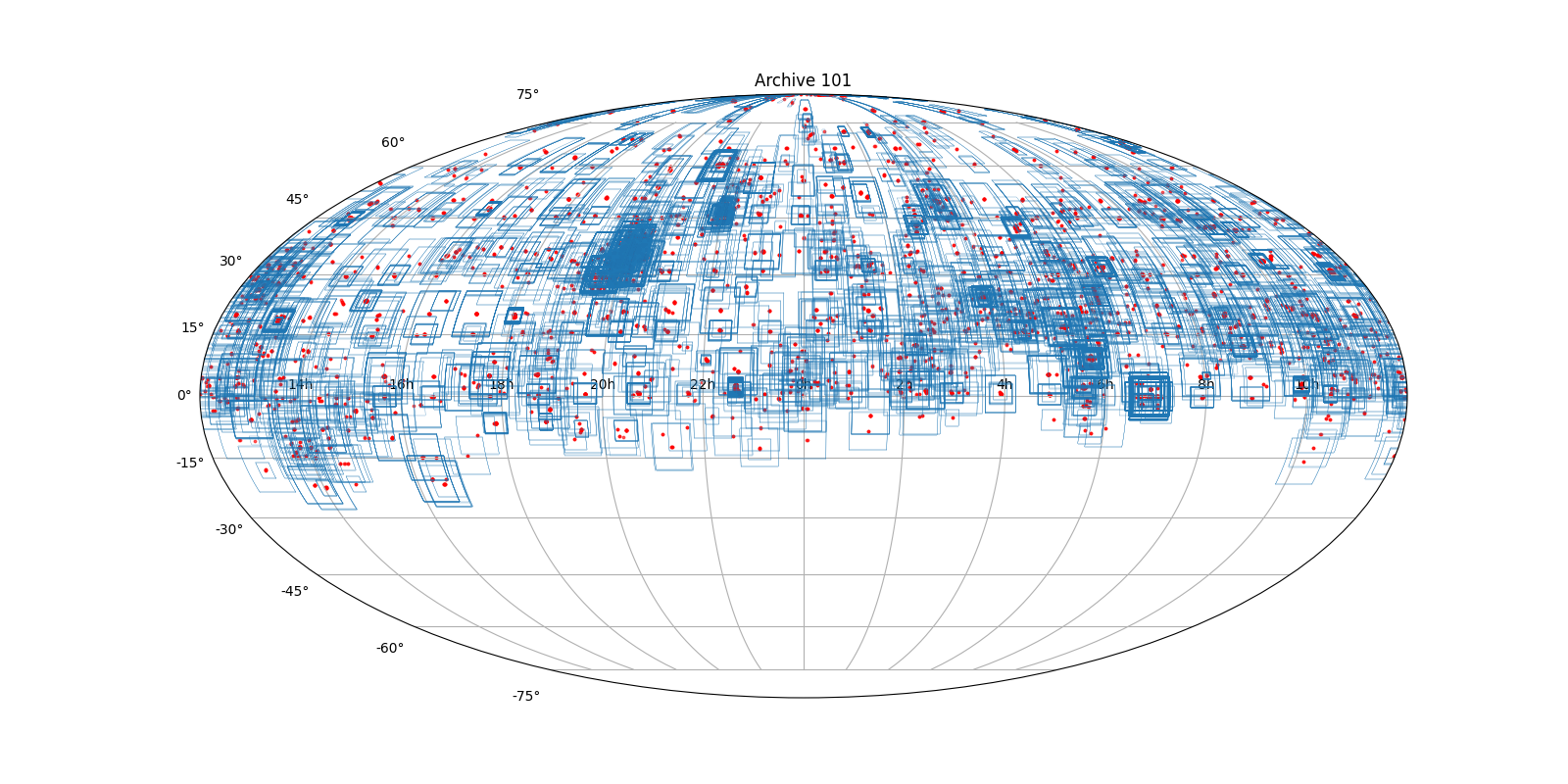}  
 \\[-11pt]
  Ross Camera, Potsdam, 64 plates (ID 5) &   Lippert-Astrograph, Hamburg, 8750 plates (ID 101)
  \\[1pt]
   \includegraphics[width=0.5\textwidth]{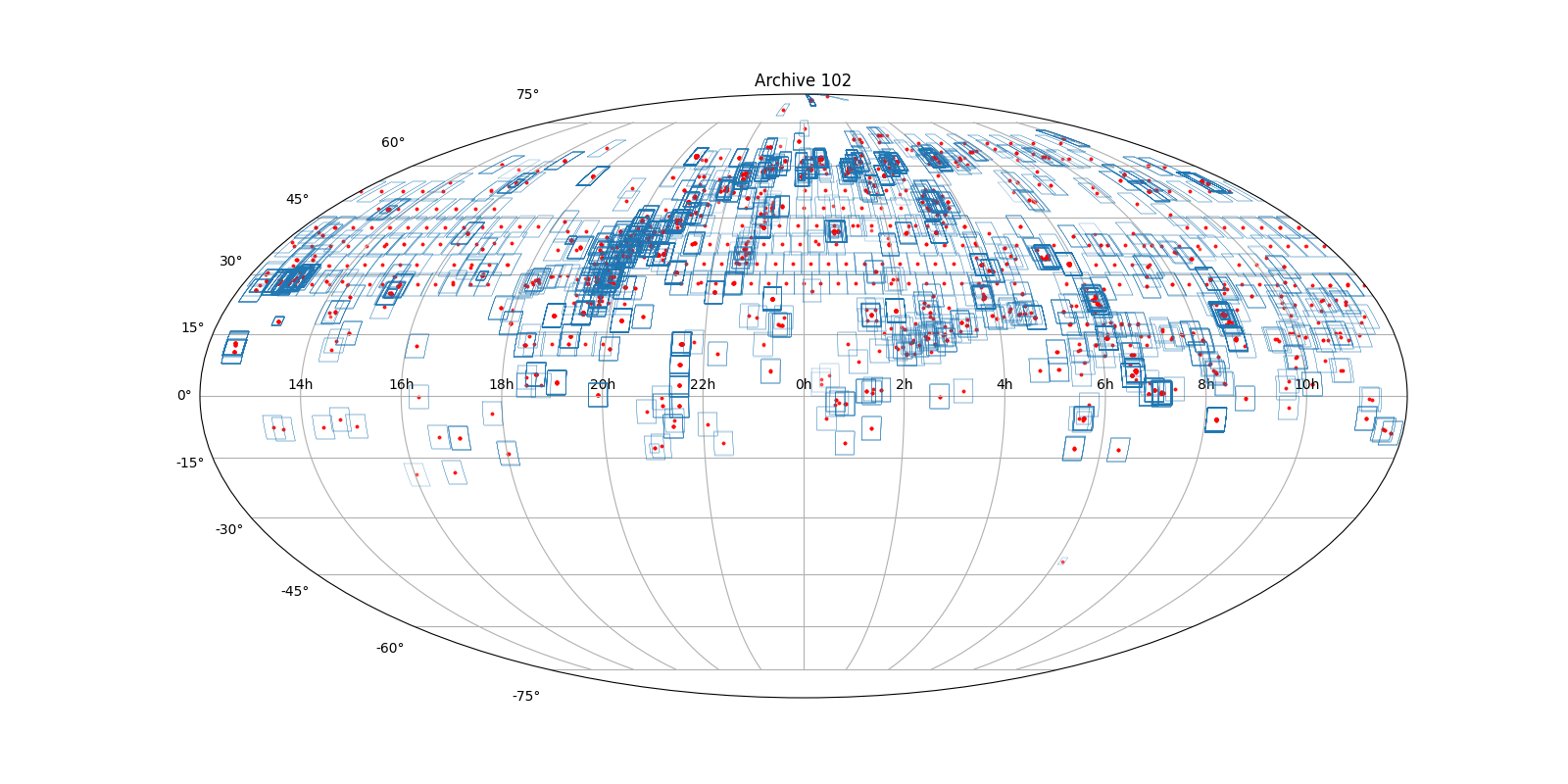} & 
   \includegraphics[width=0.5\textwidth]{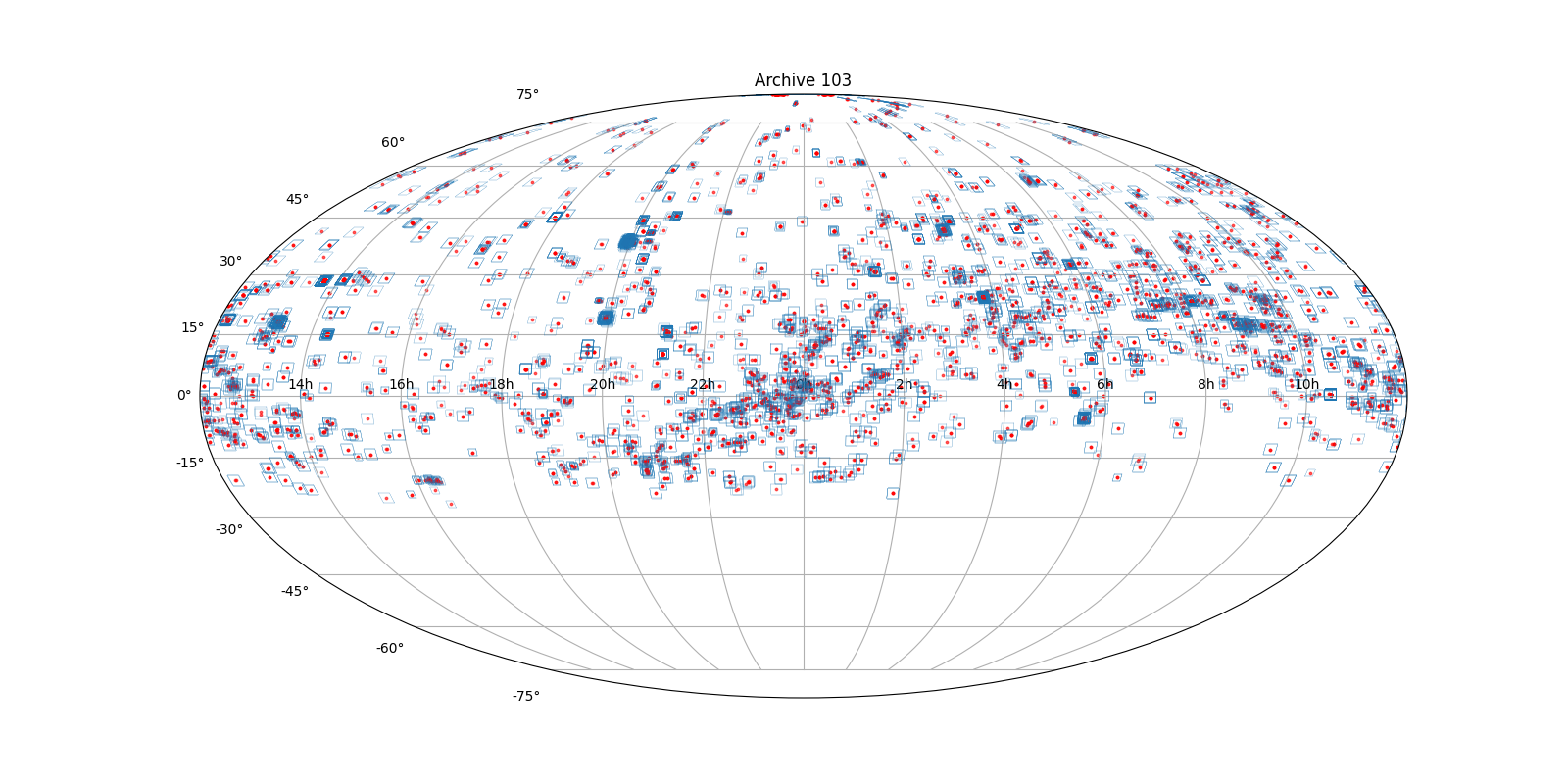} 
  \\[-11pt]
  Grosser Schmidt-Spiegel, Hamburg,  5323 plates (ID 102) & 1m-Spiegelteleskop, Hamburg, 7643 plates (ID 103)
  \\[1pt]

\end{tabular}
\end{figure}         
\pagebreak

\begin{figure}[ht]
\begin{tabular}{cc}   
   \includegraphics[width=0.5\textwidth]{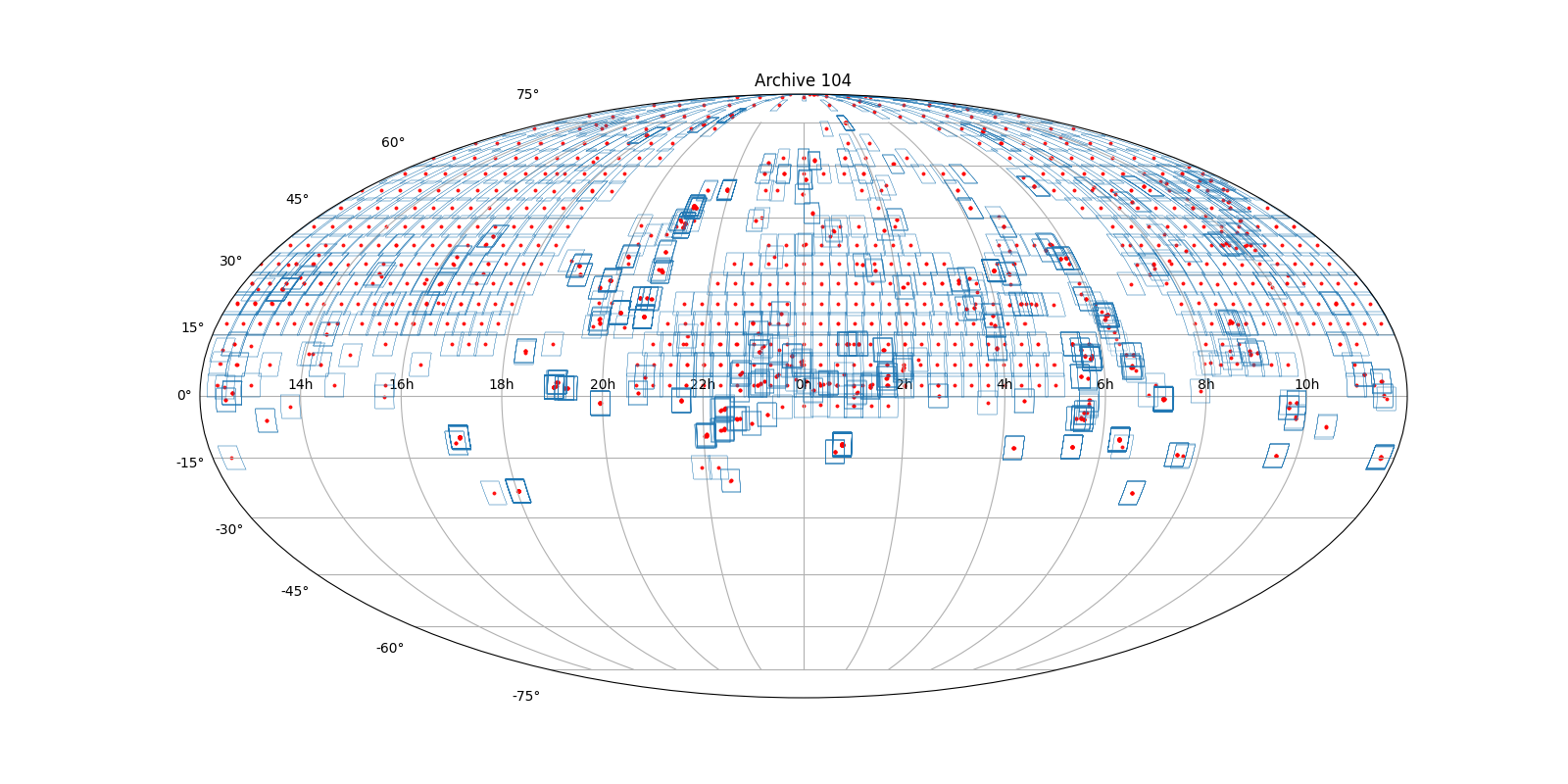} & 
   \includegraphics[width=0.5\textwidth]{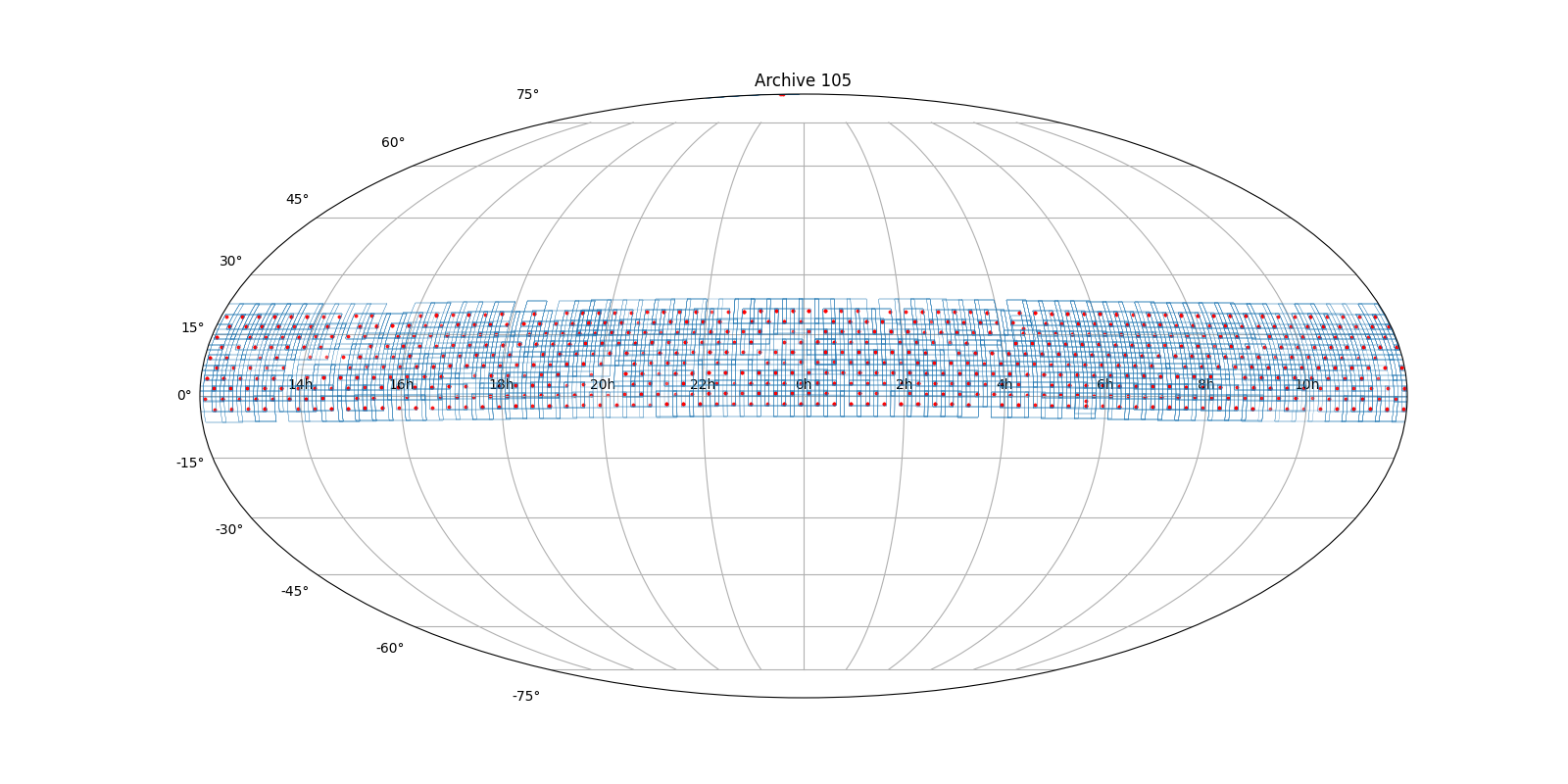} 
  \\[-11pt] 
  Hamburger Schmidt-Spiegel, Calar Alto,  3255 plates (ID 104)& AG-Teleskop, Bonn, 754 plates (ID 105)
  \\[1pt]
    \includegraphics[width=0.5\textwidth]{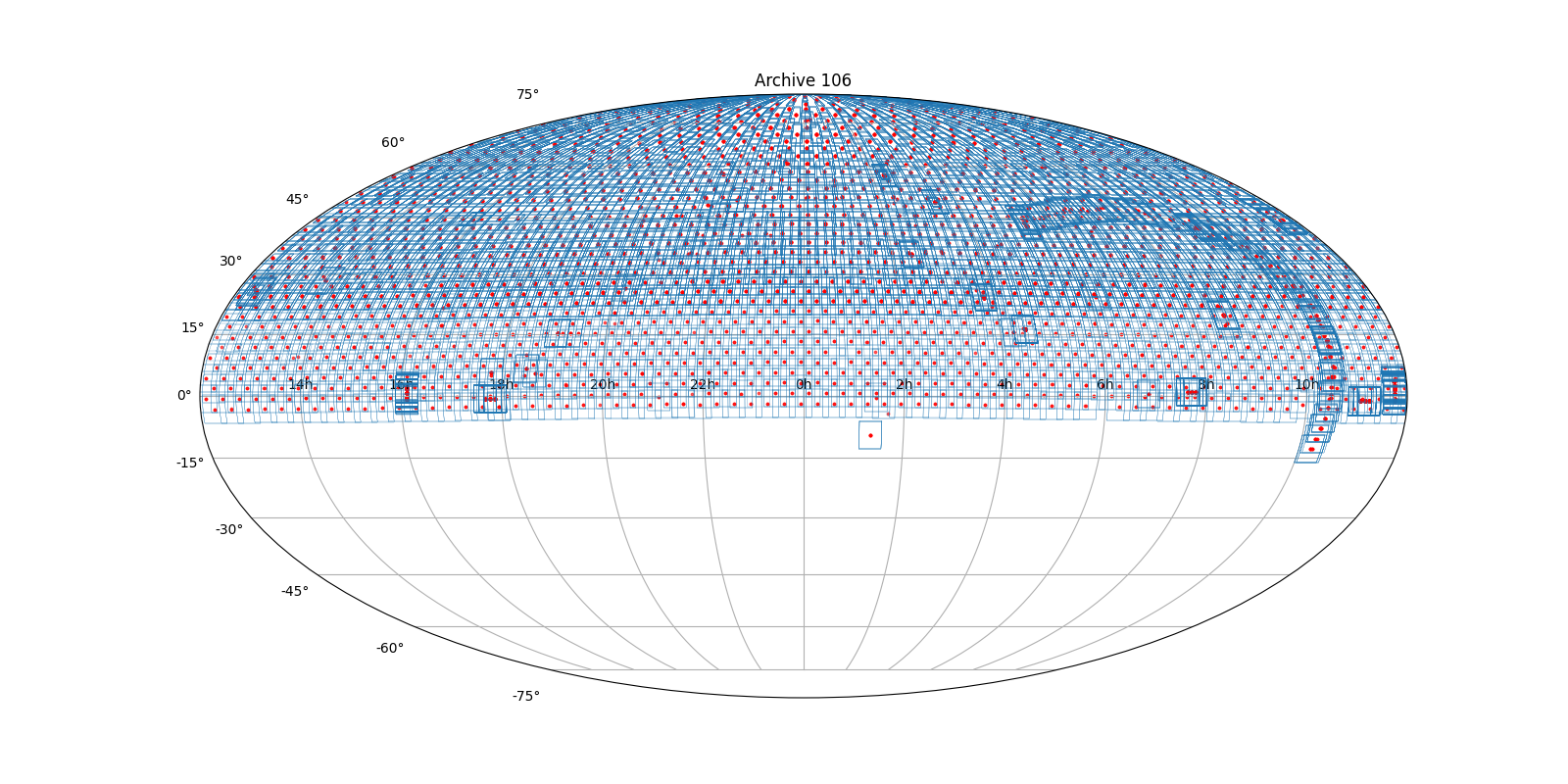} & 
    \includegraphics[width=0.5\textwidth]{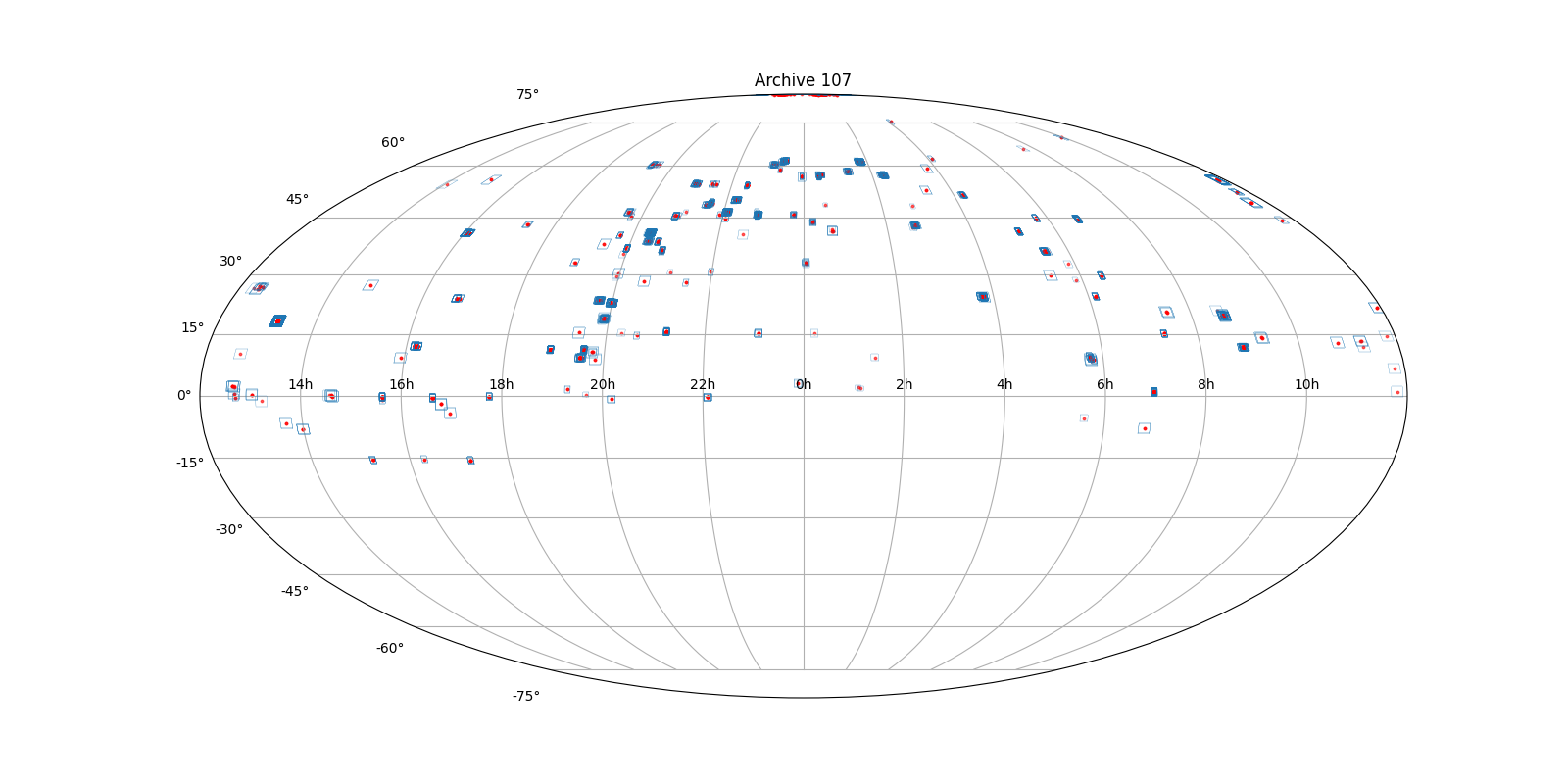} 
  \\ [-11pt]
  AG-Teleskop, Hamburg,  3529 plates (ID 106) & Doppel-Reflektor, Hamburg, 1642 plates (ID 107)
  \\[1pt]
   \includegraphics[width=0.5\textwidth]{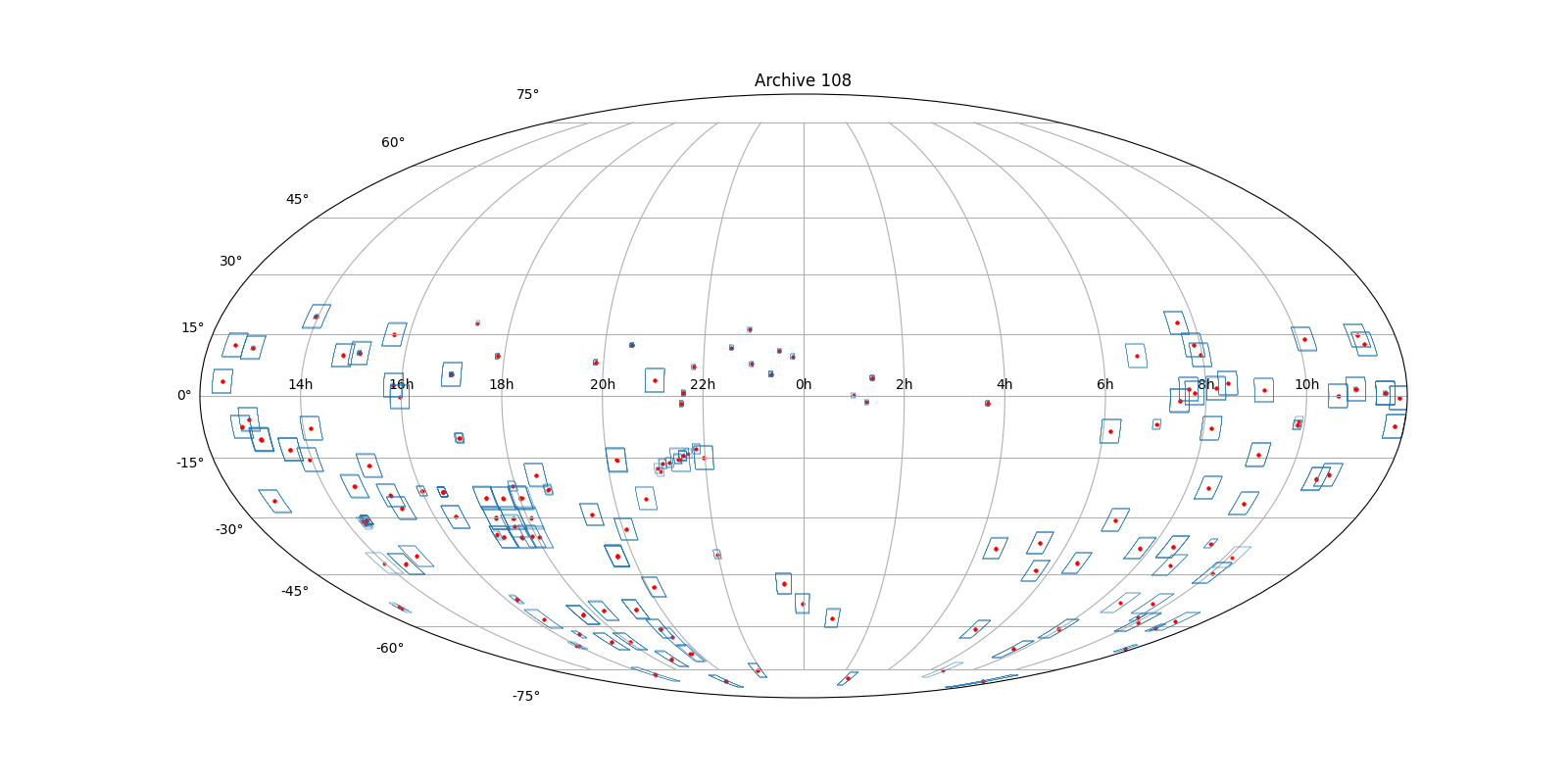} & 
   \includegraphics[width=0.5\textwidth]{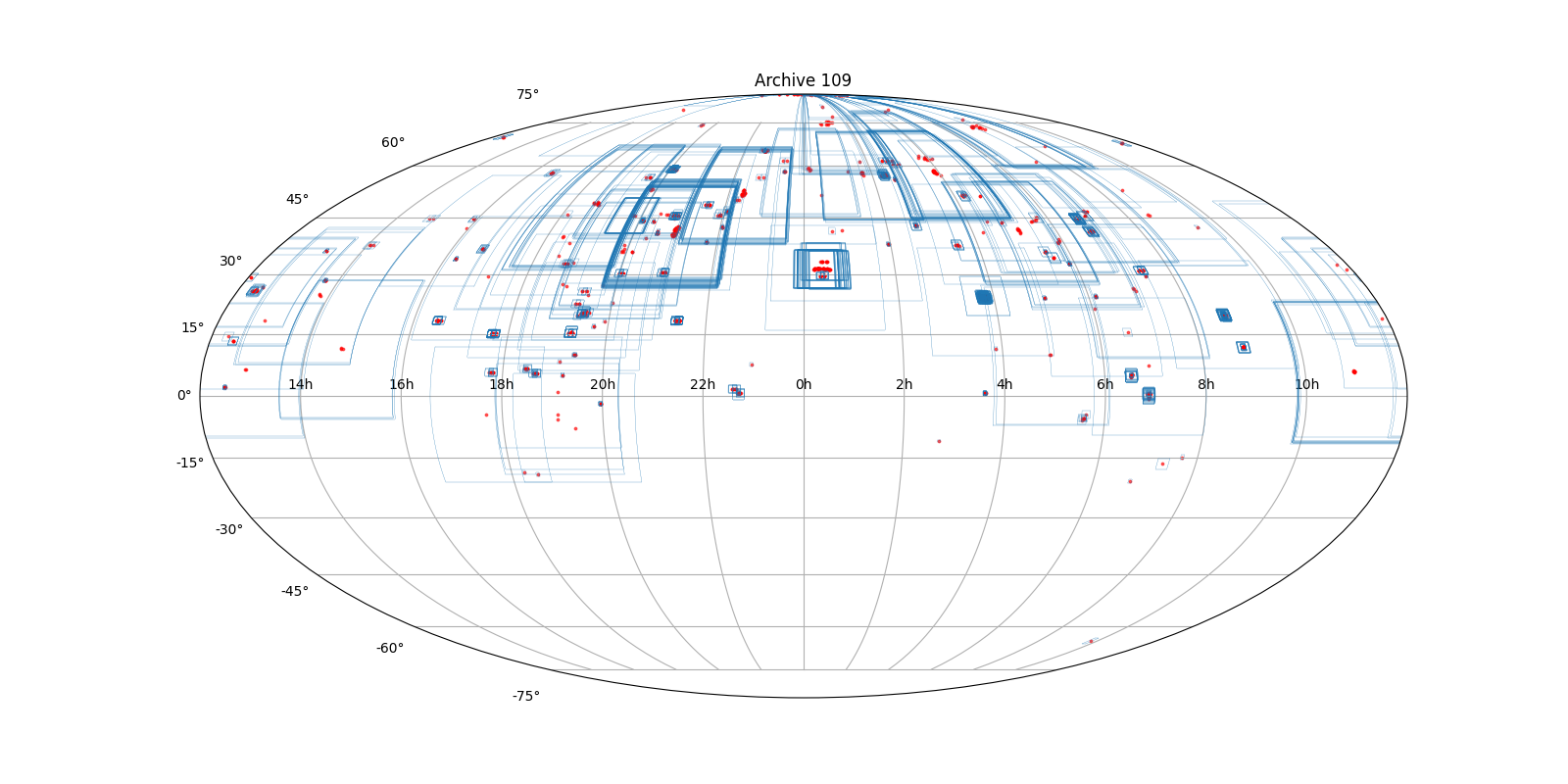} 
  \\[-11pt]
   ESO telescopes, La Silla,  2099 plates (ID 108) & Grosser Refraktor, Hamburg, 2602 plates (ID 109)
  \\[1pt]
   \includegraphics[width=0.5\textwidth]{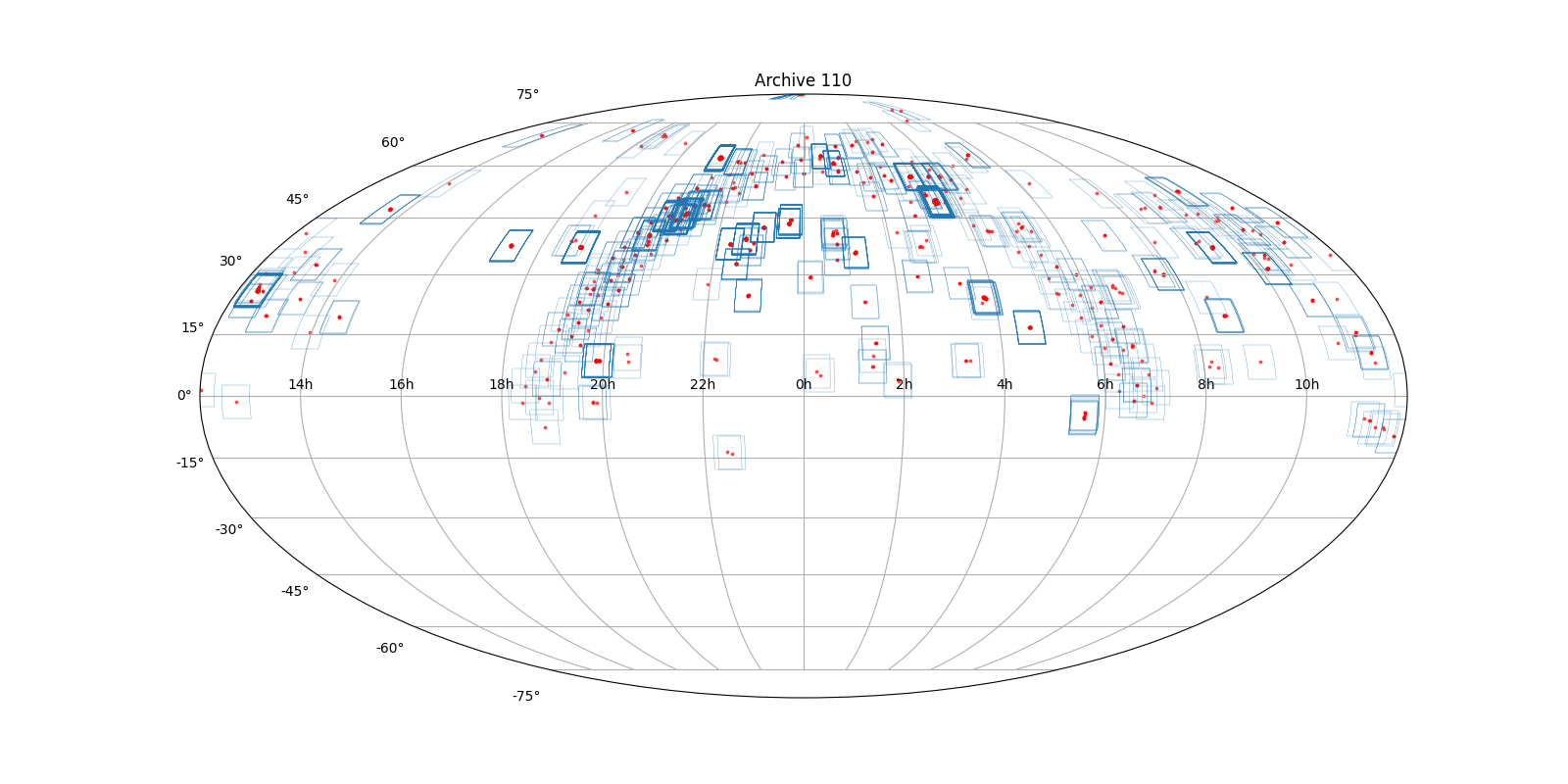} & 
   \includegraphics[width=0.5\textwidth]{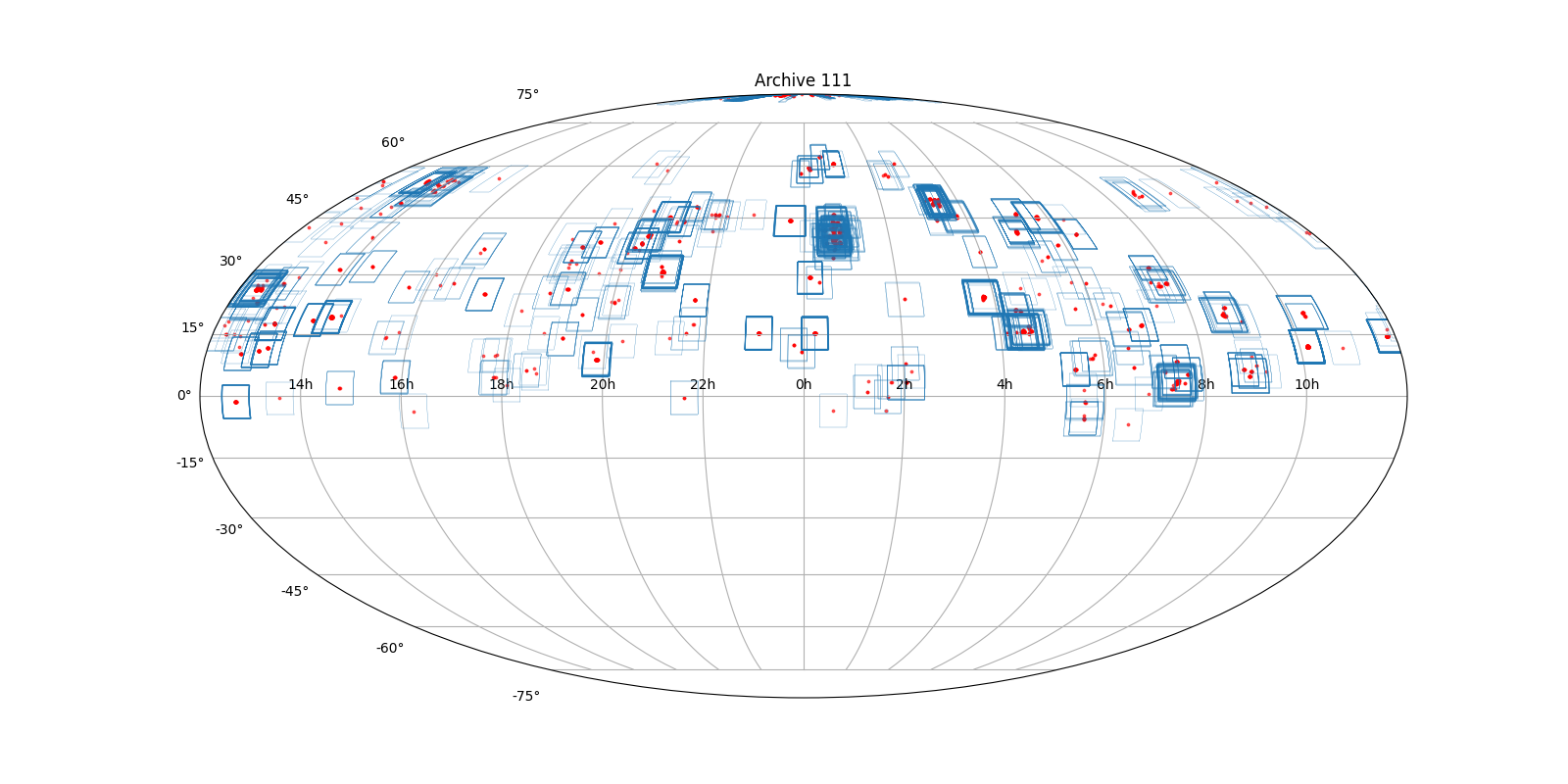} 
  \\[-11pt]
   Kleiner Schmidt-Spiegel II, Hamburg, 1566 plates (ID 110)& Schmidtsches Spiegelteleskop, Hamburg, 1746 plates (ID 111)
  \\[1pt]
   \includegraphics[width=0.5\textwidth]{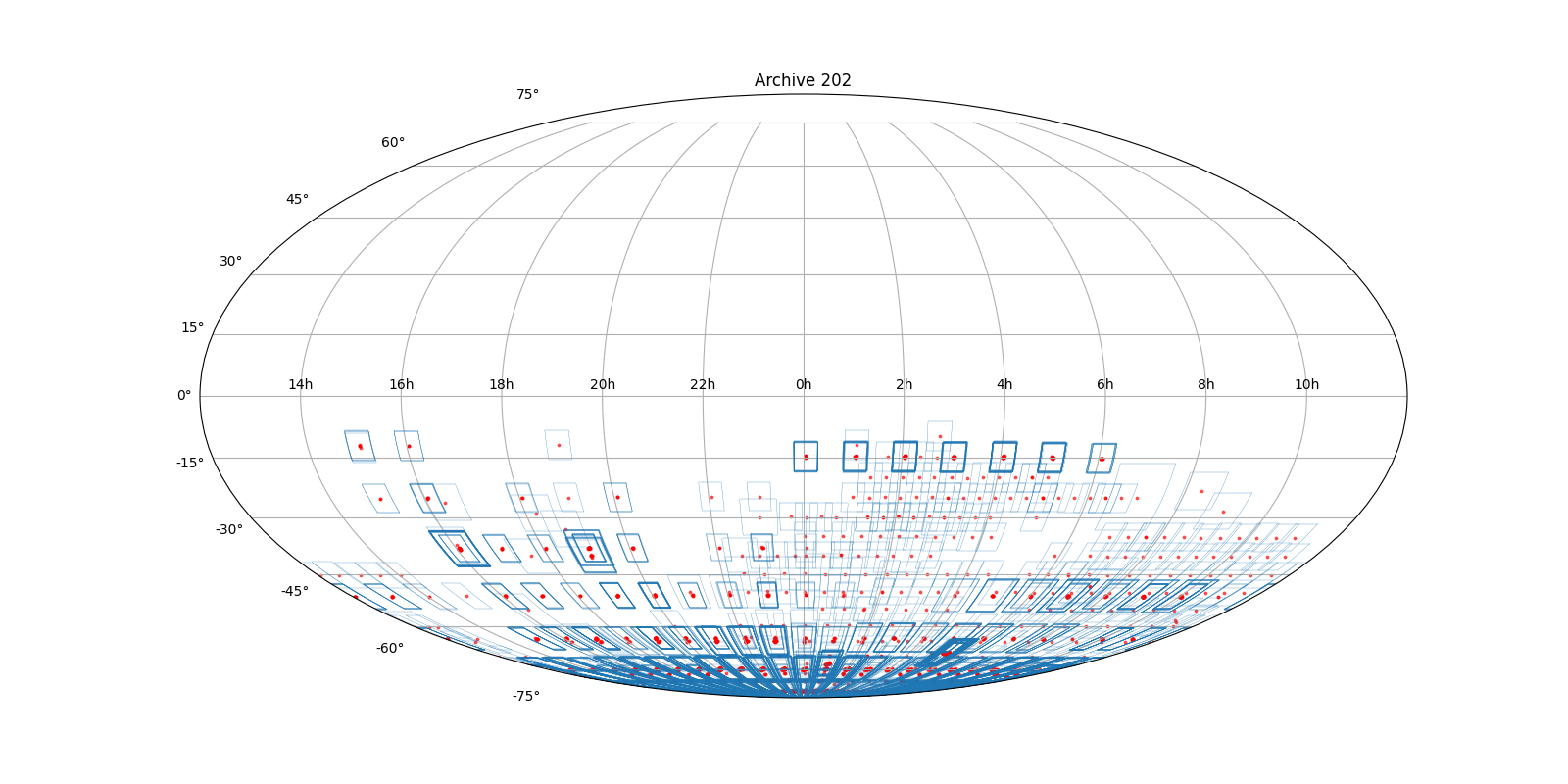} & 
   \includegraphics[width=0.5\textwidth]{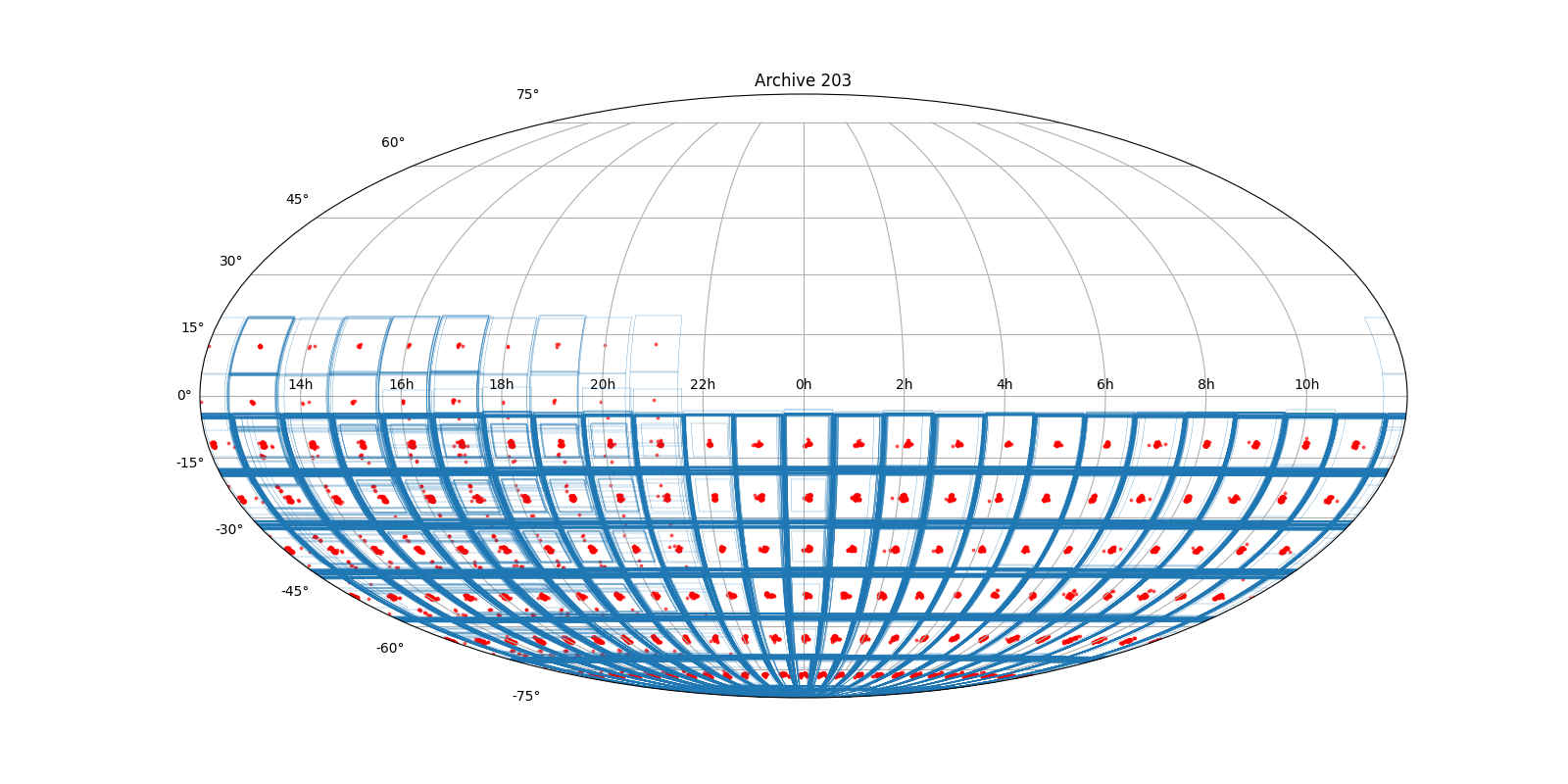} 
  \\[-11pt] 
  Metcalf Telescope, South Africa, 2718 plates (ID 202) & Bamberg Southern Sky Patrol, South Africa, 12565 plates (ID 203)
  \\[1pt]

\end{tabular}
\end{figure}         
\pagebreak

\begin{figure}[ht]
\begin{tabular}{cc}    
    \includegraphics[width=0.5\textwidth]{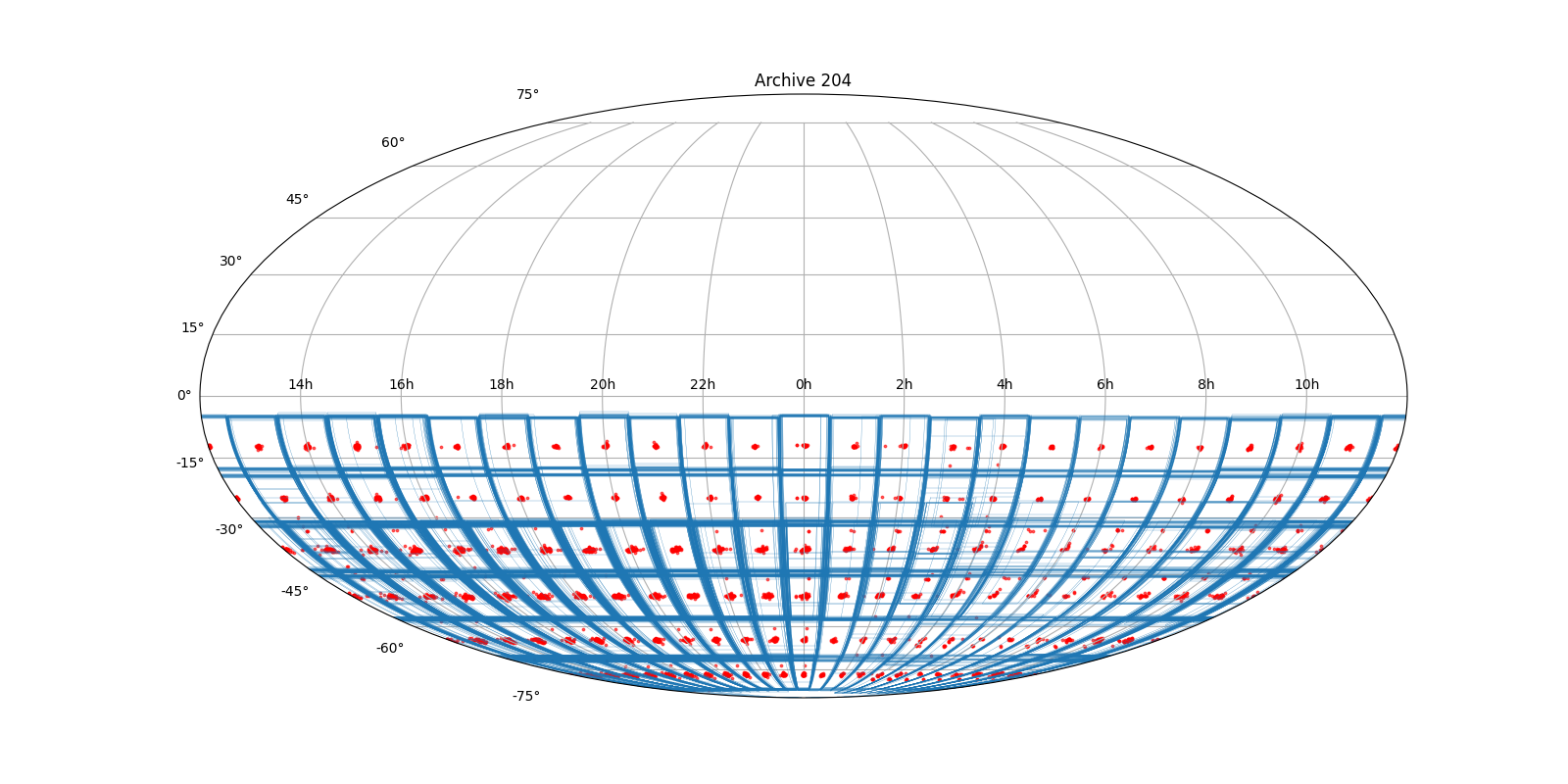} & 
    \includegraphics[width=0.5\textwidth]{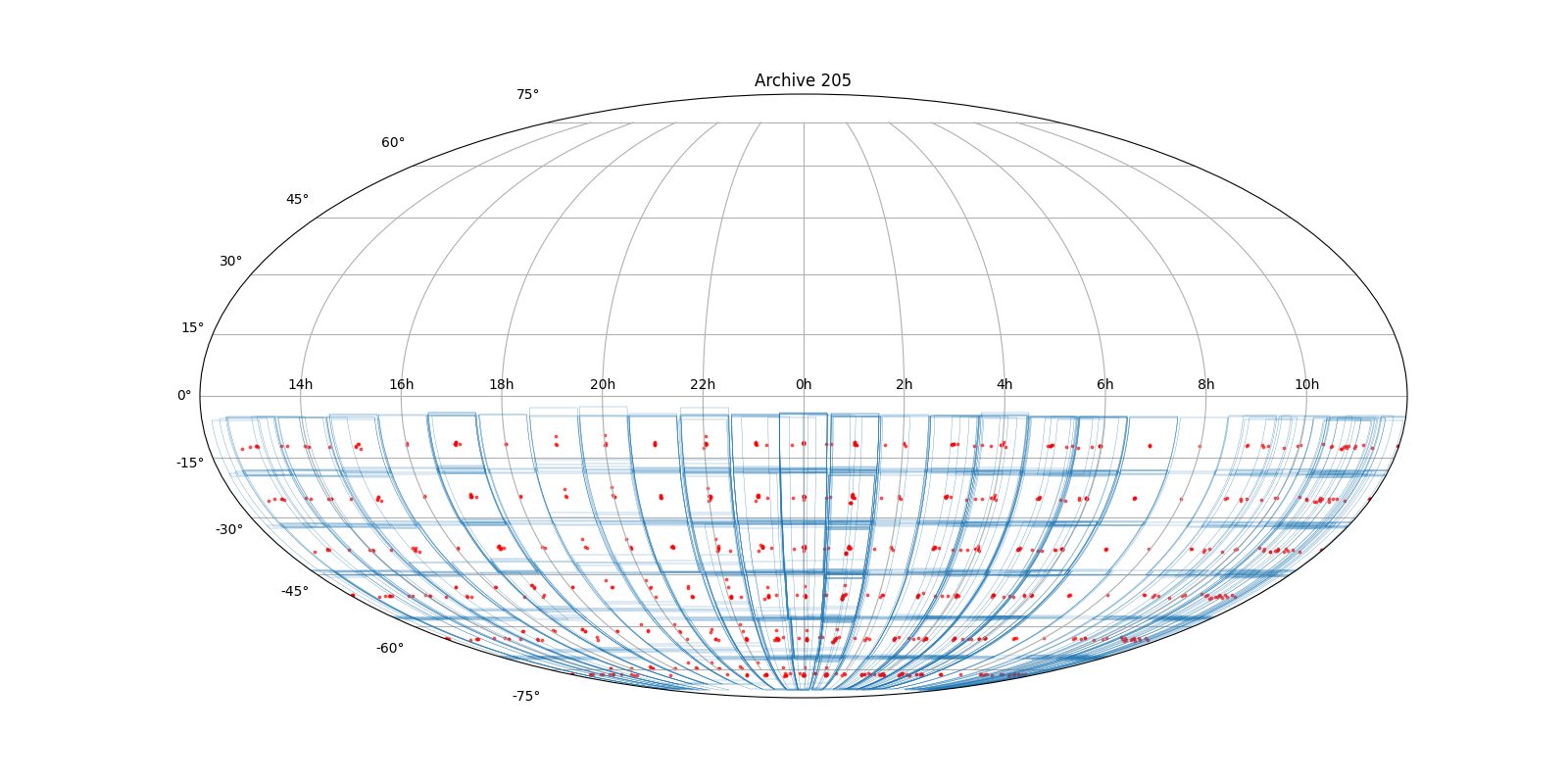} 
  \\[-11pt] 
  Bamberg Southern Sky Patrol, New Zealand, 5266 plates (ID 204) & Bamberg Southern Sky Patrol, Argentina, 870 plates (ID 205)
  \\[1pt]
    \includegraphics[width=0.5\textwidth]{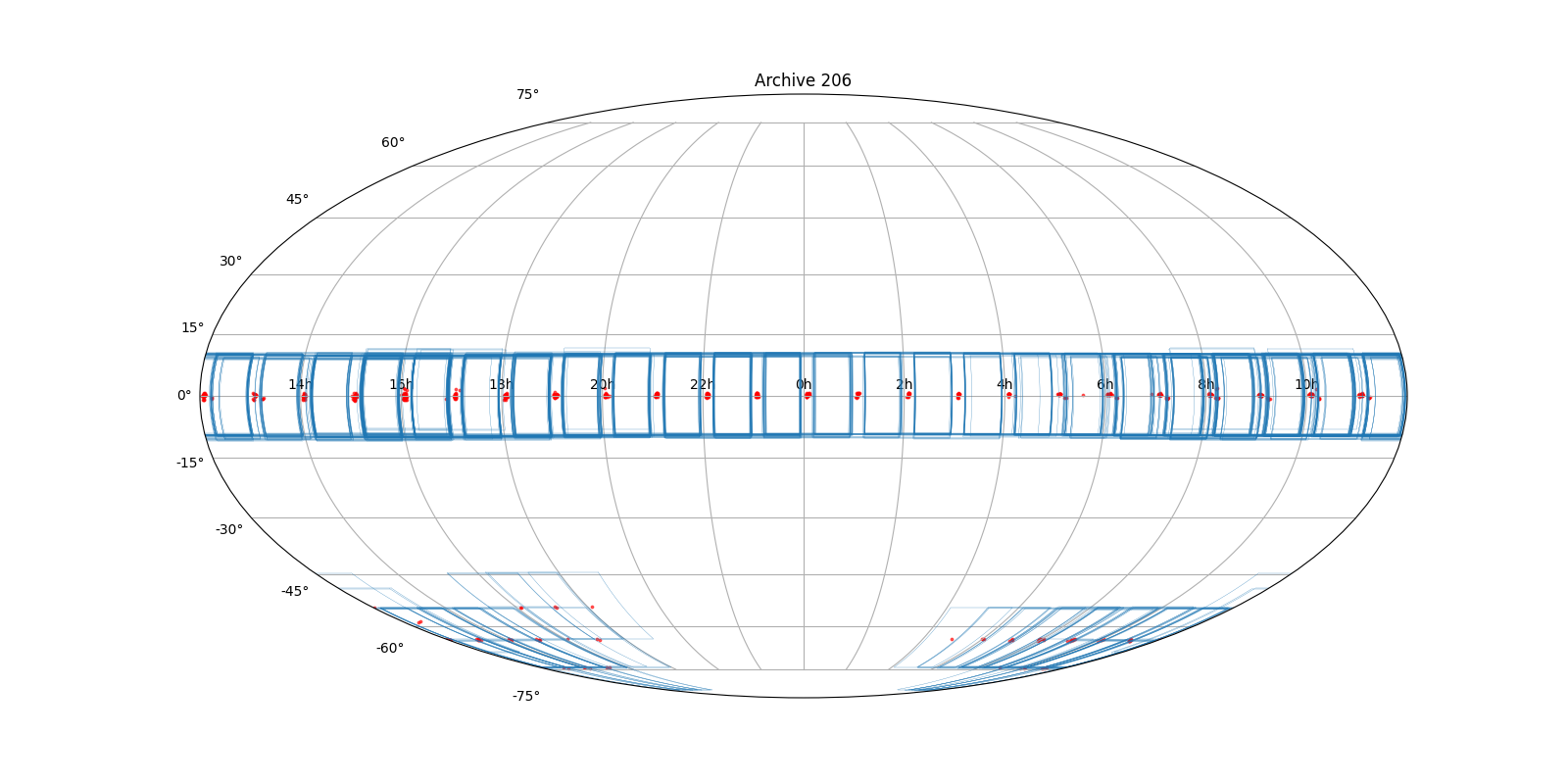} & 
    \includegraphics[width=0.5\textwidth]{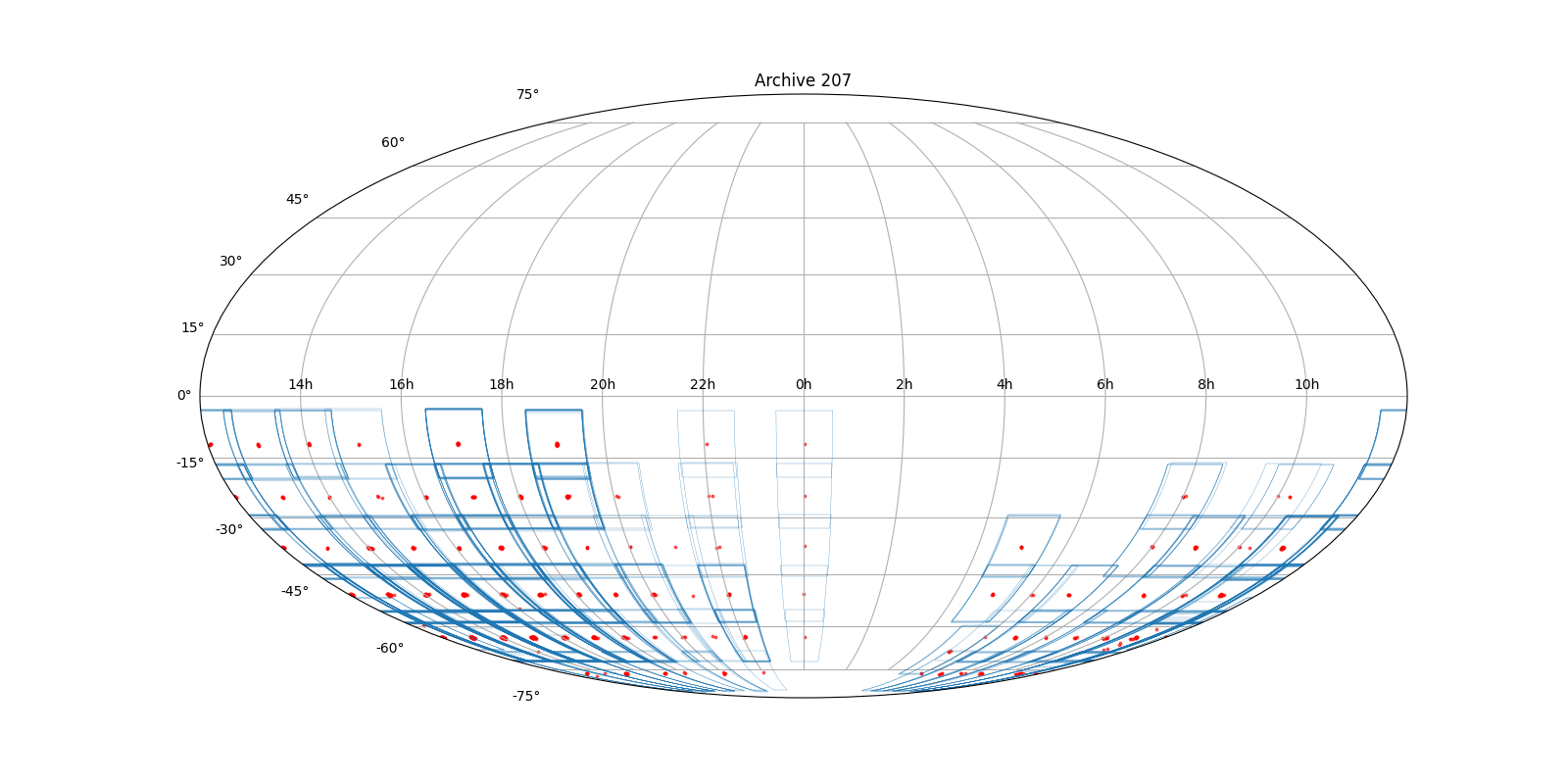} 
  \\ 
  Zeiss Objective, South Africa, 694 plates (ID 206) & Ross B Camera, South Africa, 698 plates (ID 207)
  \\[1pt]
     \includegraphics[width=0.5\textwidth]{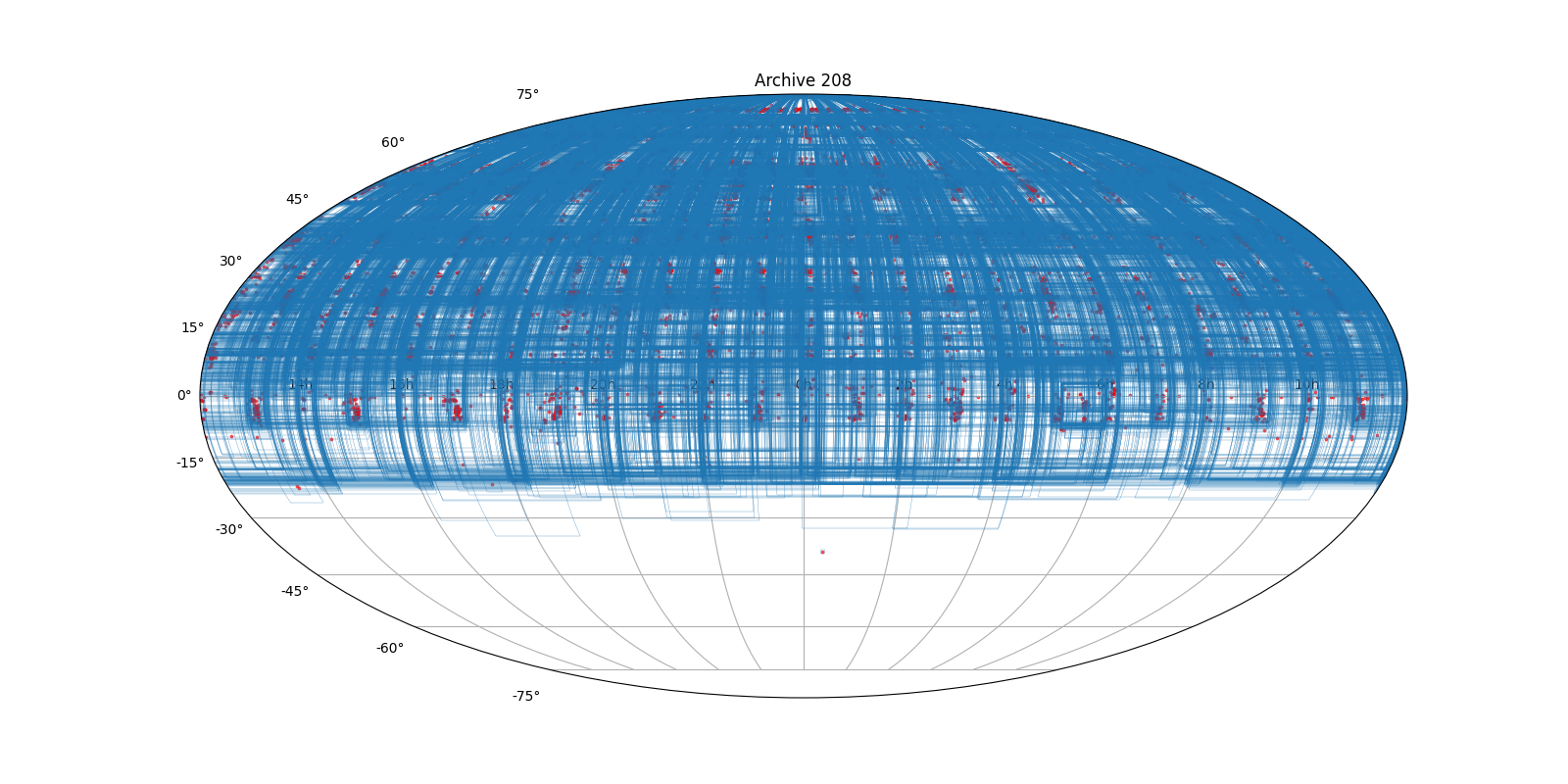} &   
         \includegraphics[width=0.5\textwidth]{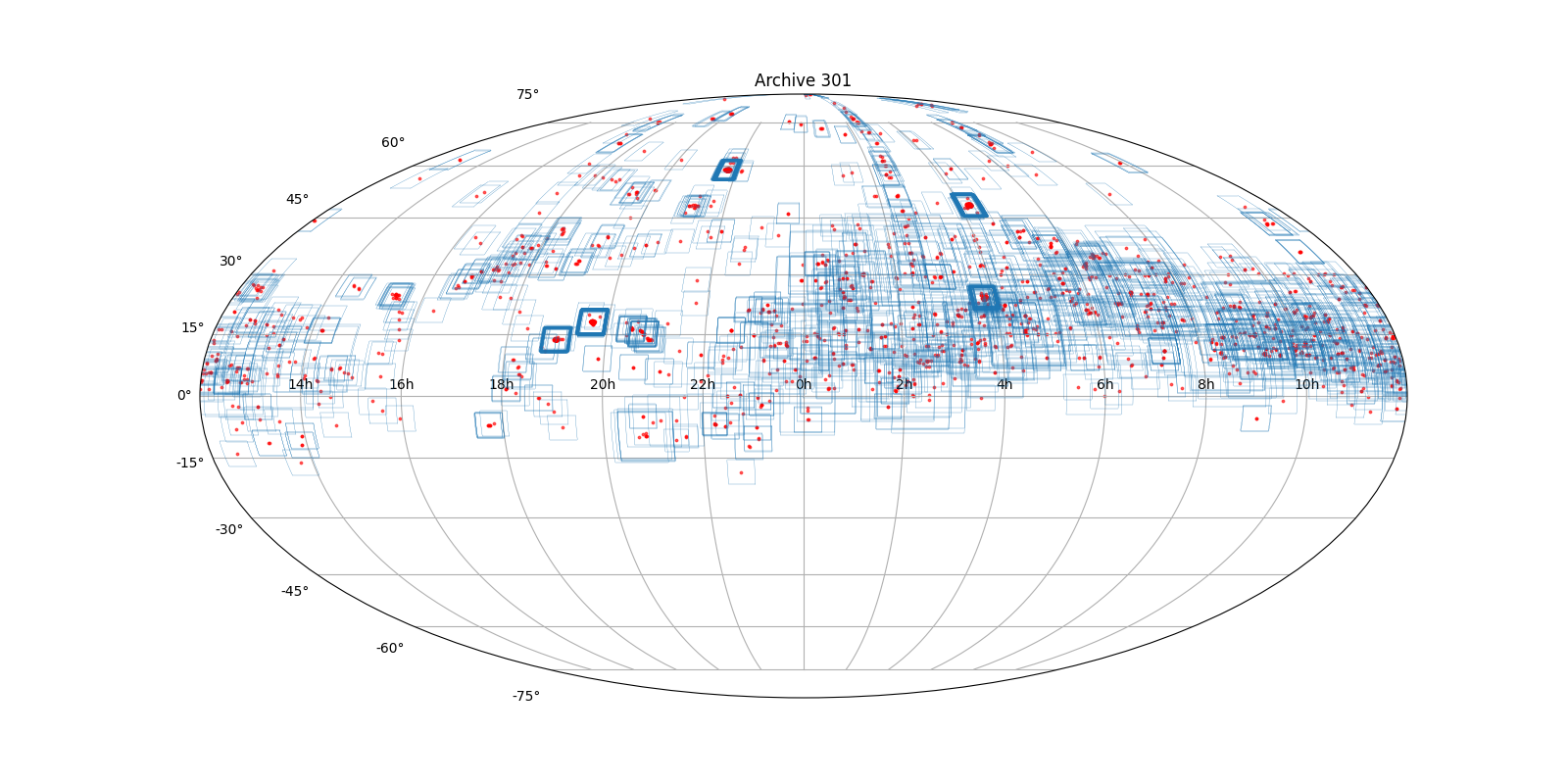} 
  \\[1pt] 
        Astrograph, Bamberg Northern Sky Patrol, 18865 plates (ID 208)& Tartu Old Observatory, Tartu, 2275 plates (ID 301)
  \\[1pt]

  \includegraphics[width=0.5\textwidth]{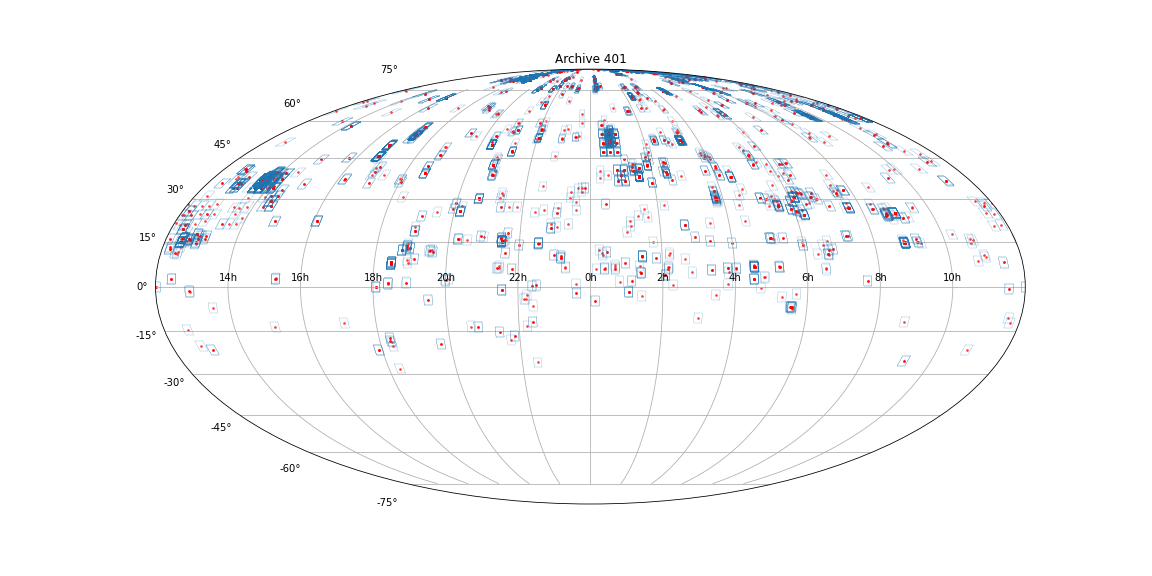} &
  \includegraphics[width=0.5\textwidth]{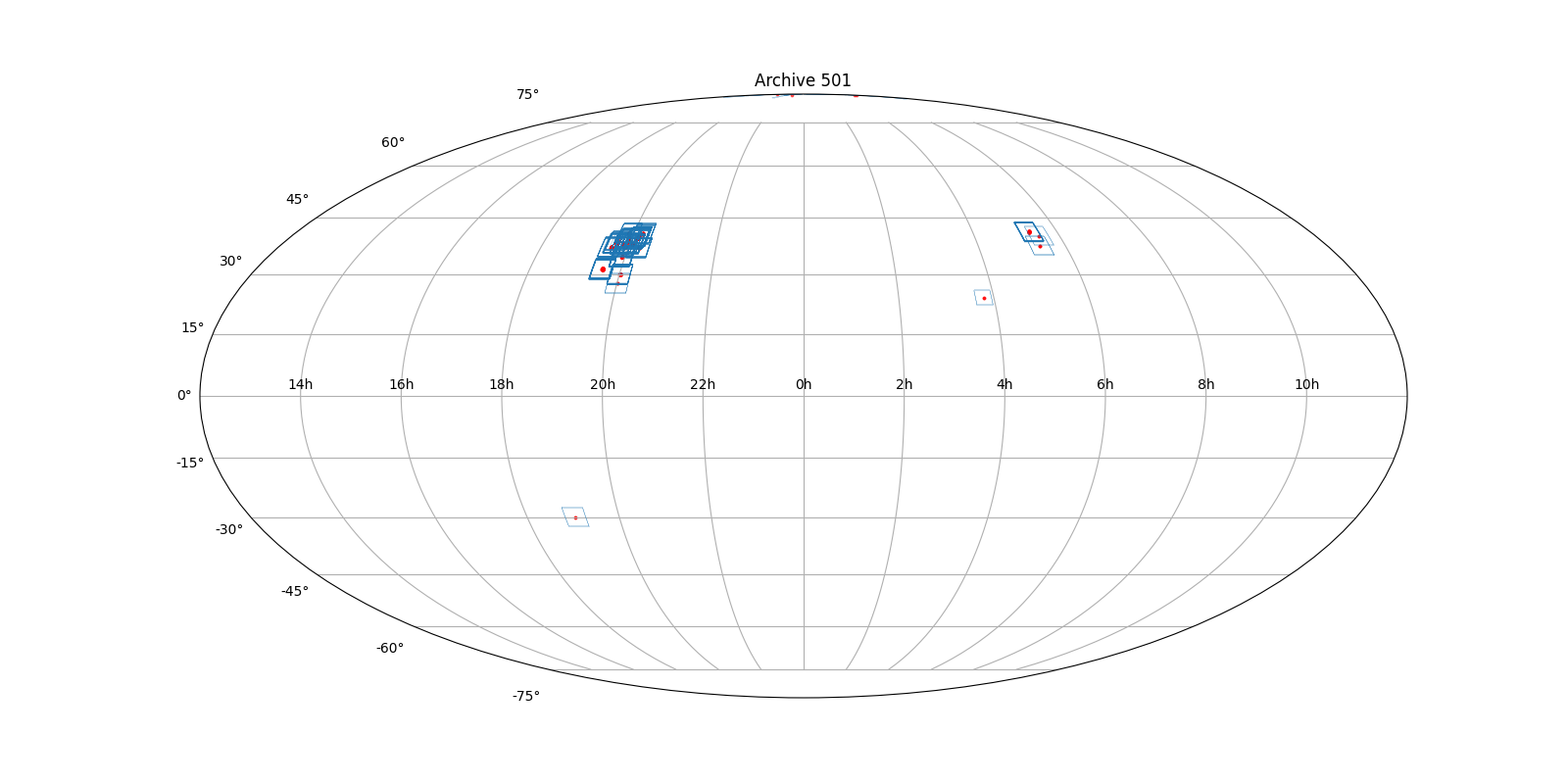}
 \\[1pt]
        Schmidt Telescope, Tautenburg, 4101 plates (ID 401) & Zeiss Double Astrograph, Vatican Observatory, 544 plates (ID 501)
 \\[1pt]

\end{tabular}
\vskip 0.5cm
\caption{Sky coverage of individual archives in APPLAUSE DR4 in equitorial coordinates (see Table \ref{dr4archives} for details). The combined sky coverage is shown in Fig. \ref{fig:equatorial}. }\label{fig:mollweide_appendix} 
\end{figure}         

\end{appendix}

\end{document}